# Centrifugation theory revisited: Understanding and modelling the centrifugation of 2D nanosheets

Stuart Goldie,[1] Steffen Ott,[2] Anthony Dawson,[3] Tamara Starke,[2] Cian Gabbett,[3] Victor Vega Mayoral,[4] Kevin Synnatschke,[2,3,5] Marilia Horn,[1,6] Jonathan N. Coleman[3,*], Claudia Backes[1,2*]

1) *Physical Chemistry of Nanomaterials and CINSaT, Kassel University, Heinrich-Plett Str. 40, 34132 Kassel, Germany*
2) *Applied Physical Chemistry, Heidelberg University, Im Neuenheimer Feld 253, 69120 Heidelberg, Germany*
3) *School of Physics and CRANN, Trinity College, Dublin 2, Ireland*
4) *Instituto Madrileño de Estudios Avanzados en Nanociencia (IMDEA), C/ Faraday 9, 28049 Madrid, Spain*
5) *Chair for Molecular Functional Materials, Dresden University of Technology, Stadtgutstr. 59, 01217 Dresden, Germany*
6) *University of Münster, Corrensstr. 3, 48149 Münster, Germany*

colemaj@tcd.ie, backes@uni-kassel.de

## Abstract

Size selection of liquid-dispersed 2D nanomaterials is a prerequisite for size-dependent studies in earlier stage research and for their targeted application in commercial settings. Centrifugation is the most widespread method for reliably sorting suspensions of polydisperse 2D nanosheets according to size. However, whilst centrifugation is effective, no *a priori* models are available to predict the outcome of centrifugation, making time consuming iterative experiments necessary. Here we present a simple model for the behaviour of 2D nanosheets during centrifugation and benchmark its predictions against experiments. This model uses simple expressions, specific to 2D particles, for the hydrodynamic radius, effective density and viscous resistance to generate the equation of motion of individual nanosheet during centrifugation. Critically, the equation of motion is then used to predict nanosheet size distributions within centrifugation products. This in turn leads to equations for easily measurable properties such as mean and maximum nanosheet sizes obtained during centrifugation-based fractionation. Comparison with experimental data demonstrates the robustness of this model for a range of 2D materials and solvent systems, and its ability to describe quite subtle effects. These results will enable more tailored size selection of nanosheets for specific applications and offer new mechanistic insights to optimise exfoliation conditions.

## Introduction

Liquid processable formulations of 2D materials hold great promise to capitalize on the remarkable properties offered by these exciting nanomaterials. By employing top-down exfoliation in liquids, a vast array of crystalline and powdered materials can be converted into nanosheets in colloidally stable liquid dispersions suitable for printing,[1,2] self-assembly,[3] chemical functionalization[4,5] and incorporation into polymer matrices.[6] Today, such methods offer variety in the size and shape of nanosheets produced: sonication-assisted liquid phase exfoliation (LPE) produces small, highly fragmented sheets;[7] shear-dominated LPE generally produce slightly larger sheets;[8] electrochemical methods can produce laterally larger and thinner sheets,[9] and chemical intercalation generally produces doped or chemically-modified nanosheets of similarly large aspect ratios.[10] The wide range of nanosheet shapes and sizes that can be produced is promising because it allows production methods to be matched with targeted applications. However, this breadth of nanosheet sizes that can be prepared, even within a single sample, can also pose many challenges in terms of accurate determination of lateral size and thickness distributions on the one hand and to the purification and isolation of the desired nanosheet size on the other.

All top-down exfoliation methods produce a poly-disperse mixture of nanosheet lengths and thicknesses, which have a correspondingly wide range of properties due to the size dependence of these materials.[11] To fully exploit such dispersions, a reliable method for targeting specific nanosheets sizes for use in specific applications is required. To-date centrifugation has remained the primary method of size selection for liquid dispersions of nanosheets because it is simple to implement, and works for all materials, regardless of their physical properties. Although simple preparative centrifuge experiments have long been known to separate larger nanosheets in the sediment from smaller nanosheets remaining in the supernatant, it has not been possible to predict the nanosheet sizes in each fraction without prior experimentation as will be outlined in more detail below.

To more precisely separate different nanosheet sizes, density gradient ultracentrifugation has been used on 1D carbon nanotubes and 2D nanosheets such as graphene and transition metal dichalcogenides (TMDCs).[12-15] In this process, the nanomaterial is dispersed in a density gradient medium and the experiment run for a long time until all nanosheets reach their neutrally buoyant positions. Since all particles have a bound hydration layer with a constant thickness, their effective density varies with particle volume and different sized particles can be extracted from different points in the density gradient. For 2D nanosheets, this allows for a sorting by thickness. It can provide exquisite control over nanosheet thickness but requires very long timescales and expensive equipment to produce relatively dilute samples of nanomaterials. For example, for monolayer $MoS_2$, a yield of 2 μg per separation has been reported.[14] Furthermore, the addition of density gradient media is required, often solutions of sugars, which need to be removed prior to further processing.

An alternative method for size separation using commonly available benchtop centrifuges is liquid cascade centrifugation (LCC).[16] This widely used method[17-24] has been discussed in recent reviews for graphene and metal chalcogenides, and reported for a range of materials including MXenes.[17,25] By sequentially increasing the centrifuge speed (or time) over multiple centrifugation steps and extracting the sediment from each step, a series of samples are obtained. These contain decreasing nanosheet size with each centrifugation run. Microscopy and spectroscopy studies of these samples clearly showed well-separated size distributions. However, different suspensions, for example with nanosheets of dissimilar density or solvents of differing viscosity, respond quantitatively differently to the same processing conditions. If these differences are not accounted for, it is problematic to benchmark new materials or production methods against established procedures. For example, particle sedimentation in viscous solvents or media is slower[26] which can lead to an overestimation of dispersed concentration after removal of unexfoliated material and hence yield and production rate.

To allow more reliable comparisons between solvent systems, density and viscosity have been used to normalise the relative centrifugal force applied to nanosheets.[17,27] A geometric mean sheet size approximation has also been applied to LCC to estimate the sheet size in each fraction.[28] However, no complete consideration of surfactant coating or the various geometric effects specific to the dimensionality of 2D nanosheet systems has been reported so that it is not possible to accurately predict the outcome of a centrifugation run, never mind size selection through a cascade of iterative runs.

Although analytical ultracentrifugation has been used to gather more information about the sedimentation of particles of various shapes, it has some limitations. For 1D nanoparticles, such as carbon nanotubes, analytical ultracentrifuge experiments have proved successful at characterising the effect of the surfactant on the buoyant density of the colloid. Specifically, the surfactant packing density and thickness around single walled carbon nanotubes in aqueous solution has been investigated by resolving the nanotube concentration as a function of position and time during centrifugation, and fitting this to a theoretical model describing nanotube sedimentation and diffusion.[29,30] However, the effect of the dimensionality of the particles on viscous friction was not accurately modelled. When applied to polydisperse samples of 2D materials, analytical ultracentrifugation has proved capable of determining the lateral sheet size distribution of monolayer graphene oxide.[31] Furthermore, a mathematical framework was developed to describe particle separation of multi-dimensional lognormal distributions[32] which lays the foundation to use analytical ultracentrifugation to determine the particle size distribution in terms of both thickness and length in a 2D nanosheet dispersion in the future. Nevertheless, while enormously powerful, such experiments require specialised ultracentrifuges capable of spatially resolving (optical) spectra from revolving centrifuge cells, that are not routinely available in research labs. Further, the determination of particle size distributions requires an accurate description of the shape-dependent sedimentation and diffusion behaviour which is currently lacking.

In the context of preparatory centrifuge experiments performed in the lab, and the development of larger scale production of 2D materials, knowledge of the sedimentation dynamics that can be easily calculated for new and existing materials would allow a more targeted approach to size selection. Especially if such predictions can be extended from the motion of an individual sheet to the change of the distribution of nanosheets within a dispersion in response to centrifugation. An improved theoretical framework that explicitly accounts for the dimensionality of 2D nanosheets would also aid size dependency studies and exfoliation comparisons for new methods and solvent systems. For example, such a framework would complement analytical ultracentrifugation, establishing it as an efficient tool to determine nanosheet size and thickness distributions in a dispersion without the need for microscopy statistics.

To solve this problem, we introduce a simple model incorporating hydrodynamic radius, effective density and viscosity, specific to nanosheets. This allows us to derive a sedimentation coefficient accounting for the particles shape and surfactant coating. A comparison of the predicted sedimentation behaviour to experimental data collected from 2D nanosheets separated by band sedimentation in a swinging bucket rotor proves the accuracy of the sedimentation coefficient. With this accurate description of the sedimentation velocity, we also show that it is possible to calculate the relative population changes in polydisperse mixtures and make predictions of average nanosheet size following a centrifugation cascade in the commonly used fixed angle rotors. Experimental data of >10 cascade centrifugation sets on a range of materials and in different solvent systems is well described by the model confirming the broader applicability.

## Results and Discussion

*Generating the equation of motion for nanosheets in a centrifugal field*

In a centrifugal field, mobile particles will experience a centrifugal force, viscous resistance from their motion through the fluid and the effects of buoyancy from the weight of solvent displaced. By assuming rapid acceleration to terminal velocity, Svedberg resolved these forces to express the sedimentation velocity, $v$, as:[26]

$$\frac{dr}{dt} = v = \frac{r\omega^2 m}{f}\left(\frac{\rho_{eff} - \rho_l}{\rho_{eff}}\right) \tag{1}$$

Here, $m$ is the particle mass, $r$ is the instantaneous distance of the particle from the centre of rotation and $\omega$ is angular velocity, such that $F_c = mr\omega^2$ is the centrifugal force. In addition, $f$ is the friction coefficient, which relates the viscous frictional force to particle velocity, $F_f = fv$.[33] The final term is the buoyancy correction factor where $\rho_{eff}$ is the effective density of the particle and $\rho_l$ is that of the liquid. The effective density should account for any tightly bound surfactant on the particle surface, slightly reducing the density of the object compared to the particle core.

Separating out the experimental conditions of centrifuge radius and speed, a sedimentation coefficient ($S$) can be defined that contains all the contributions from the material density, shape and size effects on its sedimentation velocity.

$$v = Sr\omega^2 \tag{2a}$$

Such that

$$S = \frac{m}{f}\left(\frac{\rho_{eff} - \rho_l}{\rho_{eff}}\right) \tag{2b}$$

We note that equations 1 and 2 are completely general and apply to all particles. Specifying these equations for a specific geometry requires modification of the parameters included in $S$. While sedimentation coefficients are well-defined for spheres, they are much harder to quantify for other geometries, especially when the particle is coated with a surfactant. To find $S$ for a 2D nanosheet, equation 2b must be corrected for the effect of surfactant on $\rho_{eff}$. In addition, $f$ must be modified considering the effect of geometry on both the hydrodynamic radius of the nanosheet and the extra viscous resistance felt by non-spherical objects (see below and SI). Although various papers have considered one or even two of these factors for quasi-2D particles,[13,15,31,32] none have correctly treated all three factors and, in particular, none have properly treated the extra non-spherical drag.

Here we consider all three effects to fully quantify sedimentation of 2D nanosheets. Full details of this derivation are provided in SI Section S2.1, but the key steps and assumptions are summarized here. We note that, as shown in SI Section S2.8, considering the Peclet number for nanosheets sinking in a centrifuge allows us to rule out the need to include diffusion effects in this system (except at unrealistically small values of ω).

To specify $S$ for 2D nanosheets, we first address the effective mass (mass times buoyancy correction factor). We use a simple geometric model illustrated in Figure 1A approximating the platelets as quasi-2D prisms of area $A$ and thickness, $h$, coated top and bottom with a surfactant layer of thickness $d$. Then, the effective mass of the surfactant-coated nanosheet is:

$$m\left(\frac{\rho_{eff}-\rho_l}{\rho_{eff}}\right)=\left(h(\rho_{NS}-\rho_l)+2d(\rho_S-\rho_l)\right)A \tag{3}$$

where $\rho_{NS}$, $\rho_l$ and $\rho_S$ refer to densities of nanosheet, solvent and surfactant layer respectively (see SI).

Next, we specify the effect of friction for 2D nanosheets. To do this, we note that for non-spherical particles, the frictional coefficient is given by $f=6\pi\eta c f_0$.[33,34] Here $c$ is the hydrodynamic radius of the nanosheet, $f_0$ is a geometric correction factor and $\eta$ is the viscosity of the fluid. We approximate $c$ as the radius of a sphere of equal volume to the surfactant coated nanosheet. In addition, we approximate $f_0$ by taking the expression given in refs [33,34] which describes the correction factor for an oblate spheroid and taking its limit where the spheroid lateral size far exceeds its thickness (see SI section 2.1). Combining these approximations gives an equation for the friction coefficient of a surfactant-coated nanosheet:

$$f=6\pi\eta\left(\frac{3A(h+2d)}{4\pi}\right)^{1/3}\frac{2}{\pi}\left(\frac{\sqrt{A}}{h+2d}\right)^{1/3}=12\eta\left(\frac{3}{4\pi}\right)^{1/3}A^{1/2} \tag{4}$$

The advantage of using the spheroid model for $f$ is that we obtain a relatively simple equation in contrast to the truncated approximations required by other models.[35]

Combining these contributions produces an expression for the sedimentation velocity of a 2D platelet, radially outwards away from the axis of rotation:

$$\frac{dr}{dt}=\left(\frac{(h\Delta\rho_{NS}+2d\Delta\rho_S)A^{1/2}}{12\eta(3/4\pi)^{1/3}}\right)r\omega^2=S(h,A)r\omega^2 \tag{5}$$

Where $\Delta\rho_{NS}$ is the density difference of nanosheet and liquid medium and $\Delta\rho_S$ the density difference of surfactant and liquid medium. If the distance from centre of rotation to a platelet is known at $t=0$, labelled $r_0$, solving this differential equation allows us to calculate the position of the platelet after time $t$:

$$r=r_0e^{S(h,A)\omega^2t} \tag{6}$$

Where all dependency on sheet size, shape and surfactant coating (if applicable) is included within the sedimentation coefficient $S$. For comparison, the sedimentation coefficient of spherical particles with a surfactant coating is given in Equation S2. We note that some experimental factors included within $S$, specifically the density and viscosity of the liquid medium, vary with temperature as well as composition. Rotation speed and time are explicitly included in the exponent, and $r$ and $r_0$ are the final and initial radial positions from the centre of rotation which depend on the centrifuge rotor used.

*Experimental confirmation of the equation of motion*

To experimentally confirm the validity of this model we used band (or rate zonal) sedimentation of $WS_2$, $MoS_2$ and graphene nanosheets prepared by LPE in aqueous surfactant; specific details in Methods, SI section 3. Band sedimentation is useful as it can be employed to cleanly separate materials with different sedimentation coefficients.[36-39] In this technique, a concentrated dispersion of mixed particles is floated above pure solvent, in this case a water and heavy-water mixture as illustrated in figure 1B, and the sample centrifuged at a certain speed and time so that the nanomaterial becomes spread throughout the vial after the run. Here, heavy-water establishes a subtle density gradient to stabilize the bands[40] (and facilitate the initial layering), but otherwise plays no role in the separation process. Such experiments are completed with swinging-bucket rotors so particle movement is parallel with the centrifugal field, down the tube. As such, the experiment allows the validation of equation 6, by determining the nanosheet sizes at different distances from the rotation axis (along with all other parameters encoded in $S$). Density and viscosity values of the liquid medium in the centrifuge tube were found from the average composition, discussed in SI Section 4.1.

The bands from which the technique gets its name are formed when different sized particles sediment at different rates. If the sizes are sufficiently different, horizontal 'bands' are observed down the tube, and from the distance these bands travel, sedimentation coefficients can be calculated. Typical dispersions of 2D materials do not have such discrete sizes for distinct bands to form leading to a broad spread in final nanosheet positions. However, fractions can be removed from specific radii ($r$) down the tube and the nanosheet average size or size distribution measured.

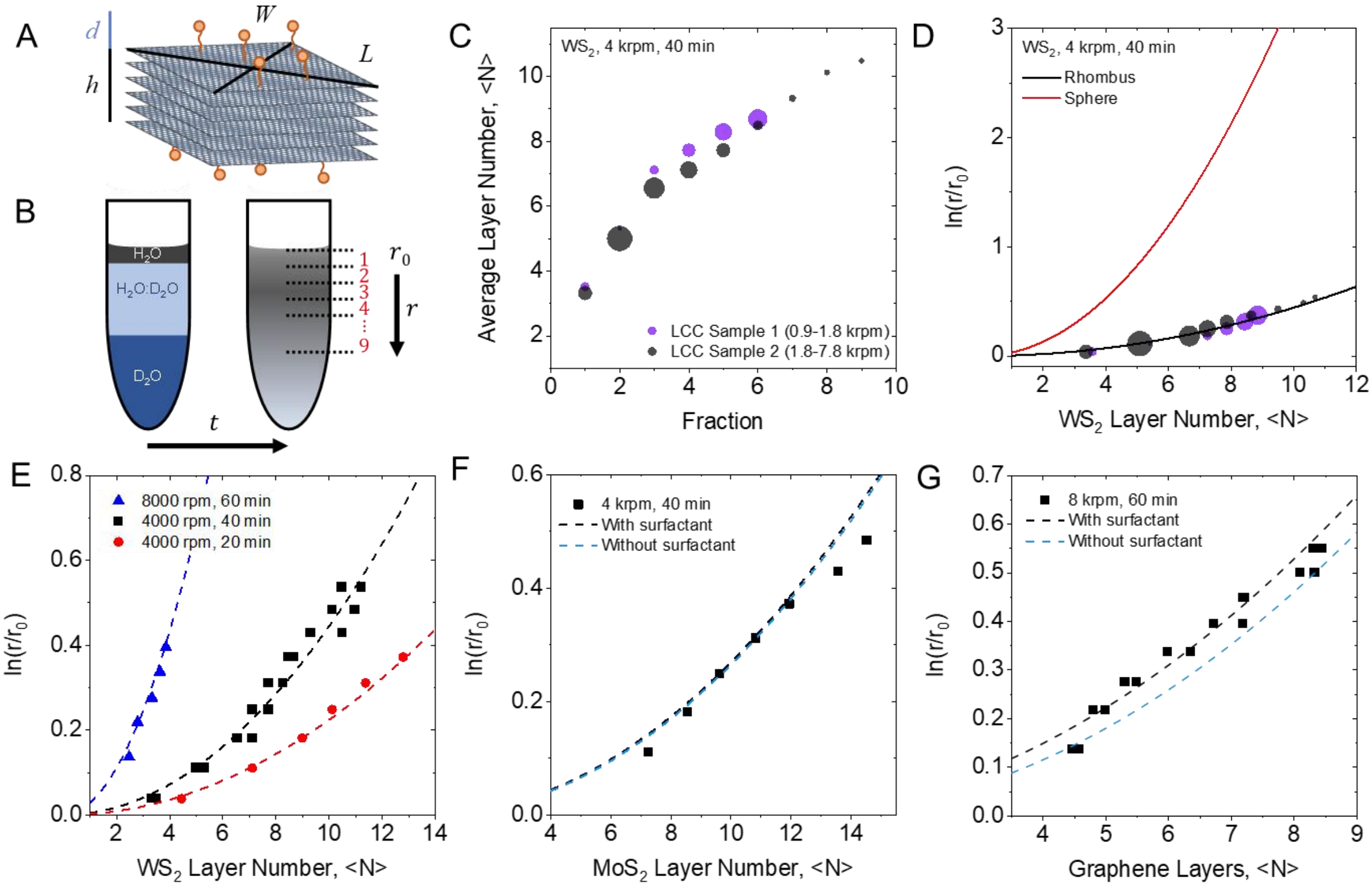

***Figure 1: Band sedimentation (swinging bucket rotor).*** *A) Schematic of the rhombohedral nanosheet with surfactant coating.*

*B) Schematic of the band sedimentation experiment indicating the withdrawal of fractions from top to bottom after centrifugation.*

*C) Plot of average layer number (of $WS_2$) of the fractions extracted at different positions in the centrifuge tube after centrifugation at 4 krpm for 40 min. The differently coloured datasets were obtained from $WS_2$ dispersions with different initial size/thickness distribution and the data point diameter is proportional to concentration.*

*D) Same data as in C presented as plot of $ln(r/r_0)$ as function of $WS_2$ layer number. The solid lines represent the expectation from theory with spherical platelets (red) and rhombus (black).*

*E-G) Plot of $ln/r/r_0$ as function of nanosheet layer number. Dashed lines show the expectation from theory. E) $WS_2$ dispersions centrifuged for different times. F) $MoS_2$. G) Graphene.*

Using established metrics from UV/VIS extinction spectroscopy,[41,42] we determined the average nanosheet thickness and concentration in each fraction (full spectra are shown in SI Section S4.2). This is visualized for $WS_2$ samples in Figure 1C where we note the steady trend of increasing sheet layer number further down the tube. Data from two different samples are shown, each subjected to different pre-processing to select different size distributions before the band sedimentation. Using different size distributions of sheets clearly has no bearing on the sedimentation dynamics, since the average thickness in each fraction is consistent between samples. But the concentrations measured, shown by the area of the data points, reflects the different make-up of the two pre-size selected samples.

For band sedimentation, the distance of the top layer from the rotation axis is $r_0$ and the distance at the midpoint of each fraction is $r$. This allows us to plot this data to compare experiment with the theoretical sedimentation coefficient derived above. However, thus far we have considered 2D materials with an area and thickness, which results in a multidimensional equation. This can be simplified using average aspect ratios that are known for common materials prepared by LPE.[43] This is because thermodynamic studies of the exfoliation process have shown that the sheet scission and exfoliation events follow principles of equipartition[44]: as energy is applied to break bonds, the proportion of inter- and intra-layer bond breakages is a function of their relative bond strengths. For this reason, LPE nanosheets (such as those investigated here) have consistent aspect ratios of length-to-thickness and length-to-width.

When discussing 2D nanosheets it is also more useful to consider the layer number, $N$, rather than absolute thickness. Using the crystallographic thickness ($d_0$), the nanosheet thickness ($h$) in the previous expression can be replaced by $h = Nd_0$. Additionally, we write the mean aspect ratios as length to thickness, $k_{Lt} = \langle L/h \rangle$, and length to width, $k_{lw} = \langle L/W \rangle$. Then we model the 2D nanosheets as rhombohedral prisms of length, $L$, and width, $W$, such that nanosheet area is $A = LW/2$. Rearranging Equation (6), and writing $A$ in terms of $h$ via the aspect ratios allows us to reduce the dimensionality of the sedimentation coefficient, such that Equation (6) becomes:

$$ln\left(\frac{r}{r_0}\right) = \frac{(Nd_0)^2(\rho_{NS} - \rho_l) + 2dNd_0(\rho_s - \rho_l)}{10.5\eta \frac{\sqrt{k_{lw}}}{k_{Lt}}}\omega^2 t \qquad (7)$$

***Table 1:*** *Relevant parameters of the materials used for band sedimentation experiments. Nanosheet aspect ratios were determined from AFM statistics and averaged over all nanosheets counted in the context of ref* [44]*. Solvent density and viscosity used were $\eta$ = 7.8 $\times 10^{-4}$ kg m$^{-1}$ s$^{-1}$ and $\rho_l$ = 1060 kg m$^{-3}$ based on average composition,*[45] *see Section S4.1 for details.*

| **Materials** | $\rho$ / kg m$^{-3}$ | $d_0$ / Å | $k_{lw}$ | $k_{Lt}$ |
|---|---|---|---|---|
| Graphene | 2260 | 3.5 | 2.4 | 180 |
| $WS_2$ | 7500 | 6.3 | 1.8 | 46 |
| $MoS_2$ | 5060 | 6.2 | 1.8 | 49 |

Plotting Equation (7) for rhombohedral nanosheets using the parameters in table 1 against a conventional hard sphere approximation (Equation S2), in Figure 1D, clearly highlights the importance of accurately describing nanoplatelet shape. The rhombohedral model was found to be accurate over a range of centrifuge speeds and conditions for $WS_2$, $MoS_2$ and graphene, shown in Figure 1E-G. As expected, at higher rotation speeds the nanosheets sink faster allowing greater separation of thinner sheets. We also note the lower density of graphene sheets required much higher rotation speeds to achieve a similar sheet separation.

When modelling low density nanosheets such as graphene (or hBN), to maximise accuracy it is important to include the surfactant layer in the effective density. Estimates for surfactant density and surfactant layer thickness ($\rho_S$ = 1595 kg m$^{-3}$, $d$ = 4.25 Å) were taken from an analytical ultracentrifugation study on single walled carbon nanotubes.[29] Since carbon nanotubes have similar surface properties to graphene we find good agreement with our experimental data. Plotting Equation (7) for graphene neglecting the surfactant, i.e. setting $\Delta\rho_S$ to zero, predicts a slightly slower sedimentation than experimentally measured, illustrated in Figure 1G. However, excluding this surfactant contribution makes little difference to the predicted movement of high density TMDCs, as illustrated for $MoS_2$ in Figure 1F where Equation (7) is plotted with and without surfactant.

*Population distributions in sediment and supernatant*

These band sedimentation experiments therefore validate our size and shape model for nanosheet motion and illustrate the important parameters for different materials. With an accurate model to describe the sedimentation coefficient of 2D nanosheets, it is possible to calculate the relative motion of a particle from a known starting position in a swinging bucket rotor. However, centrifugation is more commonly used on nanosheets homogeneously distributed throughout the centrifuge tube prior to centrifugation to separate a sediment, sometimes termed a pellet, at the bottom of the tube from the supernatant remaining above, illustrated in Figure 2A. Here, we term this homogeneous centrifugation to distinguish between different centrifugation scenarios.

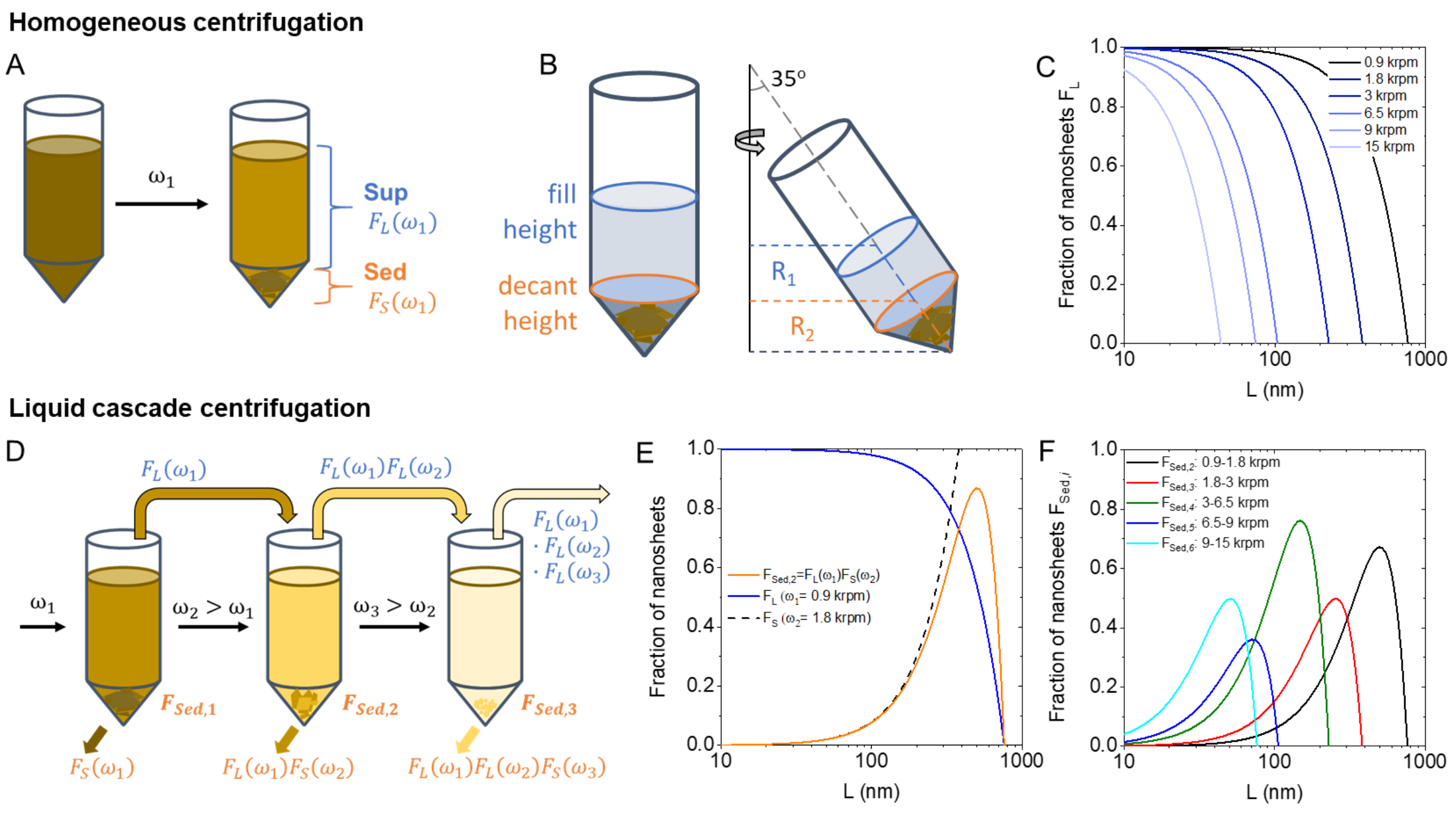

***Figure 2: Description of commonly applied centrifugation schemes in fixed angle rotors.***

*A) Schematic of a single step homogeneous centrifugation resulting in a supernatant and sediment which are a function of angular frequency and time.*

*B) Schematic of a centrifuge tube in a fixed-angle rotor with the relevant radii.*

*C) Fraction of nanosheets (as function of length) remaining in the supernatant after centrifugation ($F_L$) at angular frequencies predicted by theory ($WS_2$, $H_2O$, 10°C, 2h).*

*D) Schematic of liquid cascade centrifugation with sequentially increasing centrifugation speeds. Sediments after each step are collected for analysis and supernatants subjected to the next step at higher speeds. From the second step onward, the nanosheet size in the sediment is determined by the angular frequency of all proceeding steps. For example, the fraction of Sed3 is $F_L(\omega_1)F_L(\omega_2)F_S(\omega_3)$.*

*E) Fraction of nanosheets (as function of length) of a supernatant ($F_L$) after 0.9 krpm and a sediment ($F_S$) after 1.8 krpm compared to the fraction of nanosheets obtained in a sediment when using these angular velocities as lower and upper centrifugation boundary.*

*F) Fraction of nanosheets (as function of length) in sediments collected in a typical LCC experiments as used in this work ($WS_2$, $H_2O$, 10°C, 2h). For E) and F) the typical experimental parameters in Table 1 and 2 are used as input, as well as d=0.43 nm and $\Delta\rho_S$=530 kgm$^{-3}$.*

To describe this separation of sediment from supernatant we must define a specific height to delineate these two phases. We acknowledge that for many real samples the distinction between solid sediment and liquid supernatant can be ill-defined, but throughout many experiments we find good results are obtained when using the height of the narrower base of the tube as the decanting height, especially since this is a feature of most centrifuge tubes.

The angle of the tube relative to the plane of rotation must also be considered since many commercial centrifuges use fixed angle rotors, as these allow for higher speeds and are more stable towards vibrations. This design takes advantage of the Boycott effect, where a depleted layer of lower concentration fluid forms on the inside edge of the tube generating an additional current driving accelerated sedimentation.[46] Different corrections of the sedimentation coefficient have been proposed to account for this, however they require complex analysis of centrifuge tube size, shape and geometry and cannot be uniformly applied across experiments.[47-49] Fortunately, despite neglecting this influence here, good results are obtained for all samples tested, as will be demonstrated below. We therefore need only resolve the geometry of the centrifuge tube to ensure the distance from the radius is accurately described from vertical fill height. This simple geometry correction is shown in Figure 2B, along with the labelled radii to the top of the liquid, $R_1$, and the decanting point, $R_2$, where we delineate the supernatant from the sediment.

To find the population of sheets that sink into the sediment, we can use the equations of motion working backwards from the decanting point, $R_2$. Using the sedimentation rate defined above in Equation (6), the distance the nanoplate has travelled from an initial position, $r_0$ in time $t$ is:

$$\Delta r = r - r_0 = r_0(e^{S\omega^2 t} - 1) \tag{8}$$

Setting the final position, $r$, of the sheet to $R_2$, we can calculate the initial position (which we term $r=r_{sed}$) from which a nanosheet will migrate exactly to the decanting point by the end of the centrifuge experiment. Further, since we know $\Delta r \propto r_0$, any sheets that start at larger $r_0$, that is sheets found further down the tube, will travel an even greater distance and must finish the experiment past the decanting point, also in the sediment.

Thus, $r_{sed}$ is a threshold, above which any sheets at $t=0$ will remain in the supernatant, whilst any sheets at or below this point will terminate in the sediment. If the nanosheets are homogeneously distributed at the start of the experiment, the relative population of sheets in the sediment, $F_S$ and supernatant, $F_L$, is then found by the relative amounts of solution above and below $r_{sed}$, as given in Equation (9):

$$F_L(A, h, \omega, t) = \begin{cases} \dfrac{R_2 e^{-S\omega^2 t} - R_1}{R_2 - R_1}, & S\omega^2 t \le \ln\left(\dfrac{R_2}{R_1}\right) \\ 0, & S\omega^2 t \ge \ln\left(\dfrac{R_2}{R_1}\right) \end{cases} \tag{9}$$

We note that, by definition, $F_S + F_L = 1$. The smooth change in function allows us to account for the finite length and fill height of the centrifuge tube. In the limiting case, the maximum distance any particle can sink is from the top of the liquid, $R_1$, to the base of the centrifuge tube, $\approx R_2$; we find this condition is met when $S\omega^2 t = ln(R_2/R_1)$. For sheets sinking faster than this, the top expression predicts they will pass through the bottom of the centrifuge tube. This is clearly impossible, but we can safely assert that $F_{liq} = 0$ for any nanosheet beyond this limit since all values of $r_0$ will terminate in the sediment. A complete derivation of this expression and limiting conditions is found in Section S2.2.

To visualize this trend, we use the same mean aspect ratio approximations described for the band sedimentation experiment, except we now define $S$ exclusively in terms of sheet length, $L$. In addition, neglecting surfactant for simplicity (i.e. setting $\Delta\rho_S$ to zero) yields:

$$S(L) = \frac{\Delta\rho_{NS}L^2}{10.5\eta k_{Lt}\sqrt{k_{lw}}} \tag{10}$$

We can then combine equations 9 and 10 to plot the population fraction remaining in the supernatant, $F_L$, as a function of nanosheet length, $L$, in Figure 2C. Here we consider $WS_2$ in water at 10 $^oC$ for 2 h at various rotation speeds, modelling the effect of typical centrifuge preparation conditions reported previously. This shows that, because very small nanoplatelets have a low velocity, their population remains relatively unchanged by the centrifugation. As the platelet size increases, the fraction remaining decrease sharply to zero, the limiting case above which all nanosheets will sink completely into the sediment.

This plot also demonstrates the trend from increasing the rotation speed. As expected, at higher speeds the particles are accelerated to a higher velocity, so a greater fraction of them sink into the sediment during the time interval of the centrifugation and the curve and x-intercept move to smaller nanosheet length. This observation was the conceptual basis of liquid cascade centrifugation[16] which is illustrated in Figure 2D: a cascade of sequentially increasing rotation speeds allows selection of decreasing nanosheet sizes. While more time consuming than a single centrifuge separation, the strength of LCC is the distinct and narrow size distribution obtained in each step, allowing investigations of size-dependent properties.

Crucial to the 'cascade' of LCC is applying the sequentially increasing speeds (or times) to the supernatant from the previous step. In that way those larger sheets, which sediment at lower speeds, are removed during the initial steps while very small sheets, that require higher speeds to enter the sediment in a given time, will remain longer in the supernatant and only sink to the sediment in later steps of the cascade.

To model an LCC process, each new fraction is a product of all previous centrifuge separations that sample has been subjected to. To illustrate this, a simple three step cascade is shown in Figure 2D. In such a cascade, after the first centrifugation step (at $\omega_1$), large unwanted material is sent to the sediment while the supernatant is collected. The fraction of nanosheets in this supernatant is then $F_L(\omega_1)$. In the second step, the supernatant is centrifuged again (at $\omega_2 > \omega_1$). The resultant sediment and supernatant are then collected. The fraction of nanosheets in each are given by $F_L(\omega_1)F_S(\omega_2)$ and $F_L(\omega_1)F_L(\omega_2)$ respectively. In the third step the process is repeated again and the fraction of nanosheets in sediment and supernatant are given by $F_L(\omega_1)F_L(\omega_2)F_S(\omega_3)$ and $F_L(\omega_1)F_L(\omega_2)F_L(\omega_2)$ respectively. We note that the desired size-selected nanosheets are collected in each sediment with nanosheet size decreasing with step number. For a cascade with many steps, the fraction of nanosheets in the $i^{th}$ sediment is expressed as:

$$F_{Sed,i} = F_S(\omega_i)\prod_{j=1}^{i-1} F_L(\omega_j) \tag{11}$$

However, we note that higher speed centrifuge processes always sediment more material than lower speeds. We can then safely neglect all but the final two steps, considering only the supernatant from the second last step and the sediment obtained at the highest rotation speed:

$$F_{Sed,i} \approx F_L(\omega_{i-1})F_S(\omega_i) = F_L(\omega_{i-1})(1 - F_L(\omega_i)) \tag{12}$$

Any pair of adjacent centrifugation steps (e.g. $\omega_i$ and $\omega_{i-1}$) effectively slice the initial polydisperse nanosheet size distribution. The first step will exclude any sheets larger than the cut-off while the second step minimises the population of smaller nanoplatelets that sink too slowly to enter the sediment. This is shown in Figure 2E where the fraction of nanosheets remaining in the supernatant after a low-speed centrifugation (specifically 0.9 krpm) is plotted as a function of nanosheet length along with the fraction of nanosheets that sediment at higher speeds (specifically 1.8 krpm). The fraction of nanosheets collected as sediment trapped between this pair of centrifugation speeds is the product of both functions in equation (12).

For these illustrations we plot line graphs of the fractional population change with sheet length. However, it must be noted that the full form of this function is a multidimensional surface over all nanosheet dimensions: $L$, $W$ & $N$ (or $h$). Figure S5 illustrates this by showing a 2D surface of the fractional population change with $L$ & $N$ since any higher dimensions cannot be easily visualized. Applying the mean aspect ratio approximation is equivalent to taking a slice out of the 3D function in figure S5 along a plane defined by $h = L / k_{Lt}$. This is defined by the path of most probable nanosheet sizes found in the original distribution, as described by the thermodynamics of LPE.[44]

We note that a fully rigorous analysis requires consideration of the starting nanosheet size distribution which would be a correlated three-dimensional log-normal distribution parameterised by nanosheet length, width and height.[32] We refer to such a distribution as $P_{stock}(N,L,W)$ and the complete form of this is found in Equation S4. If this distribution was known, the exact population of all nanosheet sizes, $P$, resulting from centrifugation could be predicted by mapping the stock distribution function to the population change function (i.e. $F_S$ or $F_L$) for every $L$, $N$ and $W$:

$$P_{Sed,i}(N,L,W) = P_{stock}(N,L,W)F_{Sed,i}(N,L,W) \tag{13}$$

In principle, if the $L$, $N$ and $W$ distributions were measured in finely separated fractions, the size distribution in the stock prior to centrifugation could be reconstructed.

Unfortunately, the full population distribution of a stock directly from exfoliation has not yet been measured. Difficulties arise due to the wide range of particles sizes that must be accurately measured, in large enough numbers to be statistically significant, and without any bias towards smaller or larger sheets. In spite of advances in automated image analysis,[50,51] current microscopy methods such as atomic force microscopy (AFM) cannot resolve the smallest and largest nanosheets simultaneously to enable such an unbiased count, and spectroscopic methods are only sufficient to measure averaged values. Example images of such stock samples for $WS_2$ exfoliated in aqueous sodium cholate and N-methyl-pyrrolidone (NMP) are shown in Figures S12 and S13.

Fortunately, we find that if suitable centrifuge conditions are used – specifically small steps between rotation speed sizes, the resulting population in the sediment ($F_{sed,i}$) is much narrower than the stock distribution ($P_{stock}$), allowing us to treat the unknown stock distribution function as constant:

$$P_{Sed,i}(N,L,W) \propto F_{Sed,i}(N,L,W) \tag{14}$$

In previous investigations on LCC,[44] we have identified near optimum conditions to balance the number of steps in the cascade with well-defined sheet sizes in each fraction. The fraction of nanosheets in each sediment obtained from such a cascade is calculated using our new theoretical framework (equations 9, 10, 12) as shown in Figure 2F, again using the mean aspect ratio approximation to produce a 2D line plot of $F_{Sed,i}$ versus $L$ for $i$=2,3,4,5,6. From the plot in Figure 2F, two key features can be observed as the angular rotation speed is increased: firstly, the maximum size decreases as the nanosheets sediment faster; secondly, in every case, a well-defined peak in the distribution is observed. These offer measurable size values that will allow to predict the centrifugation process as outlined below, and thereby test the validity of our assumptions regarding the sedimentation of nanosheets. We note that, this form of the model does not account for minor effects such as shape filtering, e.g. changes in length-width aspect ratios with centrifugation that have been observed previously.[44,52]

*Calculating representative nanosheet sizes and comparing with experiment*

To test the applicability of our model to LCC data, we prepared a set of four nanosheet samples using the cascade centrifugation conditions shown in Figure 2F. To be able to assess the accuracy of the theoretical framework for sedimentation in different liquid environments, it was important to subject identical samples to centrifugation, but in different solvents. While the average nanosheet aspect ratios are primarily governed by the binding and tearing energies of the bulk material,[44] it has been reported that changes in the nanosheet population in different fractions after LCC can be observed when comparing, for example, graphite exfoliated in aqueous sodium cholate versus NMP, respectively.[44] To date, it was not clear whether this was an effect of centrifugation or the presence of different particle size distributions from exfoliation in surfactant versus solvent. To address this, $WS_2$ was exfoliated in both aqueous SC and NMP by sonication-assisted LPE. We chose $WS_2$ as model substance due to its high density that allows us to neglect contributions from the surfactant to the effective density. The as-sonicated dispersions were split into two aliquots, and each transferred to both solvents systems through high-speed centrifugation followed by redispersion, resulting in four stock dispersions that were subjected to LCC and the size distributions measured through AFM statistics (see SI section 5). Note that the scatter plots of length versus nanosheet layer number in Figure S20 demonstrate that different nanosheet size distributions and even slightly different nanosheet length-thickness aspect ratios are obtained from exfoliation in surfactant versus solvent.

To compare experiment and theory, we calculate the maximum possible nanosheet length observable within a given fraction. To do this, we use our model (see section S2.5) to calculate the nanosheet length where $F_{Sed,i}$=0. We refer to this as the upper cut-off length, $L_{cut}$, which depends on layer number, $N$, as follows:

$$L_{cut} = \frac{10.5\eta\sqrt{k_{lw}}}{\omega_{i-1}^2 t(Nd_0\Delta\rho_{NS} + 2d\Delta\rho_S)} ln\left(\frac{R_2}{R_1}\right) \quad (15)$$

Here the appropriate angular frequency is that used to generate the supernatant that was then centrifuged to give the sediment containing the nanosheets in question i.e. $\omega_{i-1}$. This cut-off function is plotted over the $L$-$N$ scatter plot of sheets measured from AFM of $WS_2$ exfoliated in aqueous SC and centrifuged in NMP in Figure 3A. Only two fractions are shown here for clarity, but full plots of all LCC fractions over the four samples mentioned above are shown in SI Figure S21. The measured sheet sizes are widely scattered below this cut but, over 12 different fractions, no data points are observed above the lines representing the predicted maximum. This supports the validity of equation 15 and the model underlying it.

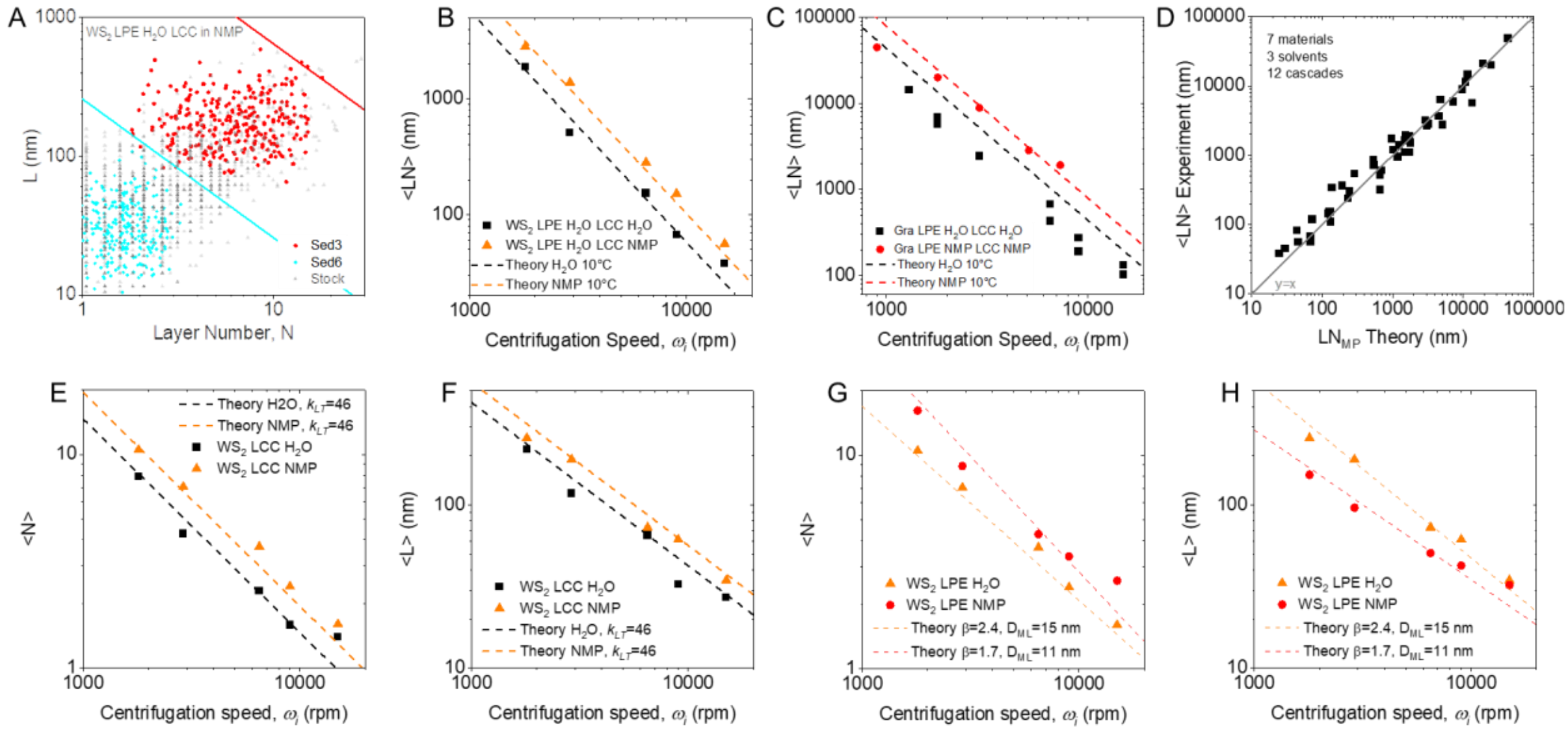

***Figure 3: Comparison of experiment and theory in LCC.***

*A) Nanosheets sizes as measured by AFM of two $WS_2$ LCC fractions, with predicted maximum sheet size limits shown as diagonal lines. $WS_2$ exfoliated in aqueous sodium cholate (SC) and centrifuged in NMP.*

*B-C) Arithmetic mean of the product of length and layer number, ⟨LN⟩, in size-selected fractions as function of sediment angular velocity for B) $WS_2$ and C) graphene centrifuged in aqueous SC and NMP, respectively. The graphene data is originally from reference [44].*

*D) Comparison of experimental and theoretical ⟨LN⟩ from all data analysed (see SI).*

*E-F) Arithmetic mean layer number (E) and length (F) as function of angular velocity of LCC size-selected $WS_2$ exfoliated in aqueous SC. Dashed lines represent theory using the mean aspect ratio approximation.*

*G, H) Arithmetic mean layer number (G) and length (H) as function of angular velocity of $WS_2$ centrifuged in NMP, exfoliated in SC and NMP, respectively showing different exponents relating ⟨N⟩ and ⟨L⟩ to the angular velocity. These can be accounted for theoretically (equations 20, 21, dashed lines) when using the actual exponents, β, relating ⟨L⟩ and ⟨N⟩ and experimental values for the characteristic monolayer length $D_{ML}$.*

The sharp peaks observed in the population distributions after cascade centrifugation in Figure 2F suggests it is also possible to calculate the most probable nanosheet size in each fraction. This parameter turns out to be a useful estimate of the mean nanosheet size for well-chosen size selection procedures. By differentiating $F_{sed}(N,L)$ with respect to $LN$ we find an expression that, so long as the difference in speed between the steps is kept small, can be solved (see SI section 2.6). This yields an equation for the most probable value of $LN$:

$$(LN)_{MP} \approx \frac{10.5\eta\sqrt{k_{lw}}}{\omega_i^2 t d_0 \Delta\rho_{NS}} ln\left(\frac{2R_2}{R_2 + R_1}\right) \approx \langle LN \rangle \quad (16)$$

This expression includes experimental parameters that could be varied to access a desired nanosheet size, most usually $\omega_i$ but also $t$. We reiterate that here, the appropriate angular frequency is that used to generate the sediment which contained the nanosheets in question (i.e. $\omega_i$, see SI Section 6.2). The material dependent information required also includes some knowledge of $k_{lw}$, the length-width aspect ratio of nanosheets. For most commonly used 2D materials prepared by LPE this has been shown to lie within a narrow range (1.5-2.5).[44,52-55]

Comparing $\langle LN \rangle$ for $WS_2$ and graphene size-separated by LCC in water and NMP, shown in Figure 3B and C, we see good agreement between the prediction of the most probable size (using input parameters from table 2) and the mean size as measured by AFM. The discrepancy found for graphene centrifuged in water is most likely due to the surfactant coating. In NMP this coating is removed, replaced by the solvent, however, as seen from band sedimentation for low density materials such as graphene, this surfactant can be a significant contribution to the effective density in aqueous dispersions (which was neglected to simplify the equations).

***Table 2:*** *Experimental parameters of the LCC, and solvent properties at 10 °C*

| **Standard LCC** | | |
|---|---|---|
| $R_1$ (mm) | 75 | |
| $R_2$ (mm) | 93 | |
| Time (s) | 7200 | |
| Temperature | ~10°C | |
| Rotation Speed (krpm) | 0.9, 1.8, 3, 6.5, 9, 15 | |
| **Solvents** | **Density* (kg m$^{-3}$)** | **Viscosity* (kg m$^{-1}$ s$^{-1}$)** |
| $H_2O$ + sodium cholate$_{(aq)}$ | 1000 | 0.0013 |
| NMP | 1030 | 0.0023 |

Considering a wider comparison of 8 different materials: $WS_2$ (this work, SI section 5), graphene[44], $MoS_2$[44], hBN[56], GaS[44], $Cu(OH)_2$[57], $NiPS_3$[55], $CrTe_3$[54]; prepared in 3 solvents: aqueous sodium cholate, NMP and N-Cyclohexyl-2-pyrrolidone (CHP), and size-selected with 12 LCC experiments from previously published literature, we show the model derived in this work accurately predicts the mean sheet sizes $\langle LN \rangle$ across the materials and experiments tested. This is summarized by the plot Figure 3D which shows the measured, experimental $\langle LN \rangle$ as function of calculated most probable nanosheet size. Here, each data point is the average of one LCC size-selected fraction (but excluding low-density materials such as graphene and hBN in aqueous surfactant solution). Full data sets and detailed comparison between experiment and theory for all materials are found in SI Section 7.

*The mean aspect ratio approximation*

An accurate prediction of $\langle LN \rangle$ is useful, but it would be even more interesting to directly predict $\langle L \rangle$ and $\langle N \rangle$ separately. We can achieve this by relating $\langle L \rangle$ and $\langle N \rangle$ via mean aspect ratio. As mentioned above, we have used $k_{Lt} = \langle L/h \rangle$. We can approximate this as $k_{Lt} \approx \langle L \rangle / d_0 \langle N \rangle$. Then approximating $\langle LN \rangle \approx \langle L \rangle \langle N \rangle$ and using the result above to replace either $\langle L \rangle$ or $\langle N \rangle$ yields equations which allow mean sheet length and layer number can also be calculated separately:

$$L_{MP} \approx \left(\frac{10.5\eta\sqrt{k_{lw}}k_{Lt}}{t\Delta\rho_{NS}} ln\left(\frac{2R_2}{R_2 + R_1}\right)\right)^{1/2} \frac{1}{\omega_i} \approx \langle L \rangle \tag{17}$$

And

$$N_{MP} \approx \left(\frac{10.5\eta\sqrt{k_{lw}}}{tk_{Lt}d_0^2\Delta\rho_{NS}} ln\left(\frac{2R_2}{R_2 + R_1}\right)\right)^{1/2} \frac{1}{\omega_i} \approx \langle N \rangle \tag{18}$$

We note that equations 17 and 18 use the approximation that the length-thickness aspect ratios of LPE nanosheets are well-defined with a narrow distribution around a constant value. To test these expressions, the experimental $\langle N \rangle$ and $\langle L \rangle$ from AFM analysis is plotted against the centrifugation speed in Figure 3E and 3F for $WS_2$ exfoliated in aqueous SC and centrifuged both in aqueous SC and NMP and shows good agreement with the theoretical prediction. However, we note that the experimental results from the NMP-exfoliated $WS_2$ deviate somewhat from the theoretical prediction (Figure S25).

However, since the decoupling of the effect of exfoliation and centrifugation on nanosheet lateral size and thickness in solvent versus surfactant media performed here clearly shows that aspect ratios can be different for exfoliation in surfactant and solvent (SI Figure S24), we expect that there are some restrictions to the mean aspect ratio approximation, i.e. restrictions to equation 17 and 18 (but not equation 16), that we will analyse in detail.

*Beyond the mean aspect ratio approximation*

This demonstrates that this mean aspect ratio approximation is not always fully accurate. This is because aspect ratios of nanosheets within a dispersion also provide information about the production and history of the sample. To address this, we now extend our model to a more accurate representation of the nanosheet size distributions. We make use of an expression we found previously which showed that that for LPE produced nanosheets there is a power-law dependence between the lateral sheet size and thickness.[44]

$$\langle LW \rangle = D_{ML}^2 \langle N \rangle^{\beta} \tag{19}$$

Where $D_{ML}$ is the characteristic monolayer size and $\beta$ is a constant in the range 1.5-3, being typically centred at ~2.8 (SI section 7). Using this relation in place of the mean aspect ratio for most probable nanosheet size, equations S10 and S11 can be derived, which show that

$$\langle N \rangle \approx \left(\frac{10.5\eta}{td_0\Delta\rho_{NS}D_{ML}} ln\left(\frac{2R_2}{R_2+R_1}\right)\right)^{1/(1+\beta/2)} \frac{1}{\omega_i^{1-(\beta-2)/(\beta+2)}} \tag{20}$$

$$\langle L \rangle \approx \sqrt{k_{wl}}D_{ML} \left(\frac{10.5\eta}{td_0\Delta\rho_{NS}D_{ML}} ln\left(\frac{2R_2}{R_2+R_1}\right)\right)^{\beta/(2+\beta)} \frac{1}{\omega_i^{1+(\beta-2)/(\beta+2)}} \tag{21}$$

We note that the values of $(\beta - 2)/(\beta + 2)$ are generally quite small such that the deviations from the behaviour predicted by equations 17 and 18, although observable, are not significant.

Exfoliating nanosheets in water and NMP has been observed to produce dispersions with different $\beta$ (but similar $D_{ML}$) values in the case of graphene.[44] We have confirmed this result here via more elaborated experiments using $WS_2$, where the effect of the solvent on exfoliation and centrifugation was decoupled in contrast to the previous work on graphene (see Figure S26A-D). Figure 3G and H plots $\langle N \rangle$ and $\langle L \rangle$ versus $\omega_i$ for the two $WS_2$ sample sets that were exfoliated in different media (but both centrifuged in NMP). These graphs clearly show angular frequency dependence that deviate slightly from the $1/\omega_i$ behaviour predicted by equations 17 and 18. However, the measured behaviour can be reproduced by plotting equations 20 and 21 over the data (dashed lines) using the experimental parameters (Table 2) and the measured values of $\beta$ and $D_{ML}$ given in the panel (see Figure S27 for measurements). We emphasise the excellent agreement between theory and data. We note that such differences in exponent have been reported on previously but were not explained.[54-56] We can now show that such differences are not due to any difference in behaviour within the centrifugal field, but rather it is a result of the difference in the statistical distribution of sheet sizes in the stock caused by different production methods.

In the cases studied here, the difference in aspect ratio was known from AFM derived statistics, but such large datasets remain time consuming to obtain. Our new understanding of the difference in exponent suggests it may be possible to estimate such mechanistic differences between exfoliation methods from centrifugation data sets in future, in addition to predicting mean nanosheet sizes.

**Conclusion**

We have derived a model for the sedimentation of 2D nanosheets within a centrifugal field, combining a rhombohedral model for sheet density with an oblate spheroid model for viscous resistance. Rate zonal sedimentation experiments have shown the reliability of sedimentation coefficients calculated from this model without needing to complete prior experiments to determine empirical correction factors. Using these sedimentation coefficients, we show it is possible to predict individual nanosheet motion, and also to extend this to predict changes in the relative populations of nanosheets in bulk dispersions. Even without knowledge of a complete size distribution function of the stock dispersion, key metrics (cut size, average nanosheet size) can be predicted as experimentally verified from a range of nanosheet dispersions size-selected through LCC. Such predictions relating to size distribution functions could also be easily applied to other particle shapes and sizes. The use of aspect ratios to reduce the dimensionality of the 2D nanosheet problem also offers information about underlying exfoliation mechanisms. In future, analytical centrifuge experiments may provide greater insights to nanosheet size distributions and production methods. The ability to predict population changes during centrifugation also allows much greater optimisation of size selection procedures, where specific sheet sizes can be tailored to applications and meaningful comparisons can be made between different material and solvent systems independent of extrinsic size effects caused by purification.

## Methods

A detailed method section is found in the SI, section 3.

*Preparation of the dispersion*

Graphite and group VI-TMD dispersions were prepared by probe sonicating the powder according to previously published methods.[44] In brief, a two-step sonication was used. The initial 1 h sonication was used to purify the commercial powder to remove impurities in the supernatant after centrifugation, while the sediment was subjected to 5 h sonication to prepare the stock dispersions used for size selection.

*Band sedimentation*

For band sedimentation, 1 g $L^{-1}$ SC was dissolved in $H_2O/D_2O$. As bottom layer, 100% $D_2O$ was used, followed by layering 50/50 vol.% $H_2O/D_2O$ on top and finally the nanomaterial dispersion. Band sedimentation was performed in swinging bucket rotors for the speeds and time indicated in the figures. After centrifuging the samples were immediately removed and bands removed in 1 mL aliquots using a needle and syringe, radius values were measured for each aliquot from the rotor and tube height. A limited initial size selection was used for all samples prior to band sedimentation experiments as detailed in the SI.

*Liquid Cascade Centrifugation*

The $WS_2$ dispersions in NMP and SC, respectively, were each split into 2 equal aliquots and centrifuged at 15 krpm for 3h. The sediments were redispersed in water and centrifuged again at 15 krpm. After this step, half of each dispersion was redispersed in SC (2 g $L^{-1}$) and half in NMP resulting in 4 stock dispersions subjected to LCC. While different centrifuges and rotors were used in this experiment, we note that relevant parameters were similar ($R_1$ = 75 mm and $R_2$ = 93 mm). For the $WS_2$ data shown in the main manuscript, all centrifugation runs were performed for 2 h (10°C).

Unexfoliated material was removed by centrifugation at 900 rpm. The supernatant was subjected to further centrifugation at 1.8 krpm. The sediment was collected in fresh surfactant, while the supernatant was centrifuged at 2.9 rpm. Again, the sediment was collected and the supernatant subjected to centrifugation at higher speeds. This procedure was repeated with the following centrifugation speeds: 6.5 krpm, 9 krpm, 15 krpm. As sample nomenclature, the lower and upper boundary of the centrifugation are indicated.

Additional LCC data for other materials analysed in the SI is published data.

*UV/VIS extinction spectroscopy*

UV/VIS extinction spectra were recorded using 0.5 nm data intervals in quartz glass cuvettes. Where required, samples were diluted into a sodium cholate solution.

*Atomic Force Microscopy*

Atomic force microscopy was carried out on a Dimension ICON3 scanning probe microscope (Bruker AXS S.A.S.) in ScanAsyst in air under ambient conditions using aluminium coated silicon cantilevers after deposition on $Si/SiO_2$ coated with (3-Aminopropyl)triethoxysilane. Typical image sizes ranged from 20x20 for larger nanosheets to 5x5 µm$^2$ at scan rates of 0.5-0.8 Hz with 1024 lines per image. Previously published step height were used to convert apparent thickness to layer number[16] and corrections were used to correct lateral dimensions from cantilever broadening.[58]

## Acknowledgement

We acknowledge funding from the European Union through the 2D-PRINTABLE project (GA-101135196) and thank Jana Zaumseil for the access to the infrastructure at the Chair of Applied Physical Chemistry.

## Author contributions

S.G. performed and evaluated band sedimentation, S.O. and T.S. performed LCC and AFM of $WS_2$, S.G, A.D., C.G., V.V-M., J.N.C. developed theoretical analysis, K.S. performed AFM, M.H. performed and evaluated rheology, J.N.C. derived the theory, C.B. evaluated AFM and analyzed LCC data, J.N.C. and C.B. conceptualized the work, S.G., J.N.C. and C.B. wrote the manuscript

## Additional information

Supplementary information is available in the online version of the paper. Correspondence and requests for materials should be addressed to C.B.

## Competing financial interests

The authors declare no competing financial interests.

# Supporting Information

# Centrifugation theory revisited: Understanding and modelling the centrifugation of 2D nanosheets

Stuart Goldie,[1] Steffen Ott,[2] Anthony Dawson,[3] Tamara Starke,[2] Cian Gabbett,[3] Victor Vega Mayoral,[4] Kevin Synnatschke,[2,3,5] Marilia Horn,[1,6] Jonathan N. Coleman,[3]* Claudia Backes[1,2]*

1) *Physical Chemistry of Nanomaterials and CINSaT, Kassel University, Heinrich-Plett Str. 40, 34132 Kassel, Germany*
2) *Applied Physical Chemistry, Heidelberg University, Im Neuenheimer Feld 253, 69120 Heidelberg, Germany*
3) *School of Physics and CRANN, Trinity College, Dublin 2, Ireland*
4) *Instituto Madrileño de Estudios Avanzados en Nanociencia (IMDEA), C/ Faraday 9, 28049 Madrid, Spain*
5) *Chair for Molecular Functional Materials, Dresden University of Technology, Stadtgutstr. 59, 01217 Dresden, Germany*
6) *University of Münster, Corrensstr. 3, 48149 Münster, Germany*

## Content

# 1 List of symbols

| | |
|---|---|
| $A$ | nanosheet area |
| $c$ | hydrodynamic radius |
| $d$ | thickness of surfactant coating |
| $d_0$ | crystallographic thickness one layer |
| $D_{ML}$ | characteristic monolayer size |
| $f$ | frictional coefficient |
| $f_0$ | shape factor |
| $F_L$ | fraction of nanosheets remaining in the liquid (supernatant) |
| $F_S$ | fraction of nanosheets entering the sediment |
| $F_{sed,i}$ | fraction of nanosheets in the $i^{th}$ sediment of an LCC |
| $h$ | nanosheet thickness |
| $k_{Lt}$ | length/thickness aspect ratio |
| $k_{lw}$ | length/width aspect ratio |
| $L$ | nanosheet length (long axis of a diamond) |
| $m$ | mass |
| $N$ | nanosheet layer number |
| $NL_{MP}$ | most probable product of nanosheet length and layer number in a size-selected fraction |
| $P$ | population size distribution of nanosheets |
| $Pe$ | Péclet number |
| $r$ | position in cent tube at certain time |
| $r_0$ | position in cent tube at time zero |
| $R_1$ | position of liquid surface relative to rotor axis |
| $R_2$ | position in tube where decantation occurs relative to rotor axis |
| $r_{sed}$ | position in tube that delineates particles that will end in supernatant and in sediment |
| $S$ | sedimentation coefficient |
| $t$ | time |
| $u$ | instantaneous velocity of particle |
| $v$ | terminal velocity of nanosheets in a centrifugal field |
| $W$ | nanosheet width perpendicular to L |
| $\beta$ | exponent relating length to thickness after exfoliation |
| $\eta$ | viscosity (of liquid) |
| $\omega$ | angular frequency of rotor rotation (rad $s^{-1}$) |
| $\rho_l$ | density (of liquid) |
| $\rho_{eff}$ | effective object density |
| $\rho_S$ | surfactant density |
| $\rho_{NS}$ | nanosheet density |
| $\Delta\rho_S$ | density difference surfactant-solvent |
| $\Delta\rho_{NS}$ | density difference nanosheet-solvent |

# 2 Modelling centrifugation of nanosheet dispersions

## *2.1 Calculating the equation of motion for nanosheets in a centrifuge*

Here we follow the standard derivation for the equation of motion of a sinking particle but include some information specific to 2D materials.

For an object (e.g. a nanosheet) moving radially through a liquid in response to a centrifugal force ($F_c = mr\omega^2$), it experiences resistance from the particle's buoyancy ($F_B = V_{NP}\rho_l\omega^2 r$) and viscosity ($F_f = f\frac{dr}{dt}$), see Figure S1. Newton's law for the resulting acceleration can be written as

$$m\frac{d^2r}{dt^2} = m\omega^2 r - \frac{m}{\rho_{eff}}\rho_l\omega^2 r - f\frac{dr}{dt}$$

or

$$m\frac{du}{dt} = m\omega^2 r\left(1 - \frac{1}{\rho_{eff}}\rho_l\right) - f\frac{dr}{dt}$$

where *m* is the object mass, ω is the angular frequency of the rotor rotation, *r* is the particles position at time, *t*, *f* is the (viscous) frictional coefficient and *u* is the instantaneous velocity at time, *t*.

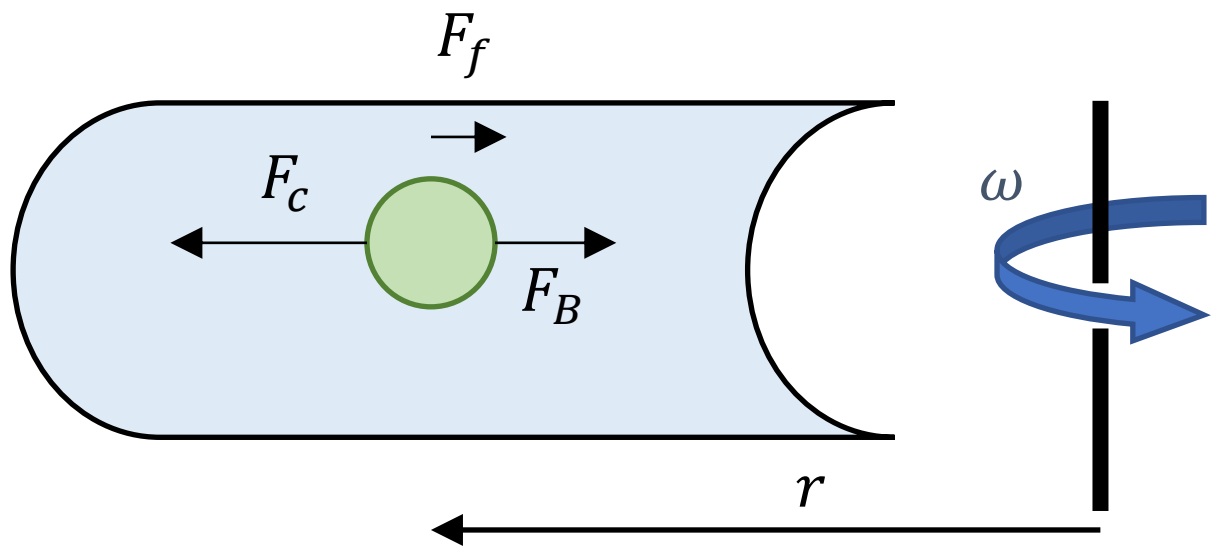

Figure S 1: Balance of forces acting on particles during centrifugation: $F_c$ - centrifugal force, $F_f$ - frictional viscous resistance, $F_B$ - buoyancy of particle from weight of liquid displaced.

The term in the bracket accounts for the buoyancy of the nanosheet, sometimes termed a buoyancy correction factor, where $\rho_l$ is the density of the liquid and $\rho_{eff}$ the effective density of the objects including effects from the bound hydration layer.

We assume that terminal velocity, *v*, is reached quickly allowing us to neglect *du/dt*. Then

$$0 = m\omega^2 r\left(\frac{\rho_{eff}-\rho_l}{\rho_{eff}}\right) - fv$$

and the terminal velocity is given by

$$v = \frac{mr\omega^2}{f}\left(\frac{\rho_{eff}-\rho_l}{\rho_{eff}}\right)$$

The frictional coefficient is given by Stoke's law:

$$f = 6\pi\eta c f_0$$

Where $c$ is the objects hydrodynamic radius and $f_0$ is a factor which corrects for shape effects. In the simplest case of a sphere, $f_0$=1.

The description above is completely general. To adapt it to describe 2D nanosheets, we must make some geometry-specific additions. For the case of a nanosheet (area, $A$, and thickness, $h$), specific expressions must be used for $m$, $\rho_{eff}$, $c$ and $f_0$ which account for its shape.

The mass of the surfactant-coated nanosheet (coated by surfactant layer of thickness, $d$) is given by

$$m = A(\rho_{NS}h + 2\rho_S d)$$

Where $\rho_S$ is the surfactant density and $\rho_{NS}$ the nanosheet density.

The effective density of the surfactant coated nanosheet (surfactant at top and bottom of sheet) is

$$\rho_{eff} = \frac{\rho_{NS}h+2\rho_S d}{h+2d}$$

This term only depends on nanosheet and surfactant thickness. At first glance this may appear strange. However, it turns out we can ignore surfactant bound to the nanosheet edges because the top and bottom dominate the available surface area. When comparing the two models we find neglecting the edges only results in ~ 0.005% error.

Inserting this into the equation for v above gives

$$v = (h(\rho_{NS} - \rho_l) + 2d(\rho_S - \rho_l))A\frac{r\omega^2}{f}$$

Which we can abbreviate to

$$v = (h\Delta\rho_{NS} + 2d\Delta\rho_S)A\frac{r\omega^2}{f}$$

For an oblate spheroid with diameter and thickness, δ and γ, $f_0$ is given by[1,2]

$$f_0 = \frac{\left(\frac{\delta^2}{\gamma^2}-1\right)^{1/2}}{\left(\frac{\delta}{\gamma}\right)^{2/3} arctan\left(\frac{\delta^2}{\gamma^2}-1\right)^{1/2}}$$

Here we utilise the fact that, for large x (x>10), arctan(x)→π/2. Then, for a 2D object, when δ>>γ, to a good approximation, this reduces to $f_0 = \frac{2}{\pi}\left(\frac{\delta}{\gamma}\right)^{1/3}$

We assume that for a liquid-suspended quasi-2D object illustrated in Figure S 2, $A = \delta^2$ and $\gamma = h + 2d$ (including the contribution of surfactant to the thickness). This can be written as

$$f_0 = \frac{2}{\pi}\left(\frac{\sqrt{A}}{h+2d}\right)^{1/3}$$

In addition, we find the hydrodynamic radius by considering the sphere of equal volume:

$\frac{4\pi}{3}c^3 = A(h + 2d)$ giving $c = \left(\frac{3A(h+2d)}{4\pi}\right)^{1/3}$

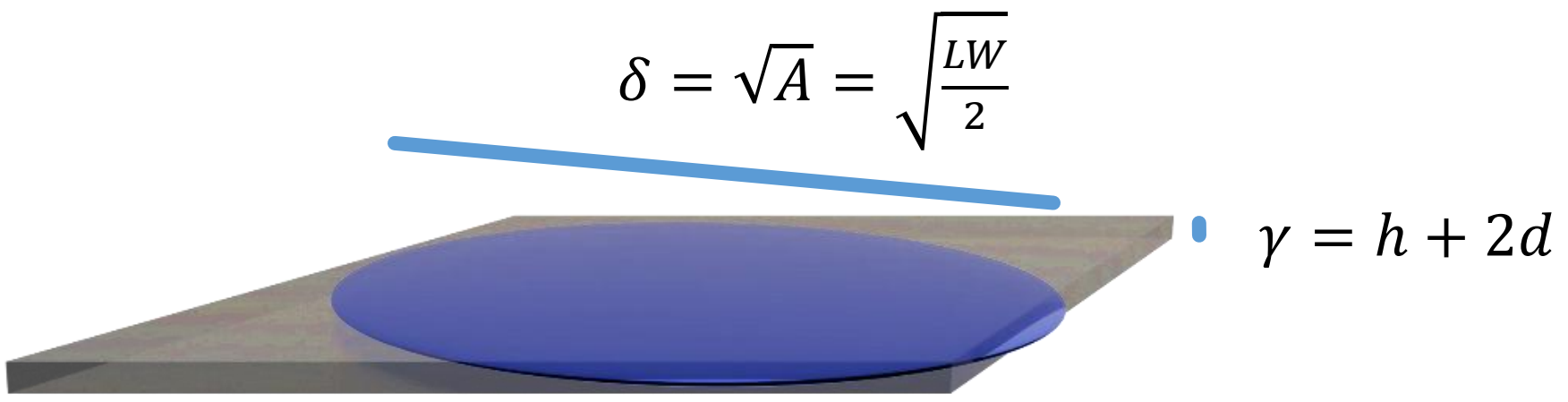

Figure S 2: Scheme of oblate-spheroid model, in dark blue, overlaid on a nanosheet, used as an approximation of frictional resistance. Shown here horizontally but nanosheets are expected to orient with the narrow face into the direction of travel.

This allows us to write

$$f = 6\pi\eta\left(\frac{3A(h+2d)}{4\pi}\right)^{1/3}\frac{2}{\pi}\left(\frac{\sqrt{A}}{h+2d}\right)^{1/3} = 12\eta\left(\frac{3}{4\pi}\right)^{1/3}A^{1/2}$$

Then the terminal velocity of drift speed depends on r and is given by

$$v = \left(\frac{(h\Delta\rho_{NS}+2d\Delta\rho_{S})A^{1/2}}{12\eta(3/4\pi)^{1/3}}\right)r\omega^2 = Sr\omega^2$$

In standard sedimentation notation, the quantity in the curved brackets is usually referred to as the sedimentation constant (or coefficient), *S*, which has units of time.[3]

Because *v=dr/dt*, this represents a differential equation:

$$\frac{dr}{dt} = r\frac{(h\Delta\rho_{NS}+2d\Delta\rho_{S})\sqrt{A}}{12\eta(3/4\pi)^{1/3}}\omega^2 .$$

The solution represents the position of the nanosheet at time t:

$$r = r_0\, exp\left(\frac{(h\Delta\rho_{NS}+2d\Delta\rho_{S})\sqrt{A}}{12\eta(3/4\pi)^{1/3}}\omega^2 t\right).$$

Where $r_0$ is the nanosheet position at time *t*=0

In the case of a nanosheet, the thickness is $h = d_0 N$ where $d_0$ is the crystallographic thickness of one layer and *N* the layer number. This gives

$$r = r_0\, exp\left(\frac{(Nd_0\Delta\rho_{NS}+2d\Delta\rho_{S})\sqrt{A}}{12\eta(3/4\pi)^{1/3}}\omega^2 t\right) = r_0 e^{S\omega^2 t}.$$

Since nanosheets are typically not quadratic, we need to find reasonable expressions for the area A. We model the nanosheets as diamond shaped with long axis, *L*, and perpendicular size, *W*. This approach is convenient, as nanosheet size is often measured with these metrics from microscopy studies.[4,5] Further, it has been shown that usually, the ratio $k_{lw} = \frac{L}{W}$ occupies a fairly narrow range for liquid-exfoliated nanosheets.[6]

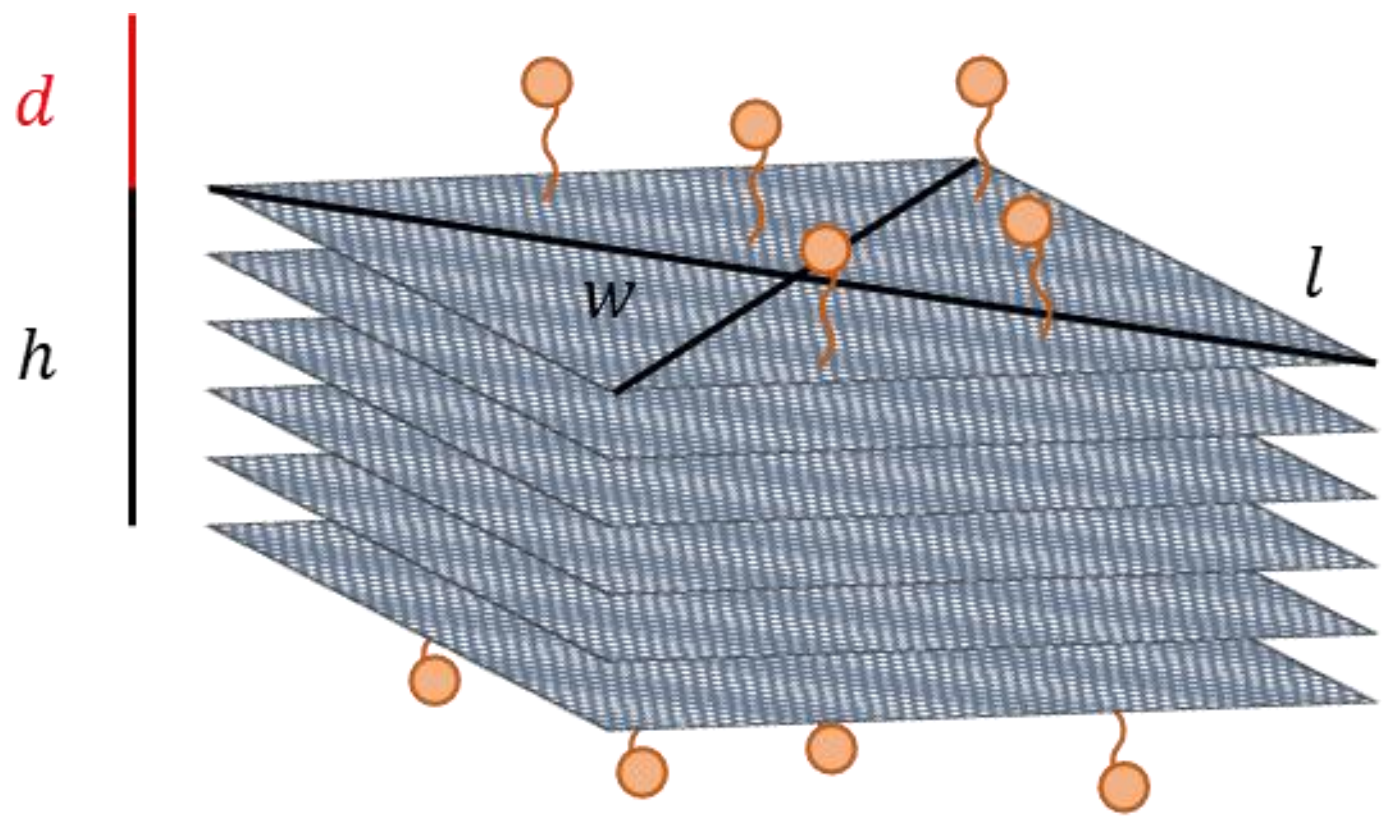

Figure S 3: Schematic of nanosheet size model, showing the length and width as defined on a rhombus nanosheet of thickness *h*, with a surfactant layer, thickness *d*, on the top and bottom.

Then we can write $A = \frac{Lw}{2} = \frac{L^2}{2k_{lw}}$, so

$$S = \frac{(h\Delta\rho_{NS} + 2d\Delta\rho_S)L}{12\eta(3/4\pi)^{1/3}\sqrt{2k_{lw}}}$$

Simplifying constants:

$$S = \frac{(h\Delta\rho_{NS} + 2d\Delta\rho_S)L}{10.5\eta\sqrt{k_{lw}}}$$

This expression still contains two variables, *L* and *N*. This can be further simplified by noting that liquid-phase exfoliation methods of producing 2D materials tend to produce length/thickness aspect ratios that also cluster around a well-defined mean value determined by nanosheet mechanics:[4,6]

$$\langle \frac{L}{h} \rangle = \langle \frac{L}{Nd_0} \rangle = k_{Lt}$$

Combining constants gives:

$$r = r_0\, exp\left(\frac{(h^2\Delta\rho_{NS} + 2dh\Delta\rho_S)}{10.5\eta\frac{\sqrt{k_{lw}}}{k_{Lt}}}\omega^2 t\right) = r_0 e^{S\omega^2 t} \qquad \text{Equation S1}$$

This is the equation of motion of a nanosheet in a centrifuge, reduced to the single parameter of layer thickness. We note that all shape/size information is contained in *S*.

This is easily modified to the case of solvent-suspended nanosheets (no surfactant) by setting *d*=0:

$$r = r_0\, exp\left(\frac{Nd_0\Delta\rho_{NS}L}{10.5\eta\sqrt{k_{lw}}}\omega^2 t\right) \quad \text{(no surfactant)}$$

Inserting the length/thickness aspect ratio into this form produces a simple expression reduced to only length:

$$r = r_0\, exp\left(\frac{\Delta\rho_{NS}L^2}{10.5\eta k_{Lt}\sqrt{k_{lw}}}\omega^2 t\right)$$

For comparison purposes, the equation of motion of a spherical particle (radius $c$) with no surfactant is:

$$r_{sphere} = r_0 \, exp\left(S_{sphere}\omega^2 t\right)$$

$$S_{sphere} = \frac{2}{9\eta}\left((c+d)^2(\rho_S - \rho_l) + \frac{c^3(\rho_{NP}-\rho_S)}{c+d}\right) \qquad \text{Equation S2}$$

$$c = \sqrt[3]{\frac{3}{8\pi}\frac{{k_{Lt}}^2}{k_{lw}}h^3}$$

## 2.2 *Calculating the fraction of nanosheets selected/rejected by centrifugation*

Above, we worked out the equation of motion for a nanosheet of length $L$, width $W=L/k_{lw}$ and layer number, $N$, in a centrifuge:

$$r = r_0 e^{S\omega^2 t}$$

The distance travelled can be expressed by:

$$r - r_0 = r_0(e^{S\omega^2 t} - 1) = \Delta r$$

where $r_0$ is the starting radial position, $t$ is the centrifugation time and the sedimentation constant $S = f(L, k, N)$:

$$S = \frac{(Nd_0 \Delta\rho_{NS} + 2d\Delta\rho_S)L}{10.5\eta\sqrt{k_{lw}}}$$

For the nanosheet to reach the sediment in a time $t$, $r(t)=R_2$ i.e. $R_2 = r_{sed}e^{S\omega^2 t}$. Under these circumstances, the distance travelled by the nanosheet is $\Delta r_{sed} = R_2 - r_{sed}$ as sketched in Figure S 4. We have introduced $r_{sed}$ as notation for the starting position, from which nanosheets will reach the sediment, $R_2$. More accurately, this is the minimum radius beyond which all nanosheets with the same sedimentation coefficient will reach the sediment.

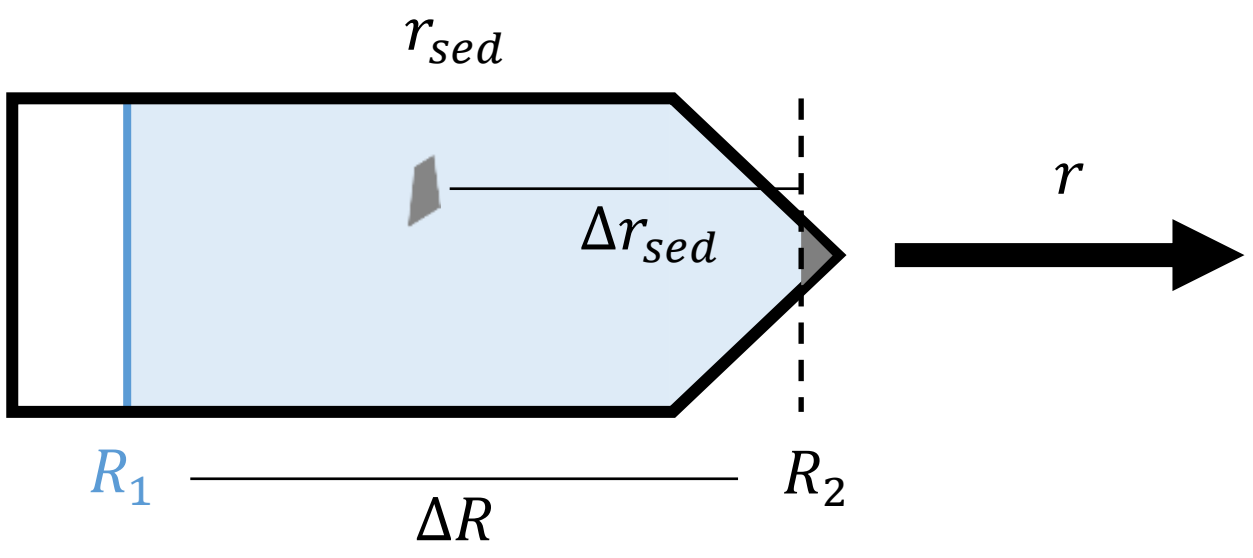

Figure S 4: Schematic of nanosheet motion during centrifugation, displayed in a horizonal plane as for swinging-bucket type rotors for simplicity. All distances are radial away from the axis of rotation as per the bold arrow. $R_1$ is the liquid surface, $R_2$ the decanting point below which is considered sediment, $r_{sed}$ is the minimum radius required for a nanosheet to reach this sediment. For a discussion of this radii, and resolving the geometry for a fixed angle rotor see Figure 2B, main manuscript.

This is best illustrated by considering the three practical cases that result for a given nanosheet size with a known S, remembering $\Delta r \propto r_0$.

1. $r_0 < r_{sed}$, in this case the distanced travelled is less than $\Delta r_{sed}$ and the nanosheet started further up the tube so its final position will not reach the sediment and remain in the supernatant at the end of the experiment.
2. $r_0 = r_{sed}$, this is the case illustrated above in which the nanosheets move to the sediment at $R_2$.
3. $r_0 > r_{sed}$, in this case the distance travelled is even greater than $\Delta r_{sed}$ and the nanosheets have sunk below the decanting point into sediment.

In this way, $r_{sed}$ becomes a dividing radius down the tube: any nanosheets initially in the range $R_1 \leq r < r_{sed}$ at $t=0$ will remain in the supernatant and any nanosheets at or below $r_{sed}$ in the range $r_0 \leq r \leq R_2$ at $t=0$ will sink into sediment. The position of this division remains a function of nanosheet size since $r_{sed} = f(L, k, N, \omega)$.

The maximum distance any nanosheet can travel is defined by the geometry of the system in Figure S 4 as $\Delta R = R_2 - R_1$. The decanting point, $R_2$, is used because everything below this is considered sediment, and since fakes cannot sink through the base of the tube, any specific distances into the sediment are irrelevant.

Assuming a homogenous distribution of nanosheets throughout the tube, the fraction that reach the sediment is

$$F_S = \frac{\Delta r_{sed}}{\Delta R} = \frac{R_2 - r_{sed}}{R_2 - R_1} = \frac{R_2\left(1 - e^{-S\omega^2 t}\right)}{R_2 - R_1} \qquad \text{Equation 3a}$$

The fraction of nanosheets remaining in the sediment is then by definition:

$$F_L = 1 - F_S = \frac{R_2 e^{-S\omega^2 t} - R_1}{R_2 - R_1} \qquad \text{Equation 3b}$$

We note that these functions are only valid when $S\omega^2 t \leq \ln\left(\frac{R_2}{R_1}\right)$. If the combination of time, speed and nanosheet size is too large, then all nanosheets have left the sediment and in reality, $F_L=0$. We therefore express the function for fractional population of the supernatant as:

$$F_L = \begin{cases} \dfrac{R_2 e^{-S\omega^2 t} - R_1}{R_2 - R_1}, & S\omega^2 t \leq \ln\left(\dfrac{R_2}{R_1}\right) \\ 0, & S\omega^2 t \geq \ln\left(\dfrac{R_2}{R_1}\right) \end{cases}$$

This describes the fractional population change of nanosheets of a specific size. For a sample of monodisperse nanosheets, i.e. all nanosheets have the same dimensions, this would be all the information required to predict the effect of any centrifuge process. For example, the concentration would be depleted according to: $c_L = F_L c_{stock}$.

## 2.3 *Real polydisperse samples of nanosheets*

However, real nanosheet dispersions are polydisperse, that is they have a relatively broad distribution of lengths, widths and thicknesses, due to the entropic contribution during their exfoliation causing disordered fragmentation. Under normal circumstances, an as-produced stock dispersion of graphene or $WS_2$ might contain nanosheets with lengths ranging from $L$~10-1000 nm and layer numbers ranging from $N$~1-30 (see section 5 and ref[6-8]).

In principle this distribution can be described by a correlated 3-dimensional log-normal distribution. For a data set of $Q$ nanosheets describing their length, width and layer number, $x_{[1,Q]} = \{l_i, w_i, N_i\}$ the correlated 3D log-normal distribution can be defined using the natural logarithm of the data set where $y_{[1,Q]} = \{ln\, l_i, ln\, w_i\,, ln\, N_i\}$.

For a specific nanosheet size, $\vec{x} = \begin{bmatrix} l \\ w \\ t \end{bmatrix}$, the population density is:[9]

$$P(\vec{x}) = \frac{1}{l \cdot w \cdot h}\left(\frac{1}{2\pi}\right)^{\frac{3}{2}}\left(\frac{1}{|\boldsymbol{\Sigma}|}\right)^{\frac{1}{2}} exp\left(-\frac{1}{2}[\ln(\vec{x}) - \vec{\mu}]^T \boldsymbol{\Sigma}^{-1}[\ln(\vec{x}) - \vec{\mu}]\right) \qquad \text{Equation S4}$$

where $\vec{\mu} = \begin{bmatrix} \mu_l \\ \mu_w \\ \mu_h \end{bmatrix}$, a vector containing the mean value of each element of $y$, $\boldsymbol{\Sigma}$ is the covariance matrix of $y$:

$$\Sigma = \begin{bmatrix} {\sigma_l}^2 & \sigma_{lw} & \sigma_{lh} \\ \sigma_{wl} & {\sigma_w}^2 & \sigma_{wh} \\ \sigma_{hl} & \sigma_{wh} & {\sigma_h}^2 \end{bmatrix}$$

And $|\boldsymbol{\Sigma}|$ is the determinant, and $\boldsymbol{\Sigma}^{-1}$ the inverse of the covariance matrix, respectively.

With such a population distribution it would in principle be possible to calculate exact changes to a sample following a centrifuge process by applying the population change functions derived above to this expression. If $P_{stock}$ represents the population distribution of each nanosheet size in the stock, the function

$$P_L(N, L, \omega, t) = P_{stock}(N, L)F_L(N, L, \omega, t)$$

represents the population distribution left in the supernatant. The same could be applied for the sediment.

Unfortunately, it has proved impossible to measure such stock samples as prepared by exfoliation (without additional processing that would alter the nanosheet size populations) with sufficient accuracy. The wide range of nanosheet sizes prevent representative microscopy images from be obtained; they would need to simultaneously resolve the smallest nanosheets whilst capturing enough of the largest to achieve statistical significance. Other techniques to probe the entire ensemble have been effective at measuring averaged values, for example UV/VIS spectroscopy or non-resonant light scattering,[10-12] but they lack the granular detail required to calculate the correlation between parameters and yield average values of the ensemble.

Nevertheless, in this work, we demonstrate approximate methods for gaining information about the size of nanosheets under various circumstances by taking such narrow "cuts" of this population distribution, that- to a first approximation - it can be ignored.

## *2.4 Modelling the outcome of liquid cascade centrifugation (LCC)*

LCC is a stepwise centrifugation protocol used to size-select nanosheets.[13,14] The stock dispersion is first centrifuged at a low angular frequency, $\omega_1$, to generate a sediment (referred to as sediment-1) which can be separated from liquid-1 (i.e. the supernatant) by decantation. Sediment-1 then contains extremely large nanosheets (and usually some unexfoliated material in sediment-1 but not subsequent sediments).

The next step is for the separated supernatant (liquid-1) to be centrifuged a second time at a higher angular frequency, $\omega_2$, to generate sediment-2 and liquid-2 after decantation. Sediment-2 contains slightly smaller nanosheets than sediment-1. This process is repeated multiple times with continually increasing angular frequencies to generate a set of sediments containing the desired size-selected nanosheets. The sizes of the nanosheets decrease as the sediment number increases. In practice the sediments are re-dispersed for further use. Note that in principle, it is possible to achieve the size selection by subsequently increasing centrifugation time rather than rotor spinning speed, since the outcome of the centrifugation will be governed by $\omega^2 t$.[15]

We can use equations S3a&b to work out the fraction of initial nanosheets going into both sediment-1, $F_S(\omega_1)$ and liquid-1, $F_L(\omega_1)$. We can then take $F_L(\omega_1)$ and use it to calculate the fraction of initial nanosheets going into sediment-2 and liquid-2. In this way, we can calculate the fraction of initial nanosheets in all sediments and liquids (see figure 2D, main manuscript). N.B. it is necessary to use the appropriate angular frequency in each case.

For example, the fraction going into sediment-3 is

$$F_{Sed,3} = F_L(\omega_1)F_L(\omega_2)F_S(\omega_3) = (1 - F_S(\omega_1))(1 - F_S(\omega_2))F_S(\omega_3)$$

We can generalise this for the nth sediment by

$$F_{Sed,i} = F_S(\omega_i) \cdot \Pi_{j=1}^{i-1} F_L(\omega_j)$$

However, because the higher speed centrifugations remove more material to the sediment than the lower speed centrifugations, this can be replaced with reasonably accuracy by the approximation:

$$F_{Sed,i} \approx F_L(\omega_{i-1})F_S(\omega_i)$$

Using equations S3a and S3b, we can write this as

$$F_{Sed,i} \approx \left(1 - \left(\frac{R_2}{R_2 - R_1}\right)\left(1 - e^{-S\omega_{i-1}^2 t_{i-1}}\right)\right)\left(\left(\frac{R_2}{R_2 - R_1}\right)\left(1 - e^{-S\omega_i^2 t_i}\right)\right) \qquad \text{Equation S5}$$

Where *n* refers to the number of the centrifugation run which yielded the sediment in question. The function is plotted as function of *L* and *N* in Figure S 5 using typical input parameters. Note that *L* and *N* are contained in the sedimentation constant.

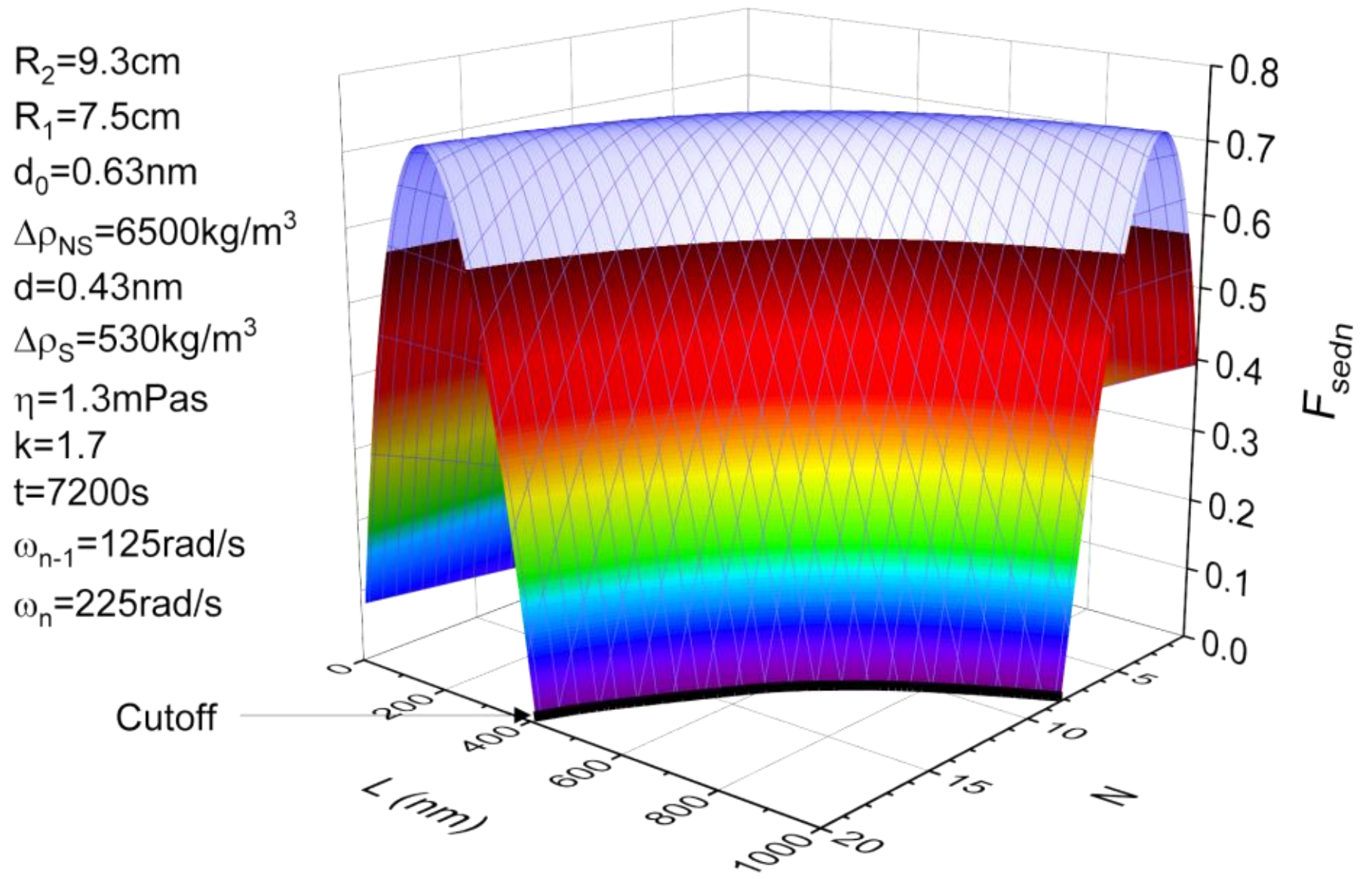

Figure S 5: Plot of $F_{sed,i}$, which is the fraction of initial nanosheets (in the stock) that end up in the nth sediment, versus nanosheet length, L, and layer number, N. This was plotted using equation S5 inputting the parameters given in the panel. Note that this function reaches zero for a well-defined set of L and N values marked by the thick black line (labelled "Cut-off"). For combinations of L and N above this cut-off line, $F_{sed,i}$=0 and no nanosheets in this size range enter the nth sediment. Thus, the cut-off line represents an upper limit to nanosheet size in the nth sediment.

We can understand $F_{sed,i}$ more simply by noting that a feature of the liquid phase exfoliation method of producing 2D materials is that the nanosheets tend to have length/thickness aspect ratios which cluster around a well-defined mean which is determined by nanosheet mechanics: $\langle L/t\rangle = \langle L/Nd_0\rangle = k_{Lt}$.[6] This means we can rewrite the sedimentation constant (neglecting the small contribution of the surfactant coating) as:

$$S \approx \frac{\Delta\rho_{NS}L^2}{10.5\eta\sqrt{k_{lw}}k_{Lt}}$$

We can use this equation to plot $F_{Sed,i}$ using Equation S5 on a 2D plot as a function of only $L$. Physically, this means focusing on a slice of the 3D plot in Figure S 5 in the region where most of the nanosheets lie. This is shown in figure 2E main manuscript (orange line) with the other lines representing $F_L(\omega_{i-1})$ and $F_S(\omega_i)$, the functions that constitute $F_{Sed,i}$ (i.e. $F_{Sed,i} \approx F_L(\omega_{i-1})F_S(\omega_i)$). The most obvious features of the graph of $F_{Sed,i}$ versus $L$ are as follows:

1. This function has a well-defined peak. This peak defines the size of the majority of nanosheets in a given sediment.
2. $F_{Sed,i}$ goes to zero at a well-defined value of $L$. This upper cut-off defined the maximum value of L found in a given sediment.

As can be seen in Figure S 6, for a given LCC cascade, the peak and cut-off values of L increase with $i$, i.e. with the angular frequency used to generate each sediment. In the example shown below, the angular velocities were chosen in such a way to result in uniform fraction distributions (unlike in the actual experiment conducted which is shown in Figure 2F in the main manuscript).

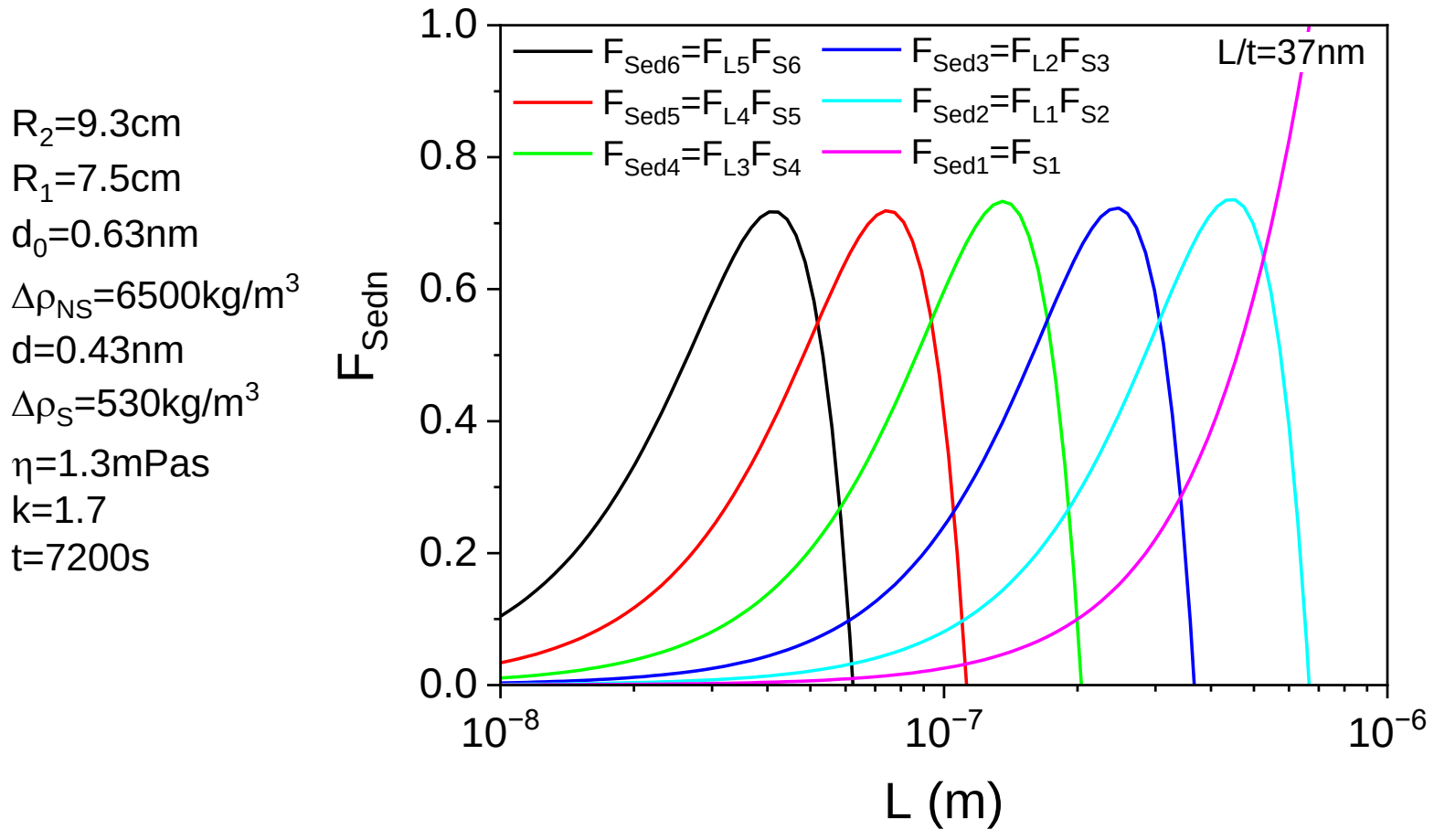

Figure S 6: Data for $F_{sed,n}$ versus L for sediments 1-6 in an LCC cascade. Angular frequencies used are: $\omega_1$=80 rad/s, $\omega_2$=145 rad/s, $\omega_3$=260 rad/s, $\omega_4$=470 rad/s, $\omega_5$=840 rad/s, $\omega_6$=1500 rad/s.

## 2.5 *The upper cut-off*

All curves have a very sharp upper cut-off which can be calculated by setting the first term in equation S5 equal to zero:

$$F_{Ln-1} = 1 - \left(\frac{R_2}{R_2 - R_1}\right)\left(1 - e^{-S\omega_{n-1}^2 t_{n-1}}\right) = 0$$

which means $S\omega_{i-1}^2 t_{i-1} = ln(R_2/R_1)$. Rearranging:

$$L_{cut} = \frac{10.5\eta\sqrt{k_{wl}}}{\omega_{i-1}^2 t(Nd_0\Delta\rho_{NS} + 2d\Delta\rho_S)} ln\left(\frac{R_2}{R_1}\right) \qquad \text{Equation S6}$$

This gives the cut-off value of L as a function of N for sediment-n.

Considering the mean aspect ratio, $\langle L/t \rangle = \langle L/Nd_0 \rangle = k_{Lt}$, and ignoring surfactant allows us to write

$$L_{cut} = \left(\frac{10.5\eta\sqrt{k_{wl}}k_{Lt}}{\omega_{i-1}^2 t\Delta\rho_{NS}} ln\left(\frac{R_2}{R_1}\right)\right)^{1/2}$$

Which is an approximate expression for the maximum nanosheet length in sediment-n for a sample with mean L/t aspect ratio of $k_{Lt}$.

## 2.6 *The most probable nanosheet size*

Similar to above, the size distribution of the nanosheets in the nth sediment is given by

$$P_{sed,i}(N, L) = P_{stock}(N, L)F_{Sed,i}(N, L)$$

Thus, in principle, to gather information about the size distribution of nanosheets in the nth sediment, one needs to know the size distribution of nanosheets in the stock. However, under normal circumstances, the size distribution of nanosheets in the stock is very broad (making it challenging to determine), certainly much broader than the peak representing $F_{Sed,i}$ (otherwise the size selection protocol is poorly chosen!). This means the function $P_{sed,i}(N,L)$ represents a peak which is similar in shape to $F_{Sed,i}(N,L)$. This allows us to compute the most probable nanosheet size by finding the peak value of $F_{Sed,i}(N,L)$. Access to equations for the most probable nanosheet size is important as we can consider them reasonable approximations to arithmetic mean nanosheet size $\langle N \rangle$.

We find the most probable nanosheet size by differentiating $F_{Sed,i}(N,L)$ with respect to nanosheet size and setting the derivative equal to zero.

In practice we note that once we neglect the surfactant coating, we get:

$$S = \frac{LNd_0 \Delta\rho_{NS}}{10.5\eta\sqrt{k_{wl}}}$$

This means that differentiating $F_{Sedn}(N,L)$ with respect to $(LN)$ will yield the most probable value of $(NL)$, $NL_{MP}$.

Finding $dF_{Sed,n}/d(NL) = 0$ and after some rearranging:

$$\frac{\omega_{i-1}^2 t_{i-1}}{(\omega_i^2 t_i + \omega_{i-1}^2 t_{i-1})} e^{S\omega_i^2 t_i} + \frac{\omega_i^2 t_i}{(\omega_i^2 t_i + \omega_{i-1}^2 t_{i-1})} \frac{R_1}{R_2} e^{S\omega_{i-1}^2 t_{i-1}} = 1$$

Solving this equation requires the approximation that $\omega_i^2 t_i \approx \omega_{i-1}^2 t_{i-1}$, i.e. that the centrifugation conditions change very little between steps in the cascade. This is the case for centrifugation scenarios that target a significant narrowing of the initial distribution. Applying this approximation yields:

$$(LN)_{MP} \approx \frac{10.5\eta\sqrt{k_{wl}}}{\omega_i^2 t d_0 \Delta\rho_{NS}} ln\left(\frac{2R_2}{R_2+R_1}\right) \approx \langle LN \rangle \qquad \text{Equation S7}$$

where we use the upper angular frequency, $\omega_i$, i.e. the one used to generate the sediment which contained the nanosheets in question. The appropriateness of this choice has been verified experimentally in Figure S 22.

This equation is important as it allows an estimation of $\langle LN \rangle$ once $k_{lw}$ is known. This is useful because $k_{lw}$ is usually in a narrow range (~1.5-2.5) for most nanosheet types meaning that all parameters in this equation are known to some degree.[6] This means this equation can be used to sanity check AFM data and the observed dependence of nanosheet size on centrifugation speed. In addition, this equation allows $N$ to be estimated from TEM measurements of L for LCC data, even if a fixed aspect ratio has not been established, for example for a new production method. Further, the expression is very useful to test the accuracy of the model using published datasets from a range of materials, solvents and cascades (see main manuscript and SI section 7).

Applying the mean aspect ratio approximation yields

$$L_{MP} \approx \left(\frac{10.5\eta\sqrt{k_{wl}}k_{Lt}}{t\Delta\rho_{NS}} ln\left(\frac{2R_2}{R_2+R_1}\right)\right)^{1/2} \frac{1}{\omega_i} \approx \langle L \rangle \qquad \text{Equation S8}$$

And

$$N_{MP} \approx \left(\frac{10.5\eta\sqrt{k_{wl}}}{tk_{Lt}d_0^2\Delta\rho_{NS}} ln\left(\frac{2R_2}{R_2+R_1}\right)\right)^{1/2} \frac{1}{\omega_i} \approx \langle N\rangle \qquad \text{Equation S9}$$

## 2.7 *Beyond the mean aspect ratio approximation*

Previous experiments have shown that for nanosheets produced by LPE, the lateral dimensions are related to the nanosheet thickness via:[6]

$$\langle LW\rangle = D_{ML}^2\langle N\rangle^{\beta}$$

Where $D_{ML}$ is the characteristic monolayer size and $\beta$ is a constant in the range 1.5-3, being typically centred at ~2.8. N.B. the $k_{Lt}$ mean aspect ratio approximation implies $\beta$=2. To reduce the dimensionality of this relationship, we approximate $\langle LW\rangle \approx \langle L\rangle\langle W\rangle$, allowing us to rearrange this into a direct expression of length:

$$\langle L\rangle \approx \sqrt{k_{wl}}D_{ML}\langle N\rangle^{\beta/2}$$

Which is approximately equal to

$$\langle LN\rangle \approx \sqrt{k_{wl}}D_{ML}\langle N\rangle^{1+\beta/2}$$

Equating this with the equation for the most probable value of $(LN)_{MP}$ yields

$$(LN)_{MP} \approx \langle LN\rangle \approx \sqrt{k_{wl}}D_{ML}\langle N\rangle^{1+\beta/2} \approx \frac{10.5\eta\sqrt{k_{wl}}}{td_0\Delta\rho_{NS}} ln\left(\frac{2R_2}{R_2+R_1}\right)\frac{1}{\omega_i^2}$$

$$so$$

$$\langle N\rangle^{1+\beta/2} \approx \frac{10.5\eta}{td_0\Delta\rho_{NS}D_{ML}} ln\left(\frac{2R_2}{R_2+R_1}\right)\frac{1}{\omega_i^2}$$

$$and$$

$$\langle N\rangle \approx \left(\frac{10.5\eta}{td_0\Delta\rho_{NS}D_{ML}} ln\left(\frac{2R_2}{R_2+R_1}\right)\right)^{1/(1+\beta/2)} \frac{1}{\omega_i^{1-(\beta-2)/(\beta+2)}} \qquad \text{Equation S10}$$

Where the exponent on $\omega_i$ has been rearranged slightly to emphasise the deviation from 1.

This means that ⟨*N*⟩ can fall with $\omega_i$ slightly more slowly than expected e.g. if beta=2.5,

$$\langle N\rangle \propto \frac{1}{\omega_i^{0.9}}$$

Similarly, $\langle L\rangle \approx \sqrt{k_{wl}}D_{ML}\langle N\rangle^{\beta/2}$, so

$$\langle L\rangle \approx \sqrt{k_{wl}}D_{ML}\left(\frac{10.5\eta}{td_0\Delta\rho_{NS}D_{ML}} ln\left(\frac{2R_2}{R_2+R_1}\right)\right)^{\beta/(2+\beta)} \frac{1}{\omega_i^{1+(\beta-2)/(\beta+2)}} \qquad \text{Equation S11}$$

This means if beta=2.5, $\langle L\rangle_{MP} \propto \frac{1}{\omega_i^{1.1}}$

N.B. when $\beta$=2, these equations become equivalent to equations S7 and 8 (because $D_{ML}/d_0 = k_{Lt}$).

Experimentally, different exponents relating ⟨*N*⟩ and ⟨*L*⟩, respectively, to rotational speed are often found.[16-18] The section above explains their origin, i.e. the exponent $\beta$ relating lateral size and thickness. Note that it is currently not yet clear which factor governs this exponent. However, being able to essentially extract it form LCC data is of great values since $\langle L\rangle$ and $\langle N\rangle$ can often be reliably determined from spectroscopic metrics such as the ones used to evaluate the band sedimentation experiments.

### *2.8 Péclet Number and Effect of Diffusion*

Throughout this work we have neglected the influence of diffusion on the nanosheet movement. We find such influences are not required for typical 2D nanomaterials under laboratory centrifuge conditions as outlined below.

The Péclet number, a ratio of the rate of diffusion to rate of advection is commonly used to determine the dominant influence (i.e. diffusion or sedimentation) within a given system. For nanosheets, this ratio can be calculated from the relative time scales of advection – in this case sedimentation, and diffusion according to Brownian motion. For $Pe \gg 1$ the advective process is dominant and diffusion can be safely neglected.

$$Pe = \frac{diffusion\ time}{sedimentation\ time}$$

Where the diffusion time is given by:

$$\frac{(r-r_0)^2}{D} = \frac{(r-r_0)^2 f}{k_B T}$$

And the sedimentation time can be rearranged from Equation S1 as:

$$t = \frac{\ln\left(\frac{r}{r_0}\right) f}{\omega^2 A(h\Delta\rho_{NS} + 2d\Delta\rho_S)}$$

The Péclet number, cancelling the friction terms, is therefore:

$$Pe = \frac{(r-r_0)^2 \omega^2 A(h\Delta\rho_{NS} + 2d\Delta\rho_S)}{k_B T \ln\left(\frac{r}{r_0}\right)} = \frac{(r-r_0)^2 \omega^2 {k_{Lt}}^2 \left(h^3\Delta\rho_{NS} + 2dh^2\Delta\rho_S\right)}{k_{lw} k_B T \ln\left(\frac{r}{r_0}\right)}$$

Plotting *Pe* as a function of angular velocity and layer number in a typical centrifuge experiment for graphene (TMDs and other materials with greater densities will have even greater *Pe* values), it is clear that, even at low speeds, the sedimentation dominates by orders of magnitude over diffusion effects.

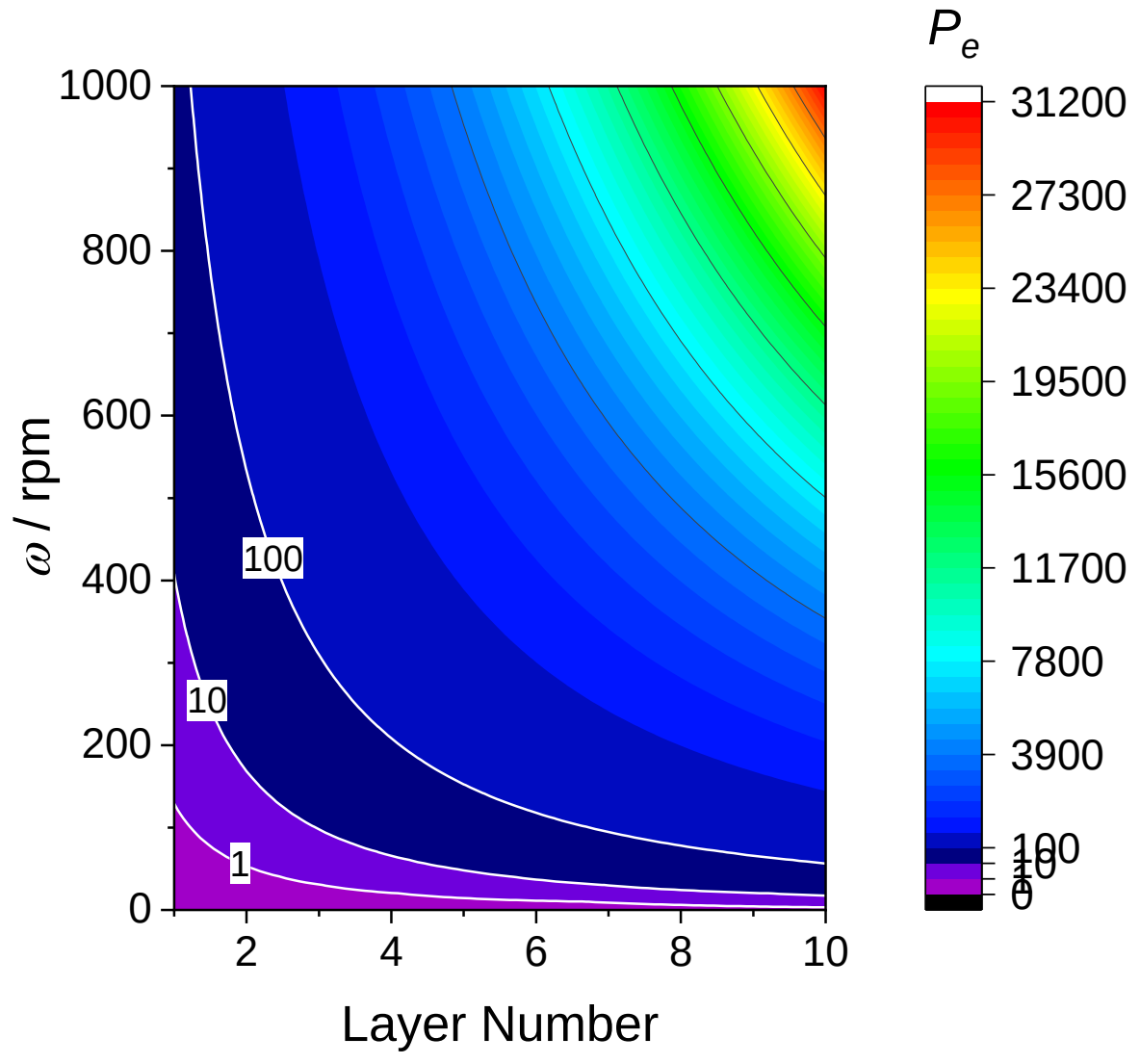

Figure S 7: Contour plot of *Pe* for graphene with typical conditions used for centrifuge experiments as detailed in Table S1 and S2.

# 3 Experimental Methods

## *3.1 Preparation of the dispersion*

Graphite and group VI-TMD dispersions were prepared by probe sonicating the powder with an initial concentration 20 g $L^{-1}$ in an aqueous sodium cholate (SC) solution. The powder was immersed in 80 mL of aqueous surfactant solution ($C_{SC}$= 6 g $L^{-1}$). The mixture was sonicated under ice-cooling in a 100 mL metal beaker by probe sonication using a solid flathead tip (Sonics VXC-500, *i.e. 500* W) for 1 h at 60 % amplitude with a pulse of 6 s on and 2 s off. During the sonication, the sonic probe was placed 1.5 cm from the bottom of the beaker. The dispersion was centrifuged in 20 mL aliquots using 50 mL vials in a Hettich Mikro 220R centrifuge equipped with a fixed-angle rotor 1016 at 5000 rpm for 1.5 h. The supernatant was discarded and the sediment collected in 80 mL of fresh surfactant ($C_{SC}$= 2 g $L^{-1}$) and subjected to a second sonication using a solid flathead tip (Sonics VX-500) for 5 h at 60 % amplitude with a pulse of 6 s on and 2 s off. From our experience, this two-step sonication procedure yields a higher concentration of exfoliated material and removes impurities.[19] $WS_2$ exfoliation in *N*-methyl-2-pyrrolidone was performed under identical conditions except for the solvent.

## *3.2 Band sedimentation*

In all cases band sedimentation experiments were prepared and conducted in 1 g $L^{-1}$ sodium cholate dispersions. A slight density gradient down the centrifuge tube was then prepared using $H_2O/D_2O$ mixtures. This gradient is used to ensure stability of the bands and remove convective effects that result from a uniform density medium. At the bottom of this gradient is a $D_2O$ solution up to 5 mL of usable volume – the very base of centrifuge tubes with a tapered or conical shape is not collected. Above this, a 5 mL layer of 50/50 vol.% $H_2O/D_2O$ was floated. 1 mL of concentrated nanomaterial dispersion in $H_2O$ was then floated on the top of this gradient and the samples immediately centrifuged to minimize diffusion mixing.

At and below 4000 rpm, band sedimentation experiments were completed in a Hettich Rotanta centrifuge using a 5094 swinging-bucket rotor and 15 mL graduated centrifuge tubes. The graduated scale printed on the tubes was used to measure sample volumes for preparation and extraction. Because this centrifuge has no temperature control, water samples were centrifuged under the same conditions in 5 min steps, with the temperature of the liquid recorded between each 5 min experiment. This showed a linear increase over 20 minutes, followed by a consistent temperature of 36 $^{o}$C which was taken as an average temperature value for the density and viscosity of the $H_2O/D_2O$ solutions.

At higher speeds above 4000 rpm, band sedimentation experiments were undertaken in a Beckman Coulter Avanti J-26S XP with a JS 13.1 swinging bucket rotor. The Beckman centrifuge tubes have no printed scale, instead an external sample holder with printed scale was used for sample preparation and extraction. Here, the temperature was controlled and set to 20 °C.

After centrifuging the samples were immediately processed and bands removed in 1 mL aliquots using a needle and syringe. The radius values were measured for each aliquot from the rotor and tube height.

A limited initial size selection was used for all samples prior to band sedimentation experiments. Freshly prepared dispersions were centrifuged for 2 hours and fractions were collected from supernatant and sediment, respectively, between the following boundaries given in rpm:

| Material | Band Sedimentation Speed | Initial Sample Selection |
|---|---|---|
| $WS_2$ | 2000 rpm | 0.9-1.8 krpm & 1.8 – 7 krpm |
| | 4000 rpm | 0.9-1.8 krpm & 1.8 – 7 krpm |
| | 8000 rpm | 7-9 krpm |
| $MoS_2$ | 4000 rpm | 0.9-1.8 krpm |
| Graphene | 8000 rpm | 1.8 – 15 krpm |

### *3.3 Liquid cascade centrifugation*

The $WS_2$ dispersions in NMP and SC, respectively, were each split into 2 equal aliquots and centrifuged at 15 krpm for 3h. The sediments were redispersed in water and centrifuged again at 15 krpm. After this step, half of each dispersion was redispersed in SC (2 g $L^{-1}$) and half in NMP resulting in 4 stock dispersions subjected to LCC.

For cascade centrifugation, two different centrifuges with three different rotors were used: For centrifugal accelerations < 5 krpm, a Hettich Mikro 220R centrifuge equipped with a fixed-angle rotor 1016 was used, for 5-10 krpm, a Beckman Coulter Avanti XP centrifuge with a JA25.50 fixed angle rotor with conical 50 mL centrifuge tubes that were filled with ~20 mL dispersion each (filling height 5 cm). For even higher speeds, the Hettich Mikro 220R centrifuge was equipped with a 1195-A rotor with 1.5 mL Eppendorf tubes. The rotor specifications of the Hettich 1016 and the Beckman JA25.50 are comparable resulting in similar $R_1$ = 75 mm and $R_2$ = 93 mm. However, the Hettich 1195-A would result in a different $R_1$ and $R_2$. This was neglected for the considerations of this manuscript. For the $WS_2$ data shown in the main manuscript, all centrifugation runs were performed for 2 h (10°C).

Unexfoliated material was removed by centrifugation at 900 rpm. The supernatant was subjected to further centrifugation at 1.8 krpm. The sediment was collected in fresh surfactant ($C_{SC}$= 0.1 g $L^{-1}$) at reduced volume (3-8 mL), while the supernatant was centrifuged at 2.9 rpm. Again, the sediment was collected and the supernatant subjected to centrifugation at higher speeds. This procedure was repeated with the following centrifugation speeds: 6.5 krpm, 9 krpm, 15 krpm. As sample nomenclature, the lower and upper boundary of the centrifugation are indicated.

Additional LCC data for other materials analysed in the SI is published data as indicated.

### *3.4 UV/VIS extinction spectroscopy*

UV/VIS extinction spectra were recorded on a Cary 60 UV-Vis Spectrophotometer (Agilent) using 0.5 nm data intervals and 0.1 s averaging times. Samples were recorded in 10 mm pathlength high performance quartz glass cuvettes. Where required, samples were diluted into a 1 g $L^{-1}$ aqueous sodium cholate solution, 50 vol.% $H_2O/D_2O$ to approximate the average solvent composition from the centrifuge tube. This 1 g $L^{-1}$ sodium cholate in 50 vol.% $H_2O/D_2O$ solution was also used for background subtraction of the UV/VIS spectra. The nIR spectra for $H_2O/D_2O$ analysis was recorded on a Cary 6000 UV-Vis-nIR Spectrophotometer but all other conditions were the same.

### *3.5 Atomic force microscopy*

Atomic force microscopy was carried out on a Dimension ICON3 scanning probe microscope (Bruker AXS S.A.S.) in ScanAsyst in air under ambient conditions using aluminium coated silicon cantilevers (OLTESPA-R3). The concentrated dispersions were diluted with isopropanol (in the case of NMP) or water (in case of the surfactant samples) to optical densities <0.1 at 500 nm. A drop of the dilute dispersions (15 μL) was deposited $Si/SiO_2$ wafers (0.5x0.5 $cm^2$) with an oxide layer of 300 nm previously coated with (3-Aminopropyl)triethoxysilane (APTS). To this end, the substrate was immersed in a solution of APTS and deionized water (1:40) for 15 minutes. Subsequently, the substrate was removed, washed repeatedly with deionized water, and finally dried with pressurized nitrogen. Once the substrate is prepared, the nanomaterial dispersion was dropped onto the substrate and held for ~20 seconds. Subsequently, the drop was discarded (e.g. blown off with pressurized nitrogen) and the sample repeatedly cleaned with deionized water and isopropanol. Typical image sizes ranged from 20x20 for larger nanosheets to 5x5 $\mu m^2$ at scan rates of 0.5-0.8 Hz with 1024 lines per image. Previously performed step height analysis was used to convert the apparent AFM thickness to layer number.[14] Previously published length corrections were used to correct lateral dimensions from cantilever broadening.[20]

# 4 Additional data band sedimentation

## *4.1 Solvent parameters*

To test the theoretical sedimentation rate by way of band sedimentation, accurate values of the solvent density and viscosity must be known. To stabilise the bands during sedimentation against convection effects, a subtle $H_2O$ to $D_2O$ gradient was employed with increasing density down the centrifuge tube. Fortunately, we find it is not required to fully include this height dependent density and viscosity, instead averaged values can be used.

To determine the average $H_2O/D_2O$ content we employ nIR extinction spectroscopy, taking bands from the centrifuge tube with the same method as employed for sedimentation studies – needle and syringe using the tube graduations for height. Using the water peak at 1196 nm (control spectra and concentration trend shown in Figure S 8A & B), the vol% of water in each fraction could be measured. We confirm that the gradient established at the start of the experiment remains mostly unchanged after 40 min, seen in Figure S 8C, so the same average vol.% of all fractions as prepared was used to calculate density and viscosity for all experiments.

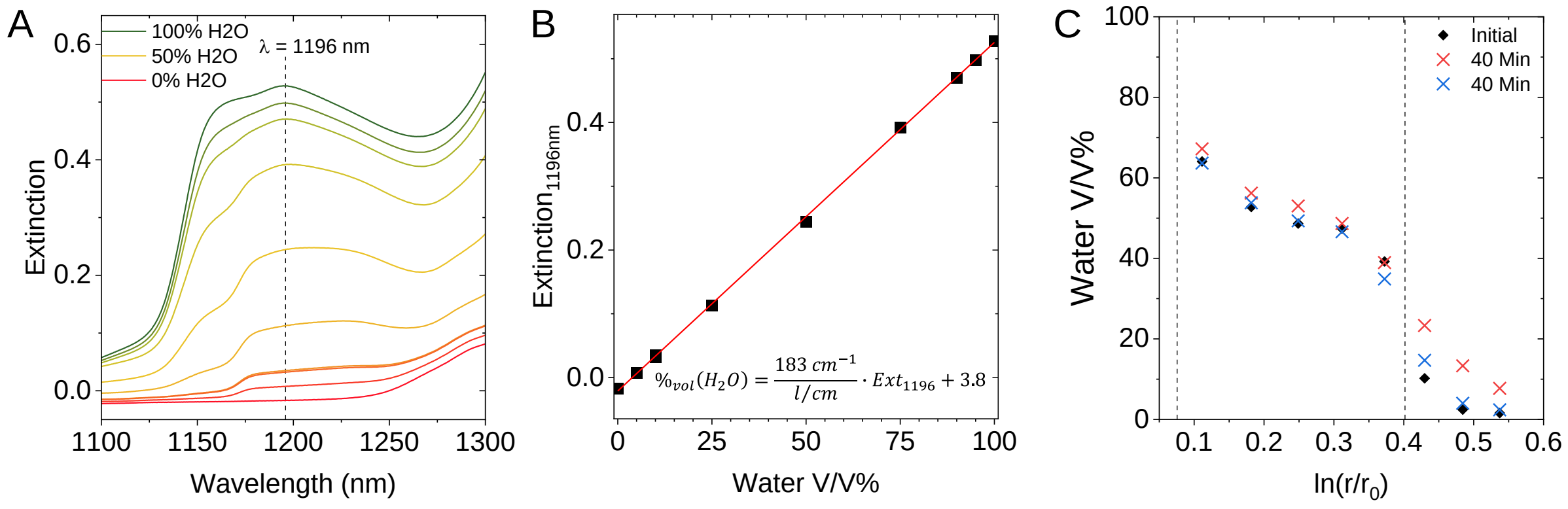

Figure S 8: Control experiments to measure the water to heavy water content. A shows nIR extinction spectra of $H_2O/D_2O$ mixtures used to establish the concentration calibration shown in B for the water absorption peak at 1196 nm. The red line shows a Beer-Lambert law linear fit. C shows the water content of different fractions, plotted using the same log scale as sediment experiments in the main manuscript, black diamonds show fractions extracted after no time, red crosses the fractions after 40 min standing under ambient conditions, and blue crosses show fractions collected after 40 mins centrifugation.

Note: the negative extinction values for low water content are not an experimental error, but an unavoidable outcome from conducting the experiment across the full composition range. Measuring 0% and 100% water is most accurately achieved using the baseline of the empty cuvette to account only for the quartz; however, when measuring 100% $D_2O$ there is more light refracted along the beam-path than observed with an empty cuvette, leading to the unusual negative y-intercept observed in our Beer-Lambert fit.

All datasets, except the 8000 rpm graphene experiment, were also conducted in a centrifuge without temperature control. It was observed that all samples increased in temperature after centrifuging, which will also affect the density and viscosity of the $H_2O/D_2O$ mixture. This was investigated by recording the temperature after centrifuging solvent samples for increasing time periods. The trend, shown in Figure S 9, clearly shows an initial approximately linear increase in temperature until a maximum, presumably when radiant heat loss compensates for the heat generated by the motor. Over 40 mins, we also take the average temperature to find an appropriate solvent density and viscosity.

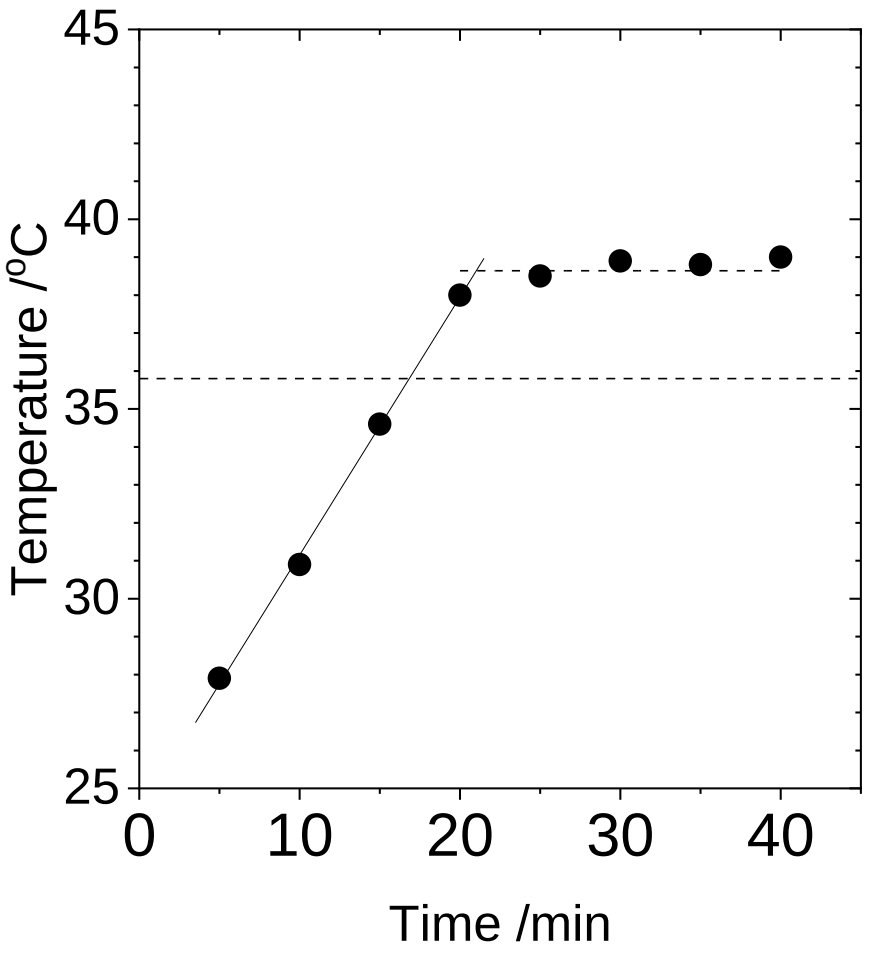

Figure S 9: Temperature gradient of water after different centrifugation times in Hettich model centrifuge without temperature control. The lines show linear fits, but the horizontal dashed line shows the average value used to calculate an appropriate solvent density and viscosity.

Based on these averaged parameters for the solvent mixture,[21] the parameters used for the model were: $\eta$ = 7.8 x $10^{-4}$ kg $m^{-1}$ $s^{-1}$ and $\rho_l$ = 1060 kg $m^{-3}$.

### *4.2 Extinction spectra of fractions*

To analyse the average nanosheet size in all the fractions collected from band sedimentation we use UV/VIS extinction spectra. The strong correlation between optical spectra and layer number for 2D materials has been well established, and for semi-conducting nanosheets we have recently produced a refined analysis algorithm to reliably handle large numbers of spectra.[11]

All fractions were collected and diluted into 1 g $L^{-1}$ sodium cholate solutions of 50 vol.% $H_2O$ to $D_2O$. This solvent was selected to mostly evenly approximate all fractions throughout the centrifuge tube to minimise any errors from the solvent baseline. Exact dilutions depended on sample concentration, but in every case the spectra are shown normalised and the layer number analysis requires only the A-exciton peak position for $WS_2$ and $MoS_2$, and relative intensities at the peak maximum and 550 nm for graphene.[22]

It is acknowledged that extinction spectra can only probe the volume fraction weighted average of the entire distribution. However, band sedimentation is known to produce narrow, Gaussian distributions of particle size, making this mean value an excellent measure of nanosheet size in each fraction as a function of distance migrated.[23]

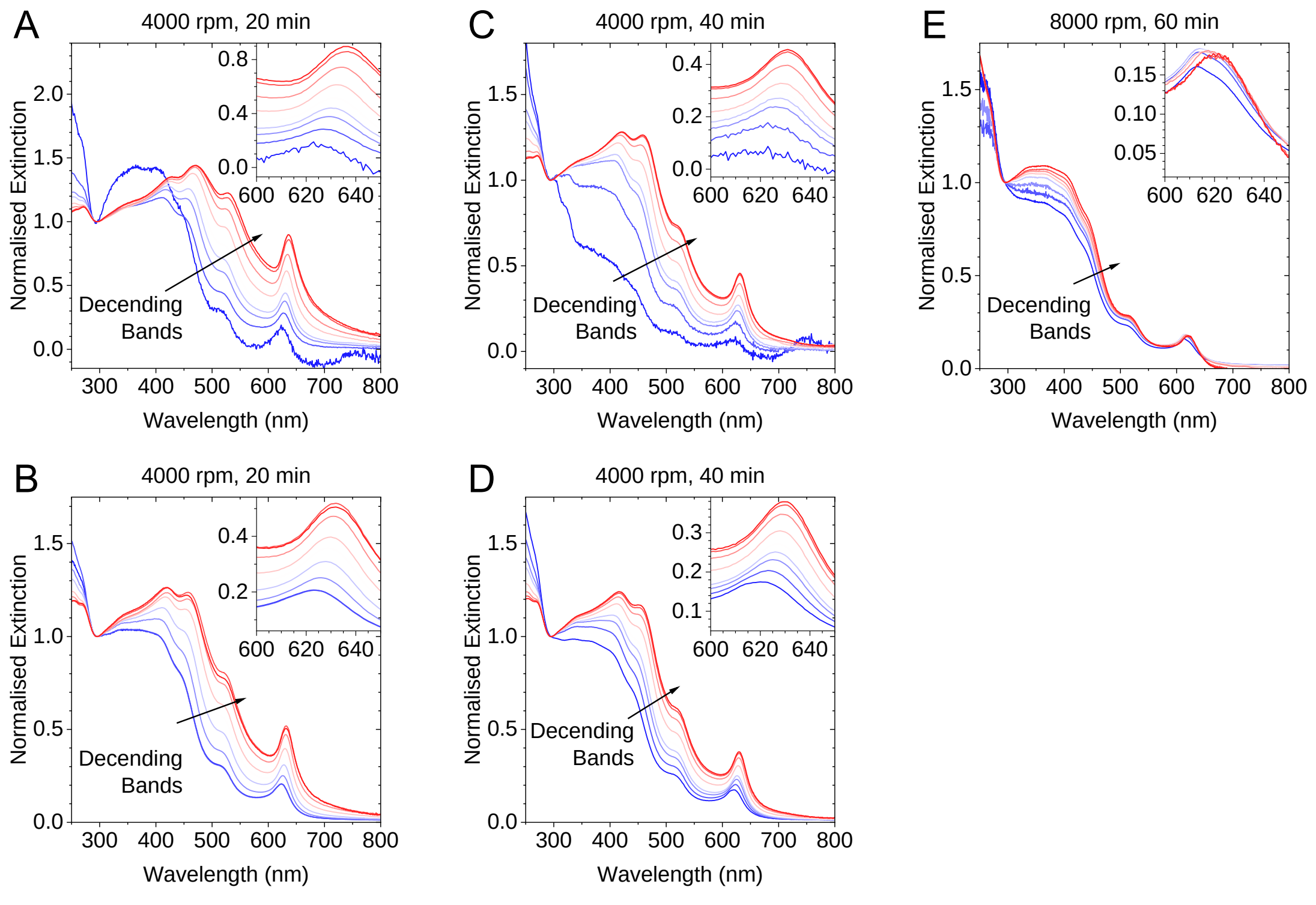

Figure S 10: Extinction spectra of $WS_2$ samples collected from descending bands down the centrifuge tube, further distance travelled noted by blue-to-red colour change and direction noted by arrows on plots. Pre-size selection was undertaken, data sets A and C are from samples pre-size selected between 0.9-1.8 krpm, B and D collected between 1.8-7 krpm, E used a starting distribution collected between 67-9 krpm.

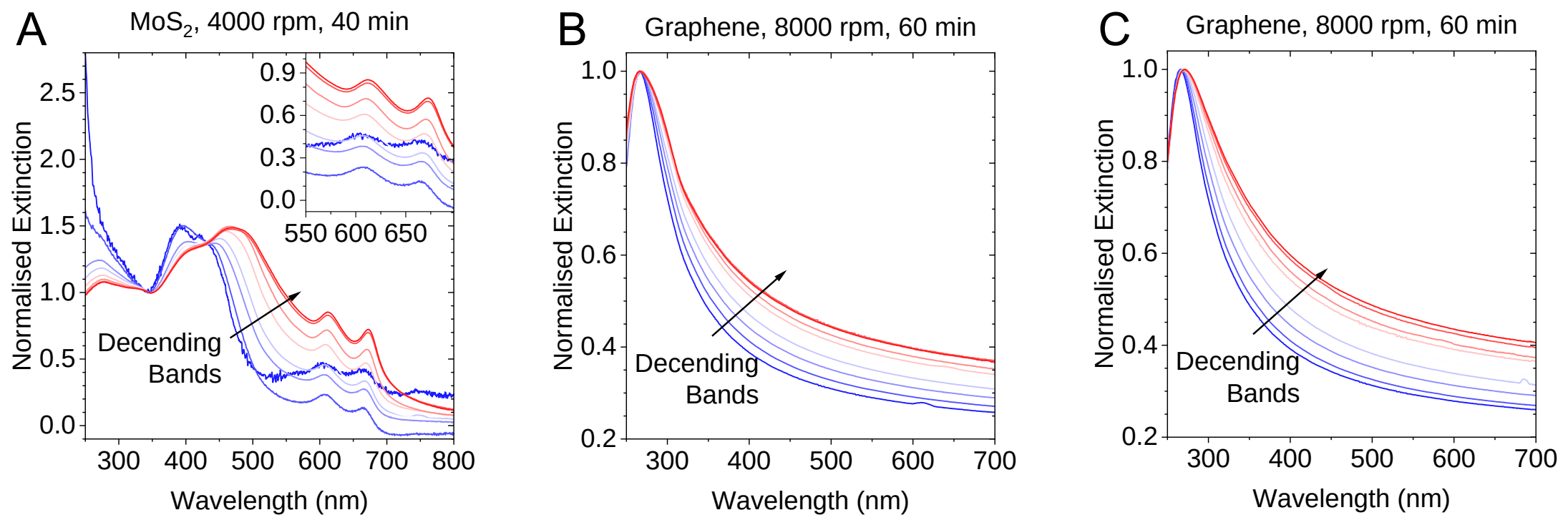

Figure S 11: A = 0.9-1.8 krpm $MoS_2$, B = 1.8-15 krpm graphene, C = 15 krpm supernatant graphene.

# 5 AFM statistics of the $WS_2$ samples produced by LCC

## *5.1 Stock dispersions*

Representative AFM images of the four $WS_2$ stock dispersions subjected to LCC are shown in Figure S 12 and Figure S 13. The deposition on APTS-coated Si/$SiO_2$ wafers (see methods) results in homogeneous coverage without stains from drying and allows for the deposition from the high boiling point solvent NMP without solvent exchange.

The samples are extremely polydisperse in lateral dimensions (L ranging from 10-1000 nm) and thickness (N ranging from 1-30 layers). In such samples, a determination of average dimensions is extremely challenging, if not impossible, as multiple images with different magnification are required (i.e. small-sized images to measure the small nanosheets and large-sized images to fully image the largest nanosheets). In spite of best attempts, the size and thickness distribution of the stock samples could not accurately be determined and the datacloud shown as faded grey datapoints in Figure 3A (main manuscript) should be rather regarded as approximation to illustrate the effect of the centrifugation.

**A) Stock LPE SC as exfoliated**

**B) Stock LPE SC transferred to NMP**

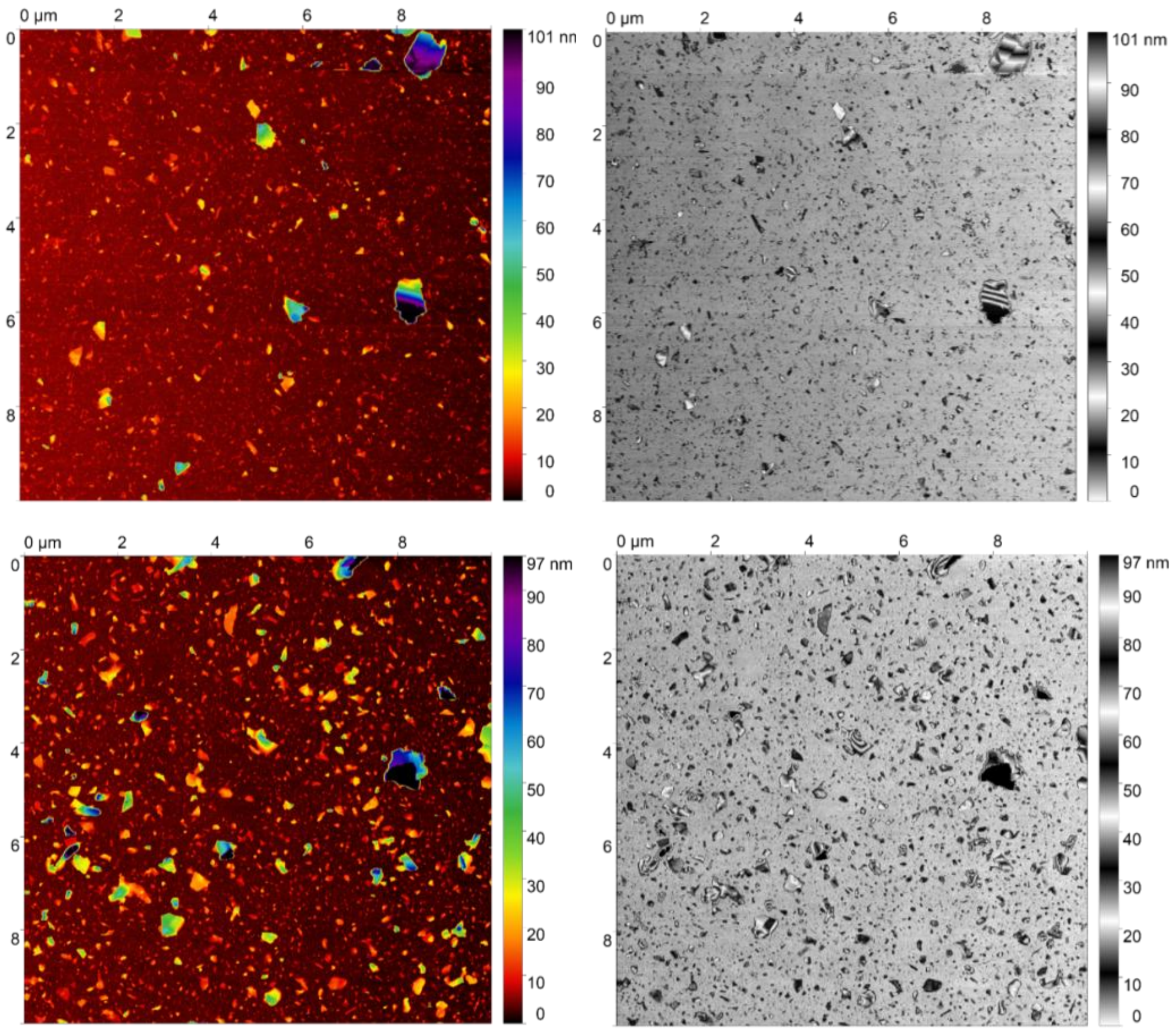

Figure S 12: AFM images in two different colour scales of $WS_2$ exfoliated in aqueous sodium cholate. A) As produced, B) sediment redispersed in NMP after centrifugation at 15 krpm.

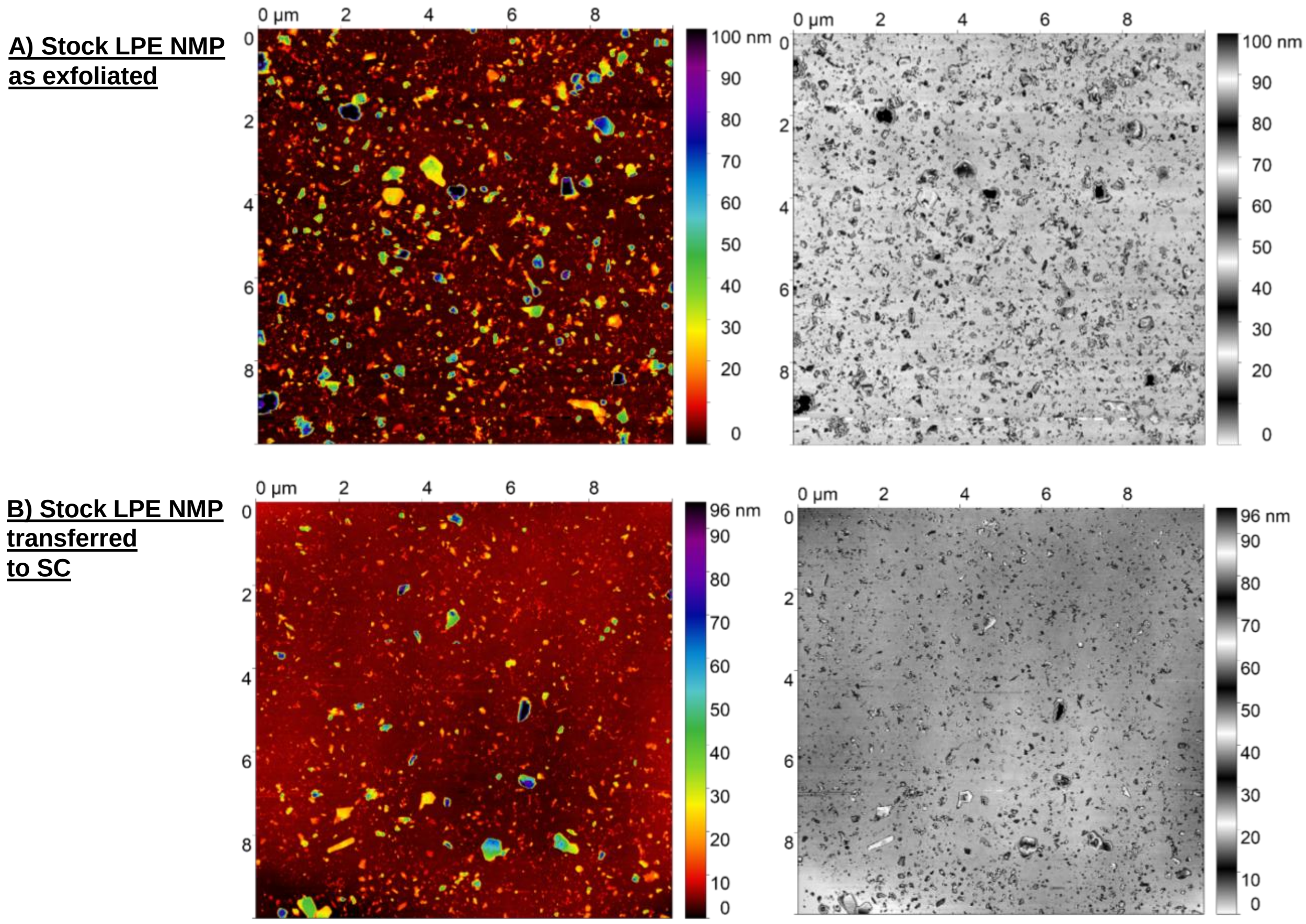

Figure S 13: AFM images in two different colour scales of $WS_2$ exfoliated in NMP. A) As produced, B) sediment redispersed in aqueous sodium cholate after centrifugation at 15 krpm.

## *5.2 LCC fractions*

Representative AFM images of the LCC size-selected fractions of the four $WS_2$ stock dispersions are shown in Figure S 14 and Figure S 15, the distribution histograms of $N$, $L$ and $W$ in Figure S 16 - Figure S 19.

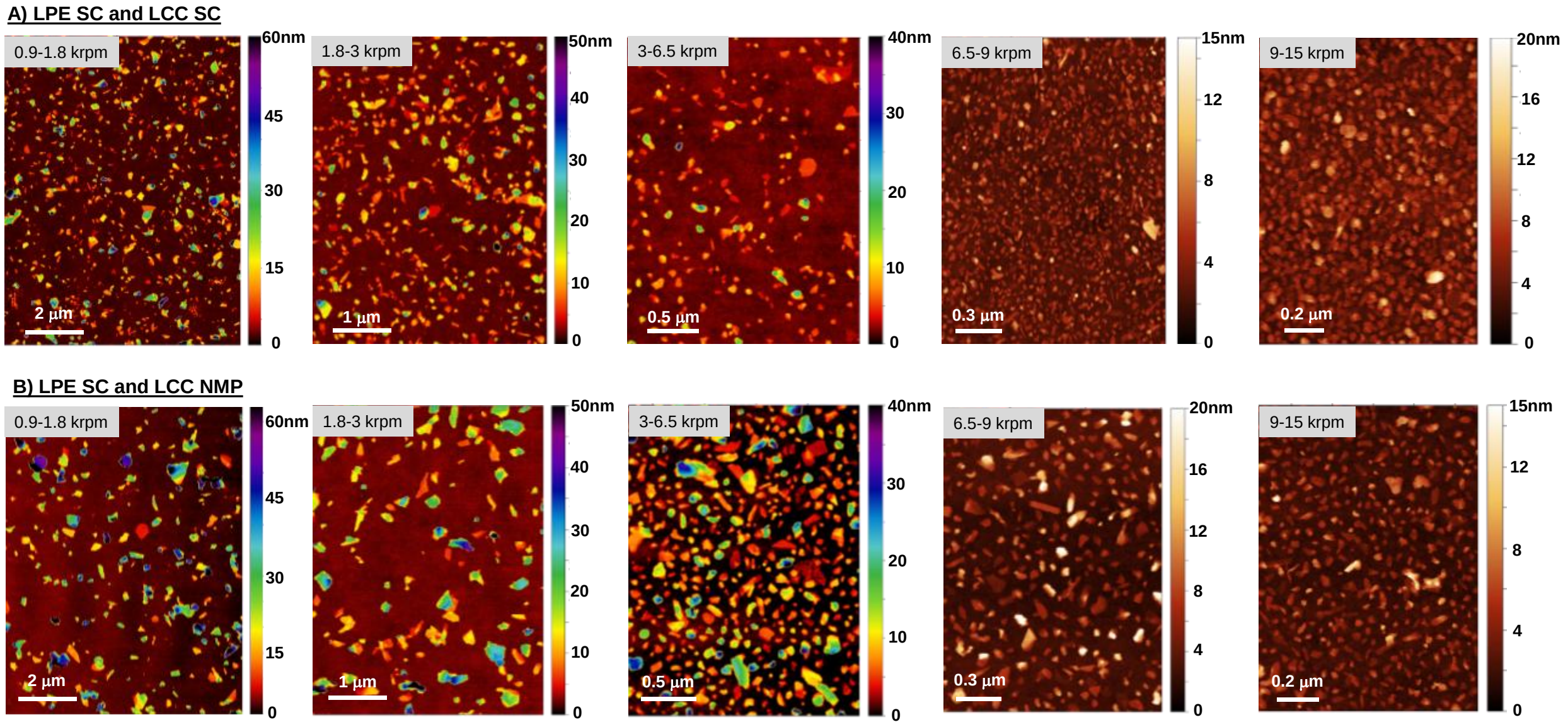

Figure S 14: AFM images of LCC size-selected $WS_2$ produced by LPE in aqueous sodium cholate. A) Centrifugation performed in aqueous sodium cholate. B) Centrifugation performed in NMP after transferring the stock dispersion to NMP.

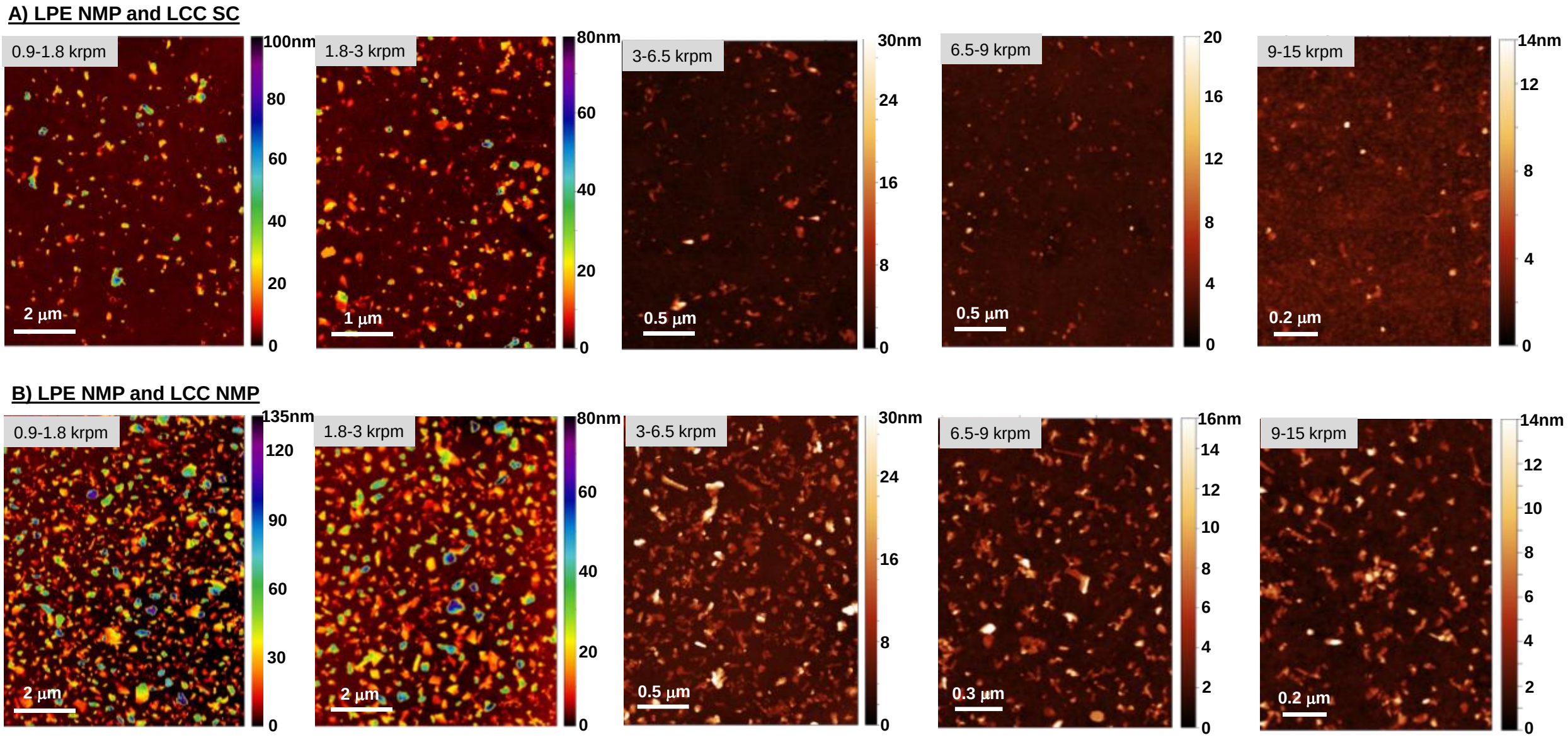

Figure S 15: AFM images of LCC size-selected $WS_2$ produced by LPE in NMP. A) Centrifugation performed in aqueous sodium cholate after transferring the stock dispersion to aqueous SC. B) Centrifugation performed in NMP.

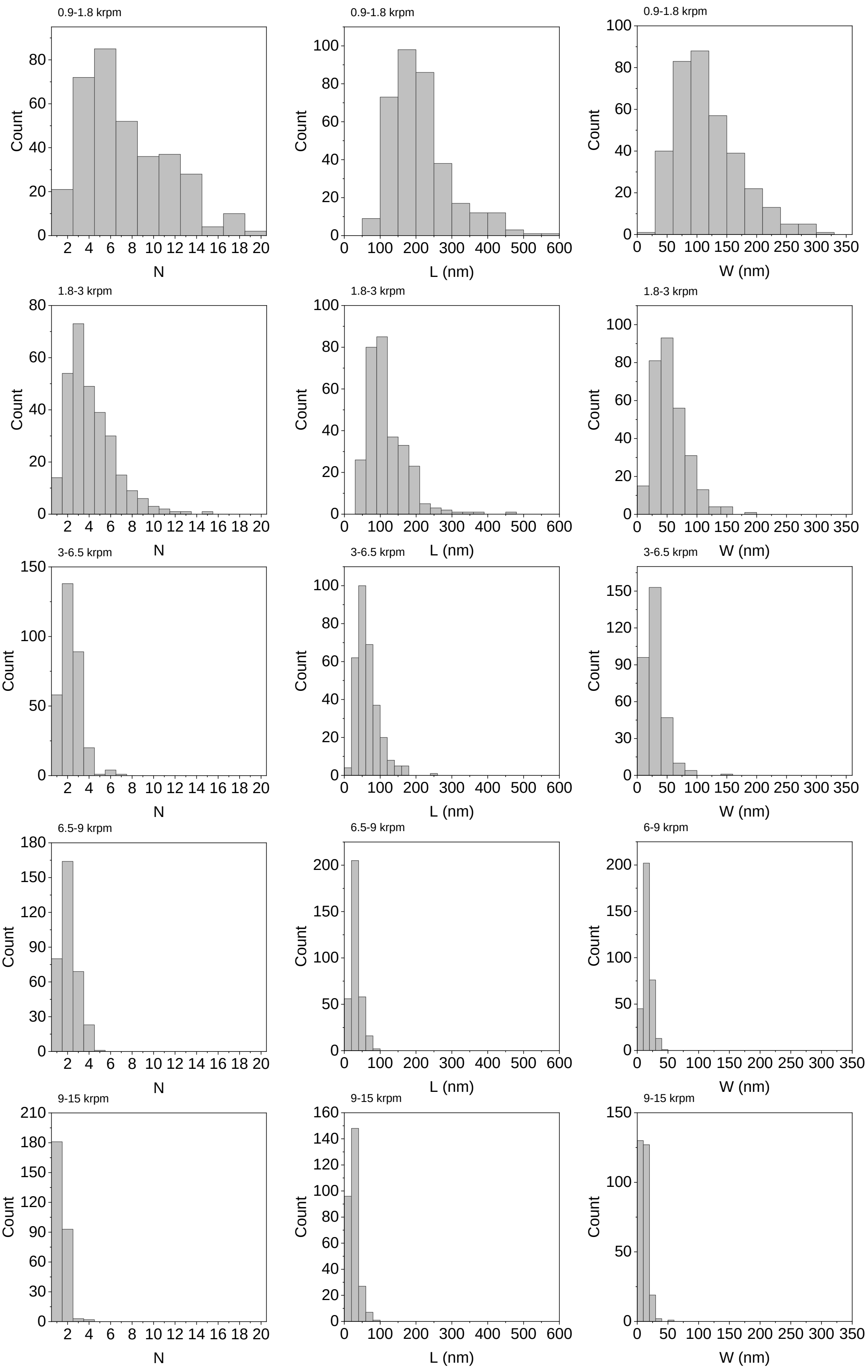

Figure S 16: Size distribution histograms of the LCC size-selected $WS_2$ produced and centrifuged in aqueous SC. From left to right: Layer number, Length, Width. Increasing centrifugation speeds and decreasing size from top to bottom.

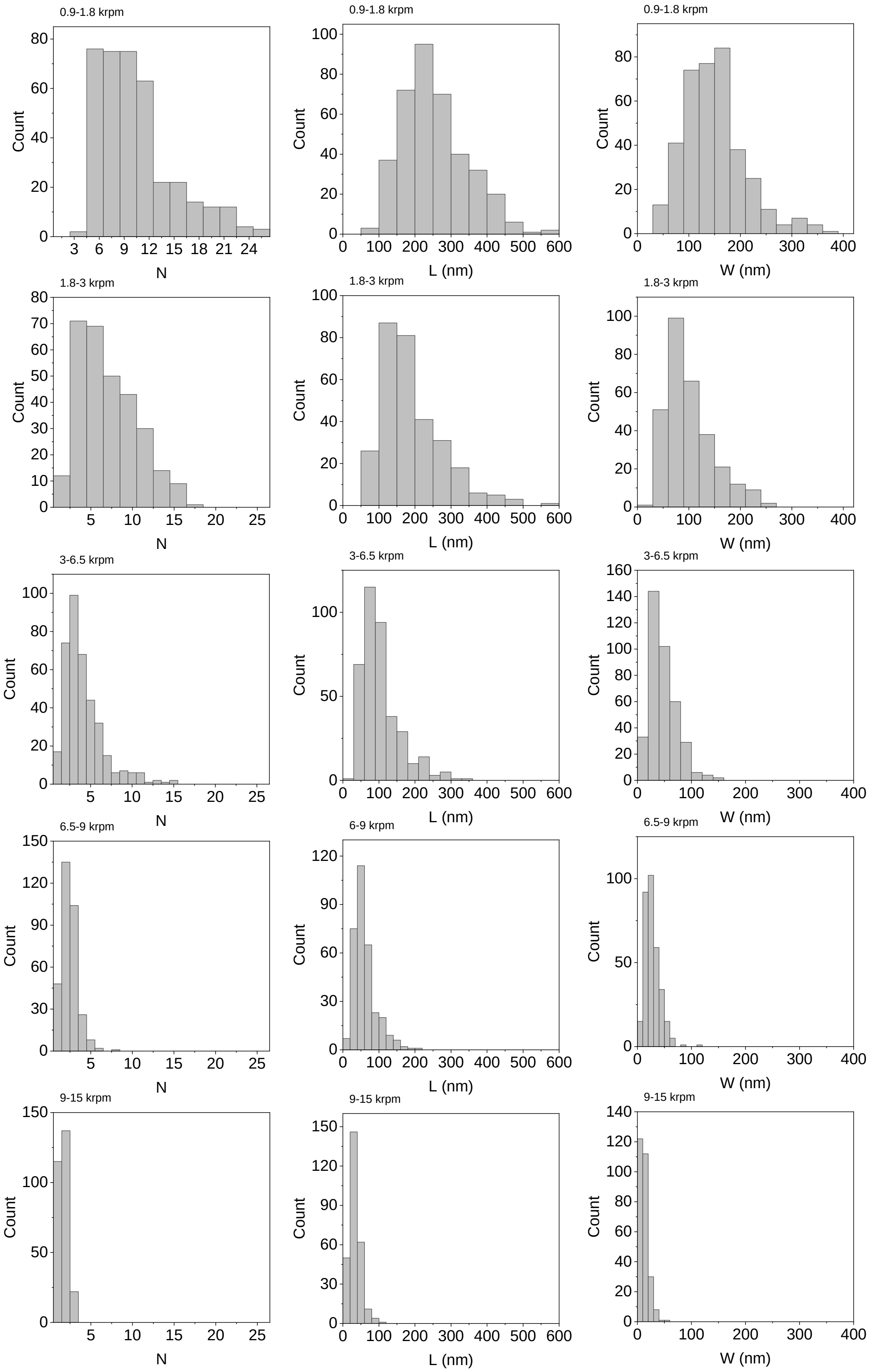

Figure S 17: Size distribution histograms of the LCC size-selected $WS_2$ produced aqueous SC and transferred to NMP for centrifugation. From left to right: Layer number, Length, Width. Increasing centrifugation speeds and decreasing size from top to bottom.

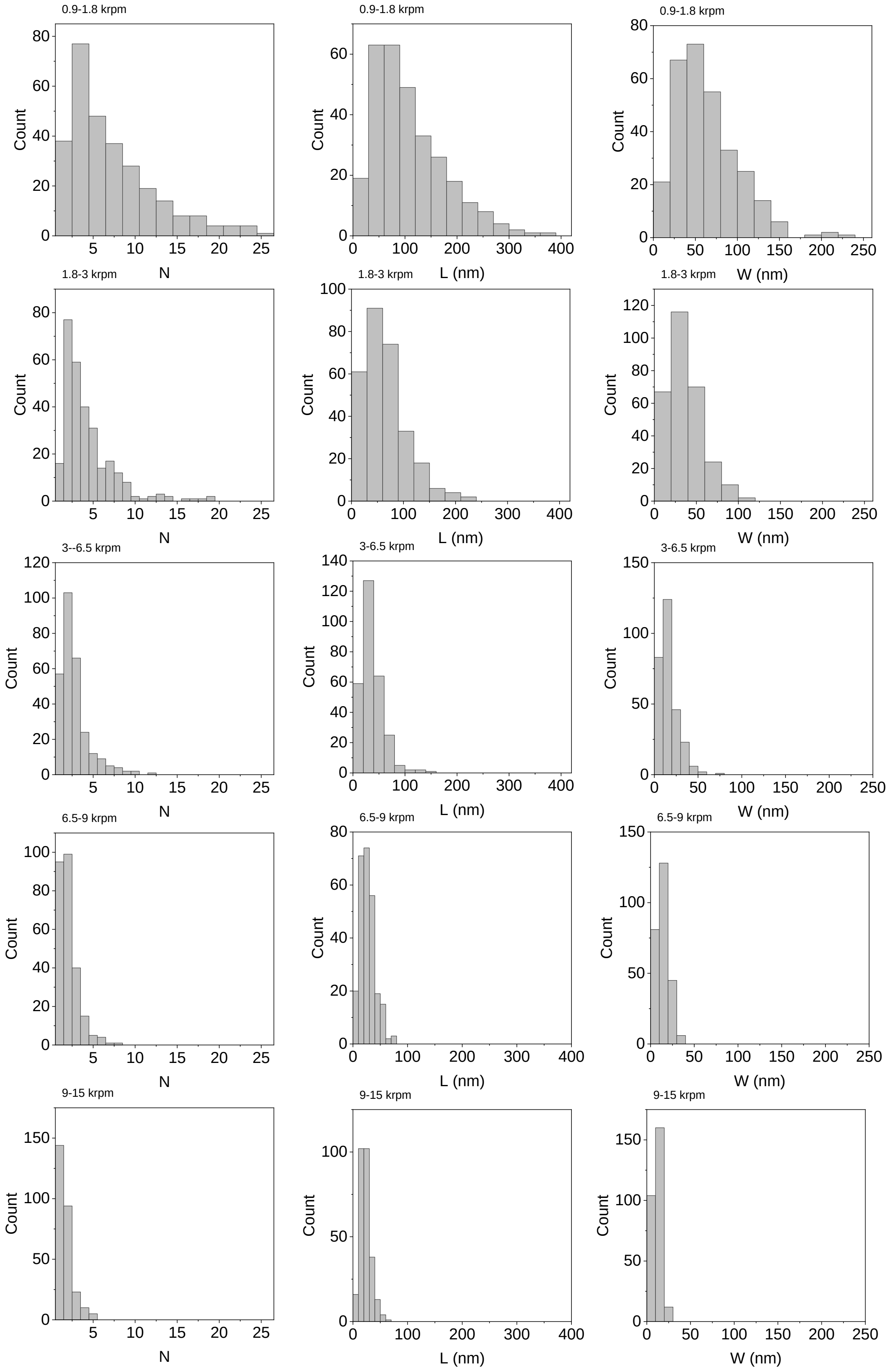

Figure S 18: Size distribution histograms of the LCC size-selected $WS_2$ produced and centrifuged in NMP. From left to right: Layer number, Length, Width. Increasing centrifugation speeds and decreasing size from top to bottom.

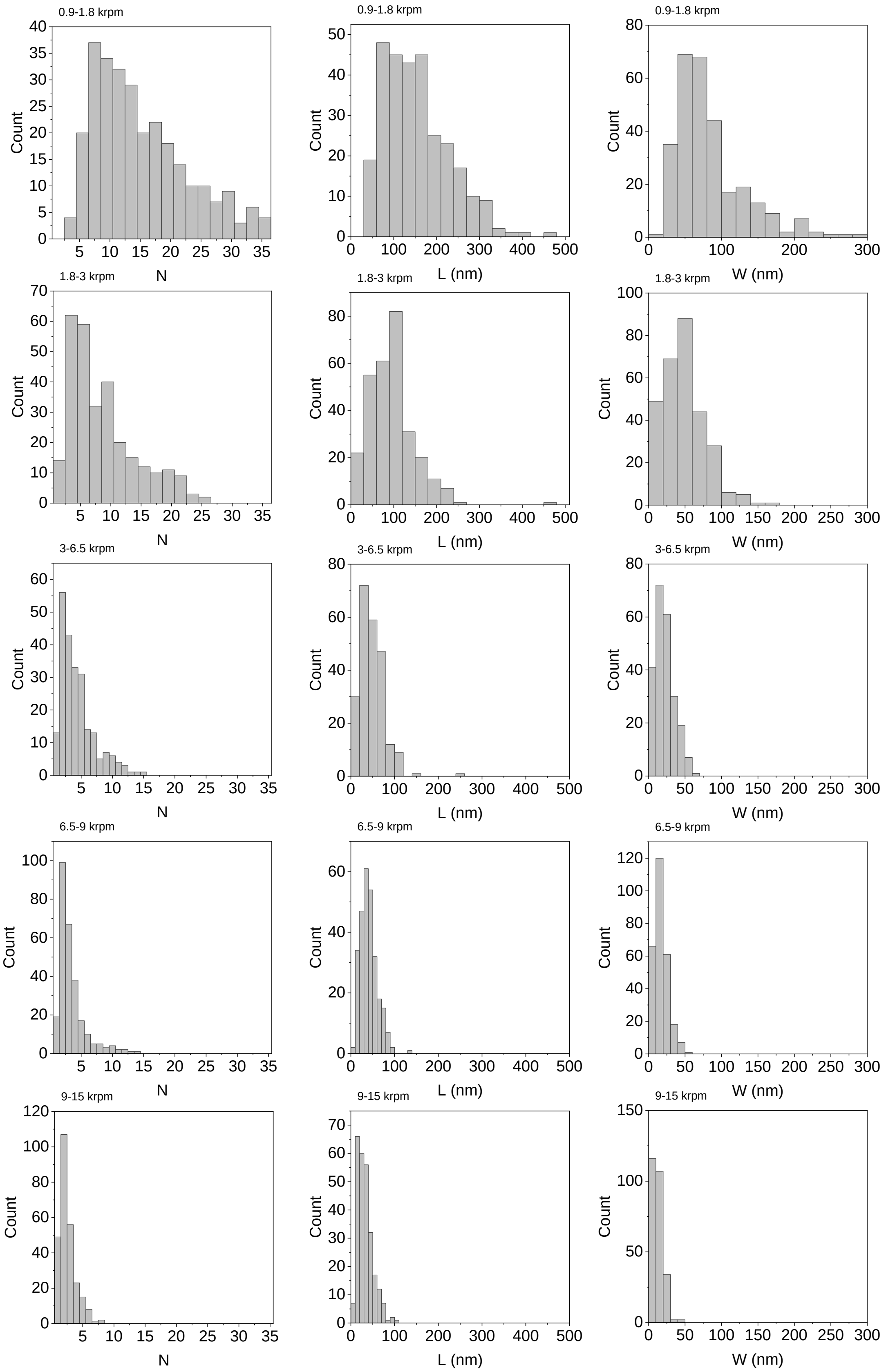

Figure S 19: Size distribution histograms of the LCC size-selected $WS_2$ produced in NMP and transferred to aqueous SC for centrifugation. From left to right: Layer number, Length, Width. Increasing centrifugation speeds and decreasing size from top to bottom.

## 5.3 *Scatter plots of lateral dimensions versus layer number*

Even though a determination of the size distribution in the stock dispersion is not accessible due to the polydispersity of the sample, we can nonetheless assess whether the four samples have a similar range of sizes and thicknesses by summing up the counts from the individual fractions and plotting the lateral dimensions of the individual nanosheets as function of layer number. These are compared in Figure S 20 approximating the area as L·W and using the square root of LW as lateral size measure.

In Figure S 20A-B, samples are compared that were exfoliated in different media, but centrifuged in the same solvent medium. The data clouds from exfoliation in NMP are vertically offset to smaller area compared to the ones from exfoliation in SC. This implies that nanosheets of a given thickness are laterally smaller when exfoliated in NMP compared to aqueous sodium cholate, i.e. that length-thickness aspect ratios slightly depend on the solvent medium used for the exfoliation. Note that this will have an impact on modelling the centrifugation as discussed in more detail below and in the main manuscript. Such behaviour was qualitatively previously observed for graphene,[6] even though the effect from different size selection in different media could not be decoupled from the effect the solvent can play in the actual exfoliation. Figure S 20C-D shows lateral size-layer number scatter plots for samples that were exfoliated in the same medium, but centrifuged in different media. Here, the data clouds perfectly overlap as expected for identical samples that were only size-selected in a slightly different way, i.e. by using a solvent with different viscosity and density.

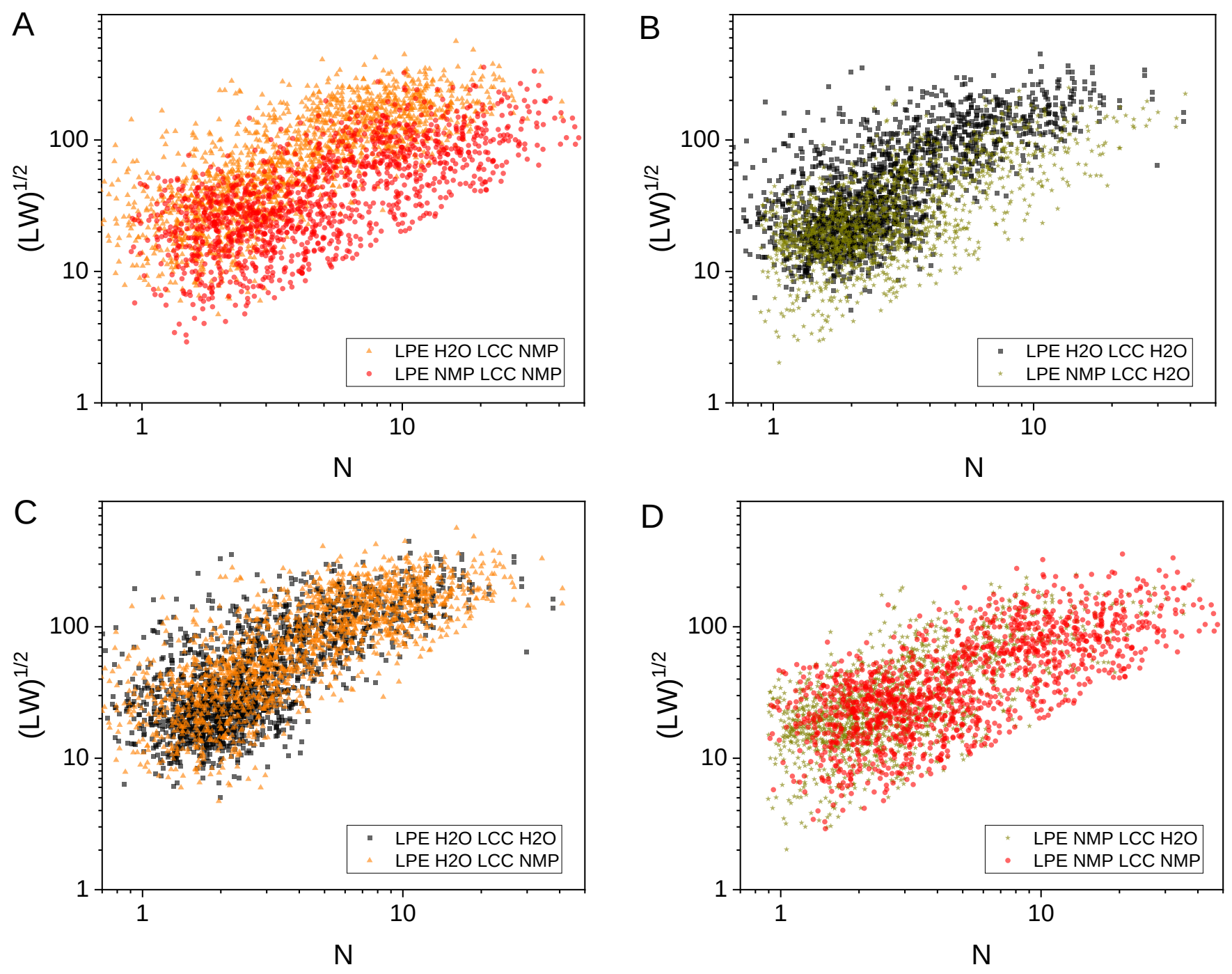

Figure S 20: Scatter plot of lateral size versus layer number for the four sample sets. Each datapoint represents an individual nanosheet measured by AFM across all size-selected fractions. A) Data from exfoliation in SC and NMP after size selection in NMP. B) Data from exfoliation in SC and NMP after size selection in SC. C) Data from exfoliation in SC after size selection in SC and NMP. D) Data from exfoliation in NMP after size selection in SC and NMP.

# 6 Detailed analysis of $WS_2$ samples produced by LCC

## 6.1 *Cut-size*

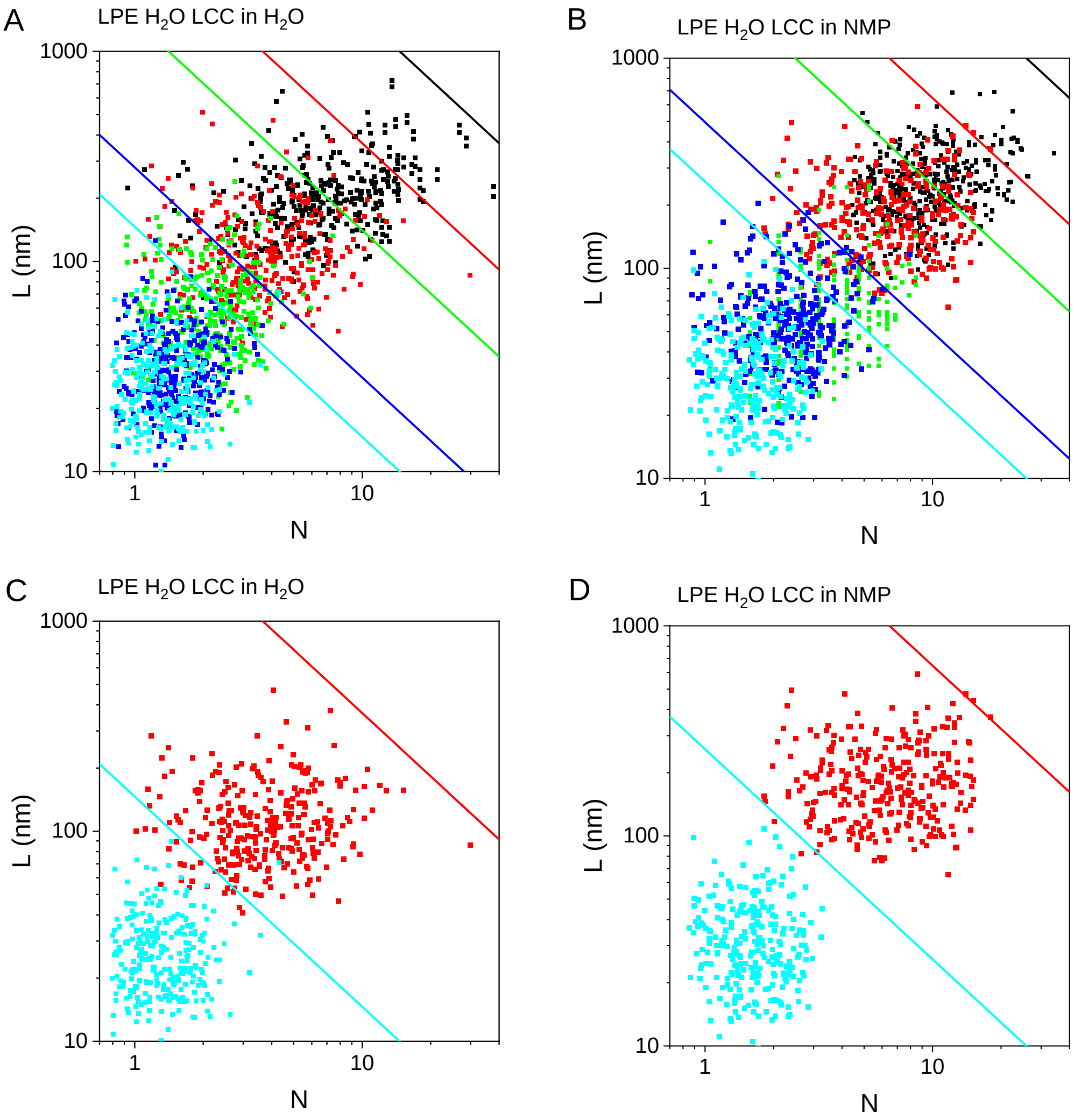

Figure S 21: Additional examples of the theoretical cut-size in scatter plots of length versus thickness. Each datapoint represents a nanosheet measured with AFM. A) $WS_2$ exfoliated in aqueous SC and size-selected in the same medium, B) $WS_2$ exfoliated in aqueous SC and transferred to NMP prior to LCC. The datapoints correspond to the data from the AFM statistics in SI section 5. C, D) Same data as in A, B) showing only two fractions for clarity (1.8-3 krpm in red and 9–15 krpm in cyan).

## 6.2 Centrifugation boundary

To derive equation S7 as estimation for the most probable nanosheet size, it was required to approximate that $\omega_i^2 t_i \approx \omega_{i-1}^2 t_{i-1}$. This has the implication, that theory can no longer distinguish between the upper and lower centrifugation boundary used to isolate certain nanosheet fractions in LCC in the experiment. The question therefore arises, which angular velocity to use to model the outcome of LCC. To decide, we took the empirical approach of testing angular velocities of the upper, lower and midpoint angular velocity and compared experiment to theory in Figure S 22. Included are theory lines at two different temperatures around the temperature that was set during centrifugation (discussion see SI section 6.3). Clearly, the best match was obtained for the upper centrifugation boundary (i.e. the one that produces the sediment that is collected). This was therefore chosen for the remainder of the study.

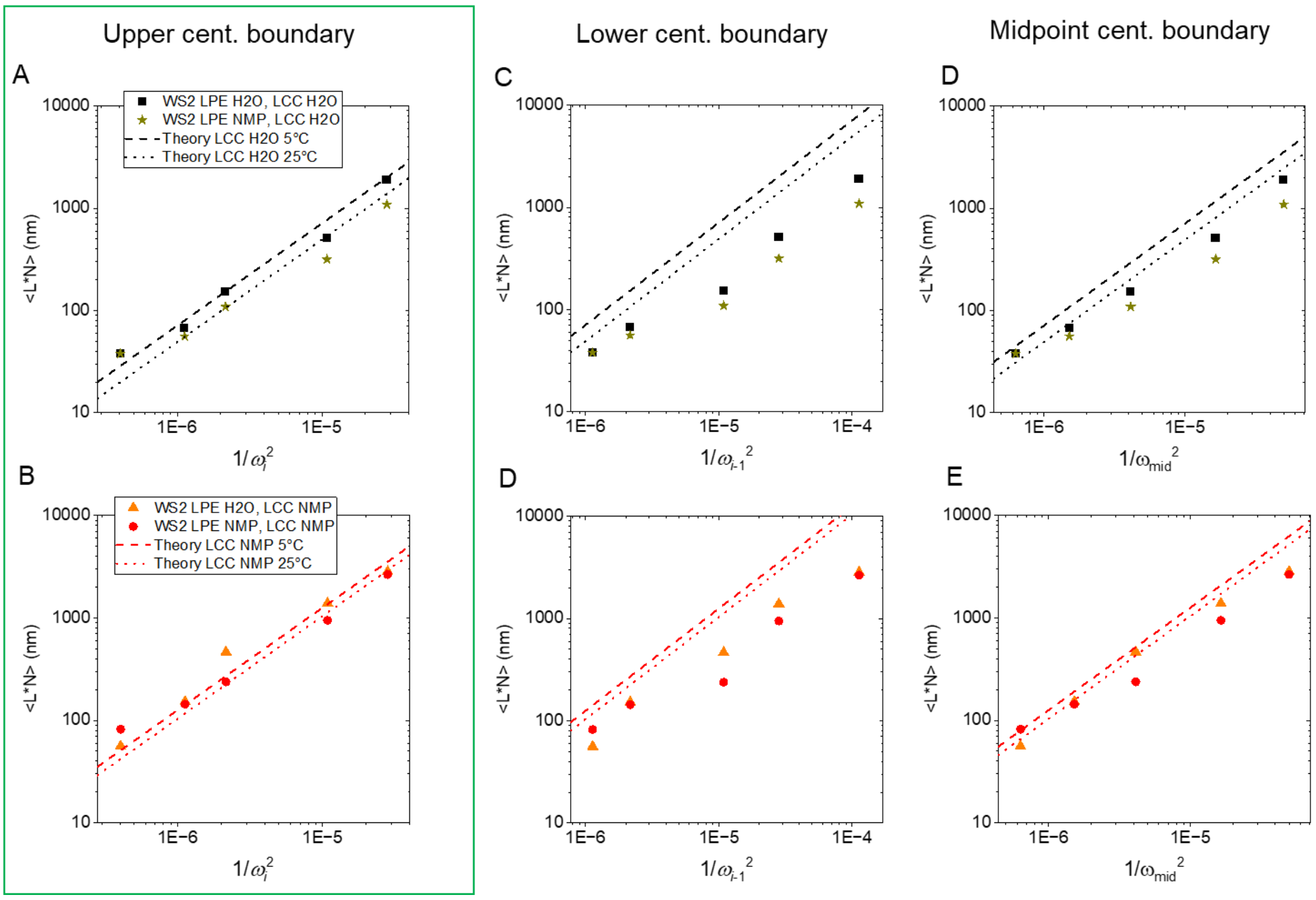

Figure S 22: Plots of experimental ⟨*LN*⟩ as function of the inverse of the square of the angular velocity following equation S7. Within the top and bottom row, respectively, the same theory lines are shown, only the experimental data points were shifted using the upper (A, B) and lower (B,C) centrifugation boundary of the experiment, as well as the midpoint between the two (D, F). A, C, D) Data of $WS_2$ exfoliated in both aqueous SC and NMP and centrifuged in aqueous SC. C, D, E) Data of $WS_2$ exfoliated in both aqueous SC and NMP and centrifuged in NMP.

## 6.3 *Discussion of experimental uncertainties*

When experimentally performing centrifugation, it is sometimes difficult to control all parameters completely:

1) Even when centrifuges are equipped with temperature control, we found that the actual sample temperature can vary. For example, when setting the temperature of the centrifuge to 10°C, the samples will initially be at room temperature and after the centrifugation experiment, the temperature in the sample often goes down to 5°C. This results in differences in solvent viscosity.
2) Even when initially using identical filling heights in the experiment of ~20 mL per 50 mL vial, the volume will decrease as the samples progresses through the cascade, as some of the volume (~1-3 mL) is removed from the sample as sediment that is re-dispersed. This results in variations in $R_1$.
3) When decanting the sample, it depends on material mass in the sediment at which position in the vial supernatant and sediment can be clearly separated. This results in variations in $R_2$.

To illustrate how much such experimental uncertainties, result in deviations between theory and experiment, we plot different scenarios in Figure S 23. Overall, the experimental error can well account for the scatter in the data that is observed.

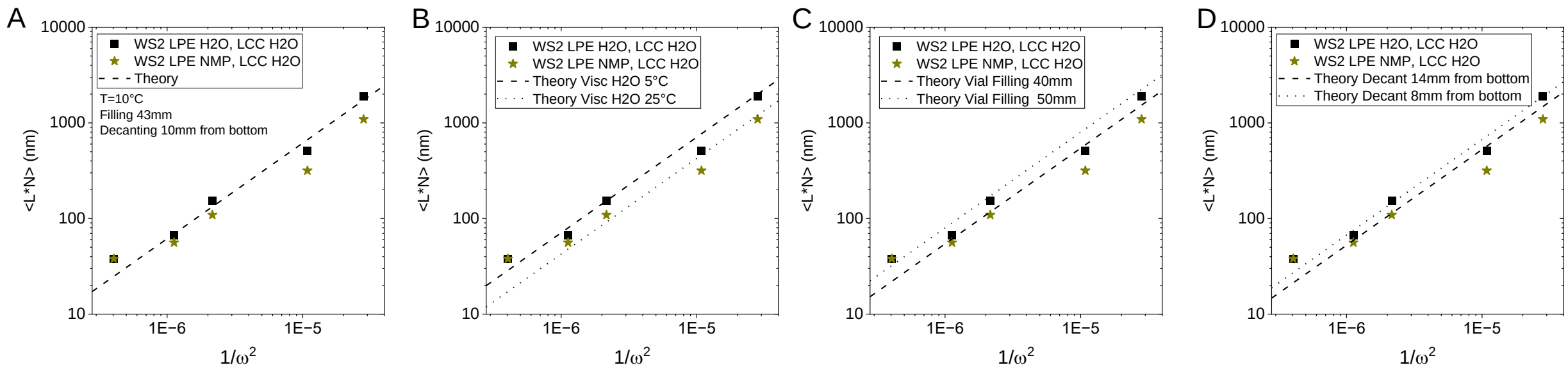

Figure S 23: Plots of experimental ⟨*LN*⟩ as function of the inverse of the square of the angular velocity following equation S7 using $WS_2$ exfoliated in both aqueous SC and NMP and centrifuged in aqueous SC. Experimental datapoints are identical, but theory lines are modified according to potential sources of experimental uncertainties. A) Average parameters of viscosity at 10°C, filling height and decanting setting $R_1$ and $R_2$ used throughout the manuscript (for comparison). B) Theory lines for different viscosities of water at 5°C and 25°C. C) Theory lines accounting for variation in the filling height. D) Theory lines accounting for variations in the decanting.

### 6.4 *Dealing with different length-thickness aspect ratio distributions*

One possibility to predict $\langle N \rangle$ and $\langle L \rangle$ independently (rather than $\langle LN \rangle$) is to use a constant length-thickness aspect ratio determined from AFM statistics for a given material exfoliated with LPE. Note that the aspect ratio is mostly governed by the in plane to out of plane binding strength of the parent crystal in (sonication-assisted) LPE.[4] However, the exfoliation medium can also have an impact, as for example shown by Read et al. and indicated by the data shown in Figure S 20A,B and Figure S 24.[24]

To test to which extend small changes in average aspect ratios across different exfoliation conditions result in a deviation between theory and experiment, we analysed our four $WS_2$ datasets in more detail using the mean aspect ratio approximation (accounting for differences between the two exfoliation media: aqueous SC and NMP), as well as the more accurate model described in SI section 2.7 which uses the actual relationship between $\langle L \rangle$ and $\langle N \rangle$. A subset of the data is shown in figure 3 in the main manuscript. All $WS_2$ datasets are displayed in Figure S 25, where the experimental datapoints are overlaid with theoretical predictions based on the mean aspect ratio approximation. The same experimental data is shown in Figure S 26, but this time overlaid with the theoretical prediction using the experimental $\langle L \rangle$-$\langle N \rangle$ relationships resulting in the input values of $D_{ML}$ and $\beta$ for equations S10 and S11.

In summary, we find that it can be important to go beyond the mean aspect ratio approximation for sample sets where the exponent $\beta$ relating $\langle L \rangle$ and $\langle N \rangle$ and the characteristic monolayer size, $D_{ML}$ vary across samples. Both parameters can be determined experimentally, when fitting $\langle L \rangle$ vs $\langle N \rangle$ as a power law, see Figure S 26 A,B,E,F. While the data clearly shows that changes in $\beta$ and $D_{ML}$ can result in deviations from the expected behaviour that $\langle L \rangle$ and $\langle N \rangle$ scale linearly with $1/\omega_i$, it remains unclear which factors control $\beta$ and $D_{ML}$. Further, it should be noted that the current theoretical framework does not allow for a decoupled prediction of $\langle N \rangle$ and $\langle L \rangle$ for samples where length-thickness aspect ratios are not constant such as electrochemically-exfoliated nanosheets. In such cases, only the cut size and $\langle LN \rangle_{MP}$ can be estimated.

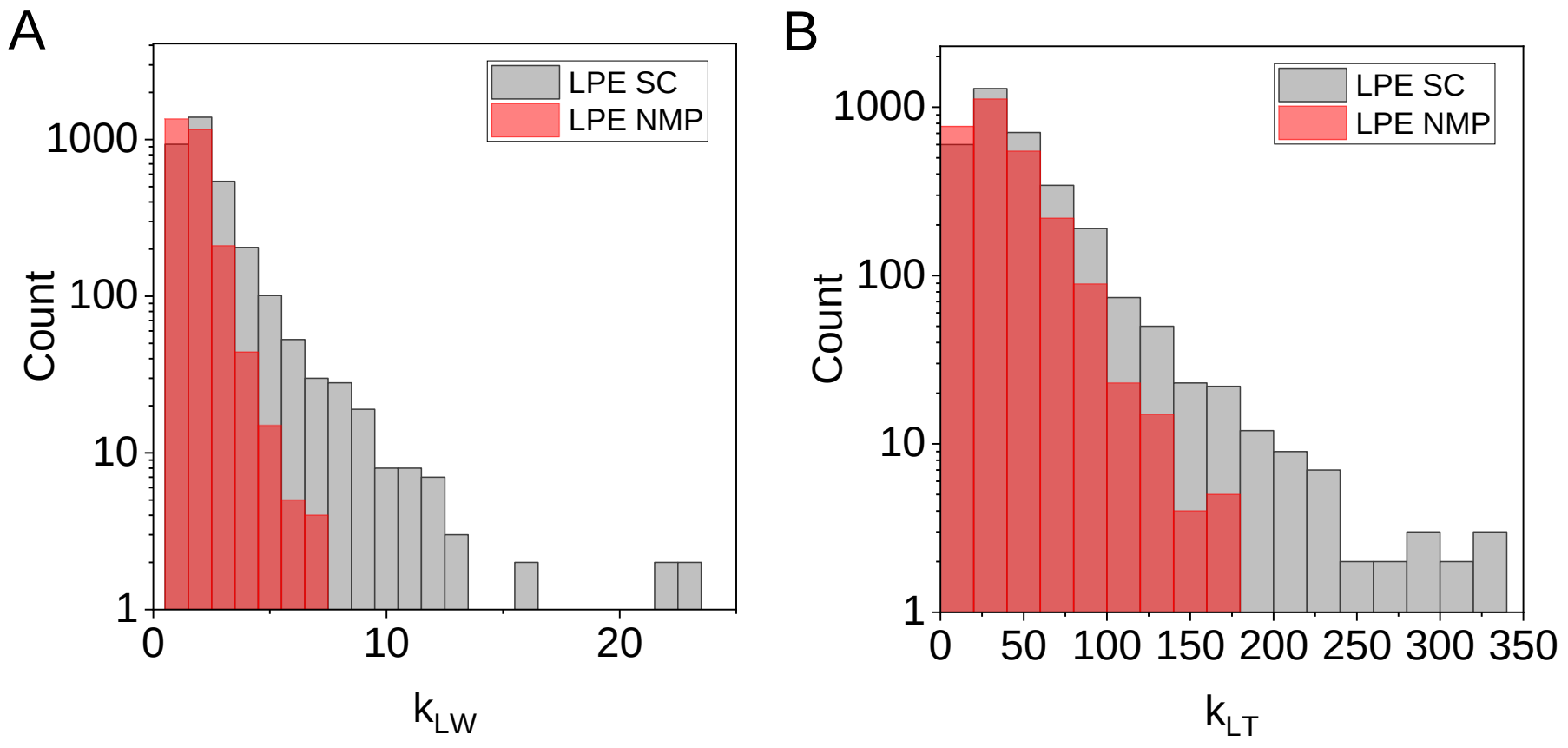

Figure S 24: Aspect ratio distributions of nanosheets measured with AFM of $WS_2$ exfoliated in aqueous SC and NMP. A) Length-width aspect ratio, B) Length-thickness aspect ratio. Both aspect ratio distributions are narrower from exfoliation in NMP compared to aqueous SC. The arithmetic mean $k_{Lt}$ from exfoliation in aqueous SC is 46, while it is 35 in NMP. The average over all nanosheets is 40 which is for example used as input parameter for $WS_2$ in Figure 2 in the main manuscript.

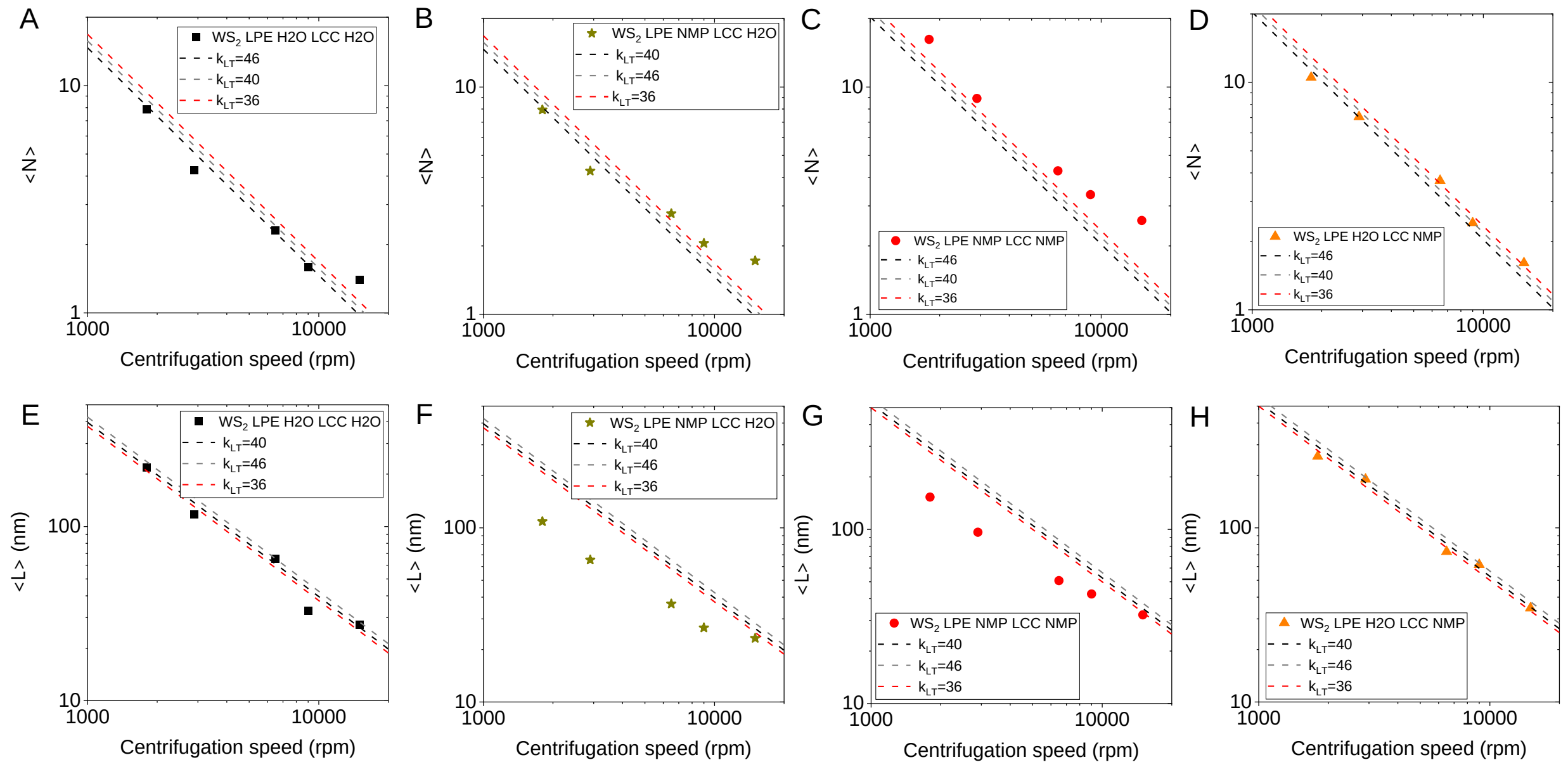

Figure S 25: Plots of ⟨*N*⟩ (A-D) and ⟨*L*⟩ as function of angular velocity of the four $WS_2$ LCC datasets in comparison to theory using the mean aspect ratio approximation. A, E) $WS_2$ exfoliated and centrifuged in aqueous SC. B,F) $WS_2$ exfoliated in NMP and centrifuged in aqueous SC. C,G) $WS_2$ exfoliated and centrifuged in aqueous NMP, D,H) $WS_2$ exfoliated in aqueous SC and centrifuged in NMP. The mean aspect ratio approximation (equations S8,9) works well to predict ⟨*N*⟩ in all cases, but only describes ⟨*L*⟩ sufficiently for exfoliation in aqueous SC.

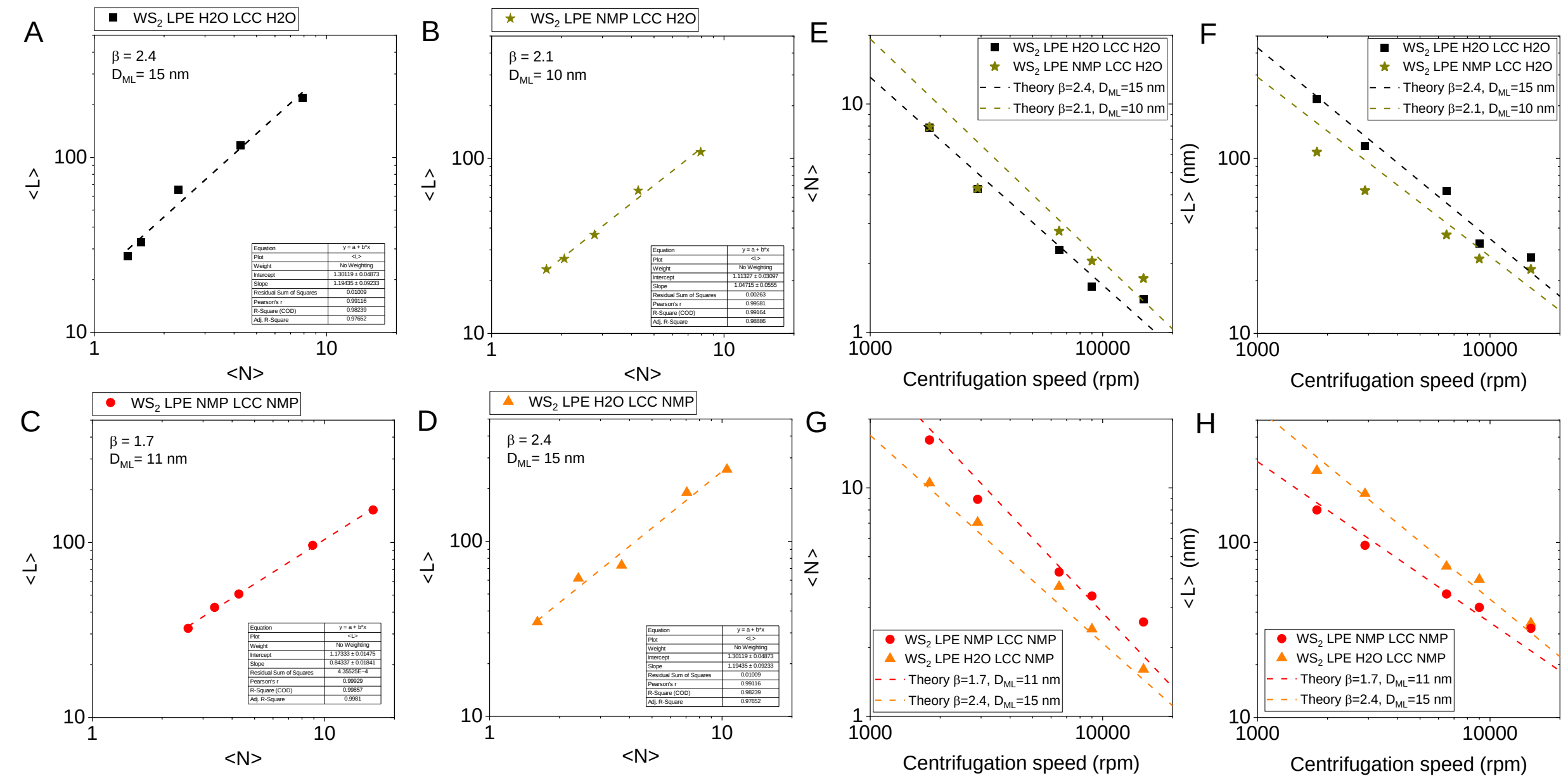

Figure S 26: A-D) Plots of ⟨*L*⟩ versus ⟨*N*⟩ from AFM statistics of the four $WS_2$ datasets on a double logarithmic scale including fits to a power-law to extract the exponent relating ⟨*L*⟩ and ⟨*N*⟩ ($\beta$) and intercept with 1 layer corresponding to the characteristic monolayer length ($D_{ML}$). These parameters were used as input for the modelling data (dashed lines) in E-H) according to equations S10,11. E, G) Plots of ⟨*N*⟩ versus angular velocity and F,H) ⟨*L*⟩ versus angular velocity of E,F) the two samples sets centrifuged in aqueous SC, but exfoliated in different

media and G,H) the two samples sets centrifuged in NMP, but exfoliated in different media. In spite of scatter in the experimental data, the data is better described using the model based on equation S10 and S11 compared to the mean aspect ratio approximation shown in Figure S 25.

# 7 Analysis of other material and solvent systems from literature

## *7.1 Overview*

To assess the robustness of the centrifugation theory, we applied it to previously produced data in our laboratories. In total, we modelled the LCC data from 8 materials, in three solvent systems and in total 12 different cascades. Table S1 gives an overview of the relevant material parameters, table S2 of the solvent parameters. The (temperature dependent) viscosity of CHP was determined through rheological measurements described in section S7.2. In Figure S 27 we provide an overview of the $\langle LN \rangle$ data as function of the inverse of the square of the angular velocity following equation S7. A more detailed analysis based on equations S7-11 is presented in the following subchapters of the SI material by material.

Table S 1: Relevant parameters of the materials used for centrifugation. Nanosheet aspect ratios were determined from AFM statistics and averaged over all nanosheets counted (in all fractions analysed).

| Material | Density (kg m$^{-3}$) | $d_0$ (nm) | $k_{lw}$ | $k_{Lt}$ | Solvent† |
|---|---|---|---|---|---|
| $WS_2$ | 7500 | 0.63 | 1.7 | 46 (SC), 35(NMP) | SC and NMP |
| Graphene | 2260 | 0.35 | 2.4 | 180(SC), 125(NMP) | SC and NMP |
| $MoS_2$ | 5060 | 0.62 | 1.8 | 49 | SC |
| hBN | 2100 | 0.34 | 1.5 | 140 | SC |
| GaS | 3860 | 0.75 | 1.8 | 22.9 | NMP |
| $Cu(OH)_2$ | 3370 | 0.5 | 4.5 | 21 | SC |
| $NiPS_3$ | 3180 | 0.66 | 1.6 | 35 | CHP |
| $CrTe_3$ | 4700 | 1.1 | 1.6 | 14 | CHP |

† SC = 2 g L$^{-1}$ sodium cholate surfactant in water; NMP = N-Methyl-2-pyrrolidone, exfoliation under ambient conditions; CHP = N-Cyclohexyl-2-pyrrolidone solvent, exfoliation under protective inert atmosphere

Table S 2: Relevant parameters of the solvents used for centrifugation

| Solvent | Density (kg m$^{-3}$) | Viscosity (5$^{o}$C) / kg ms$^{-1}$ | Viscosity (10$^{o}$C) / kg ms$^{-1}$ | Viscosity (25$^{o}$C) / kg ms$^{-1}$ |
|---|---|---|---|---|
| Water | 1000 | 0.0015 | 0.0013 | 0.0009 |
| NMP | 1030 | 0.0024 | 0.0023 | 0.0019 |
| CHP | 1000 | 0.027 | 0.021 | 0.012 (at 20 $^{o}$C) |

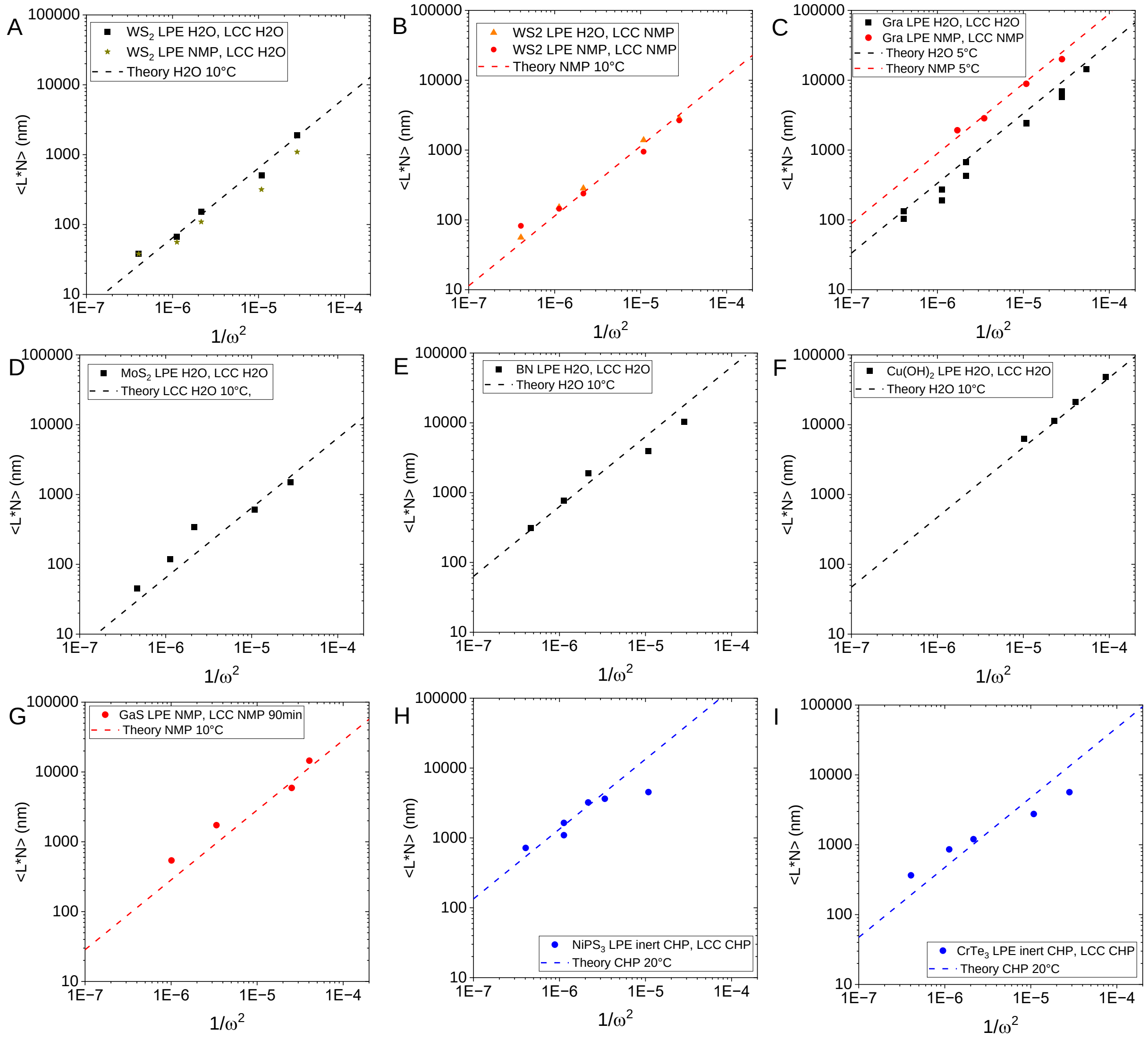

Figure S 27: Plots of experimental ⟨*LN*⟩ as function of the inverse of the square of the angular velocity following equation S7 for all LCC data analysed in this work. Data is colour coded for solvent used. Dashed lines are the theory prediction of the most probable $\langle LN \rangle_{MP}$ (slightly adjusted to experimental conditions such as filling height in the vial or temperature during centrifugation). A, B) $WS_2$ (original data). C) Graphene[6] D) $MoS_2$[10] E) h-BN[16] F) $Cu(OH)_2$[12] G) GaS[12] H) $NiPS_3$ inert exfoliation[18] I) $CrTe_3$[17]

## *7.2 Rheological data for CHP to assess viscosity*

The viscosity-temperature dependence of the CHP solvent was determined in a stress-controlled MCR 302 Rheometer (Anton Paar, Austria) coupled with a Peltier temperature control unit set. The steady shear flow properties were measured at a shear rate of 0.1 to 100 $s^{-1}$ at different temperatures (5, 10, 15, 20, and 25 °C) using a 50 mm parallel plate geometry.. This resulted in a linear relationship between shear stress and shear rate which allowed to calculate the dynamic viscosity. This is plotted as function of shear rate in Figure S28A showing a constant viscosity over the range of shear rates used characteristic for a Newtonian fluid. The average dynamic viscosity is plotted as function of temperature in figure S28B.

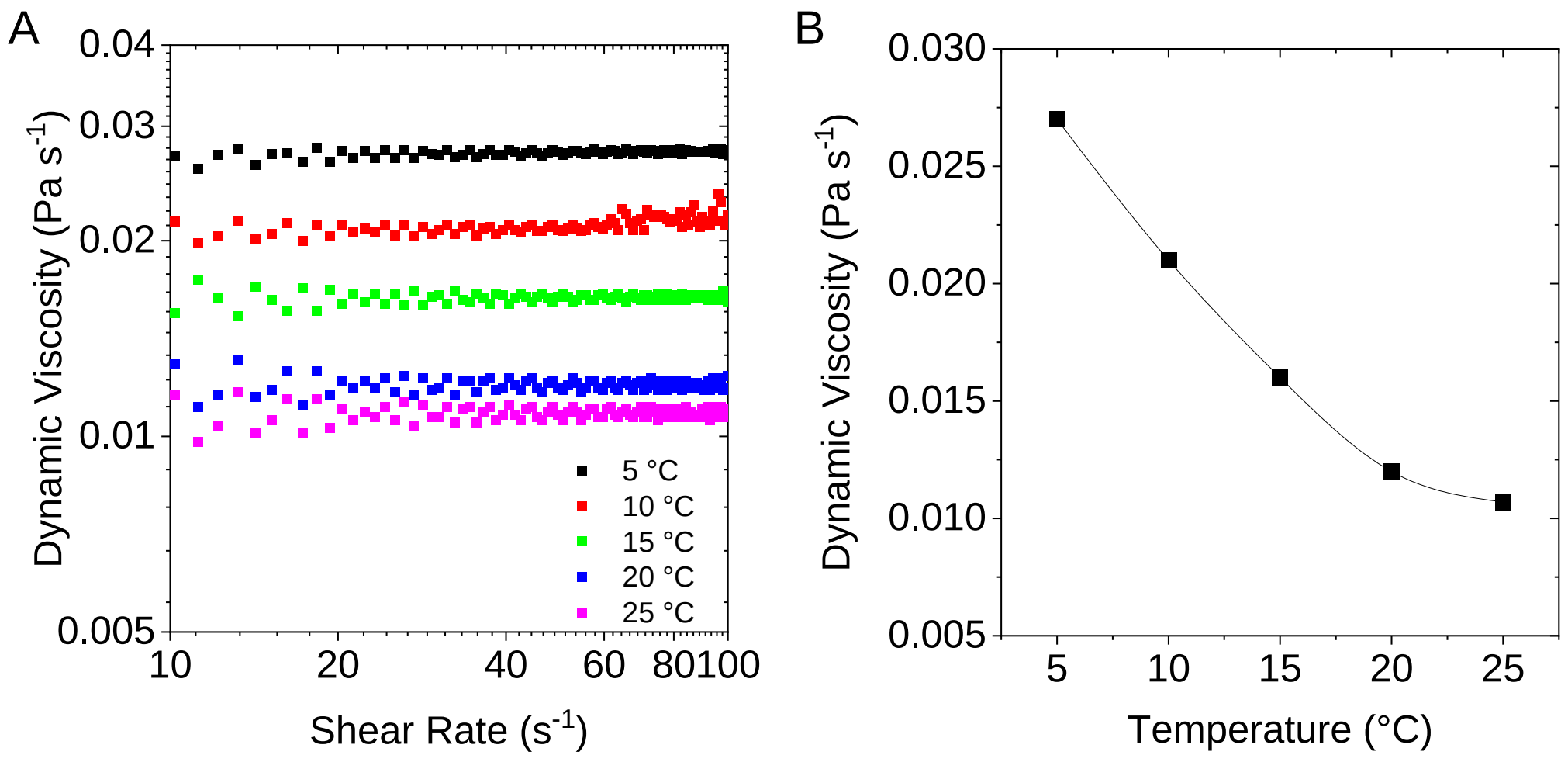

Figure S 28: Rheology data for CHP. A) Dynamic viscosity as function of shear rate measured at different temperatures. B) Average dynamic viscosity as function of temperature.

## 7.3 Graphene (SC and NMP)

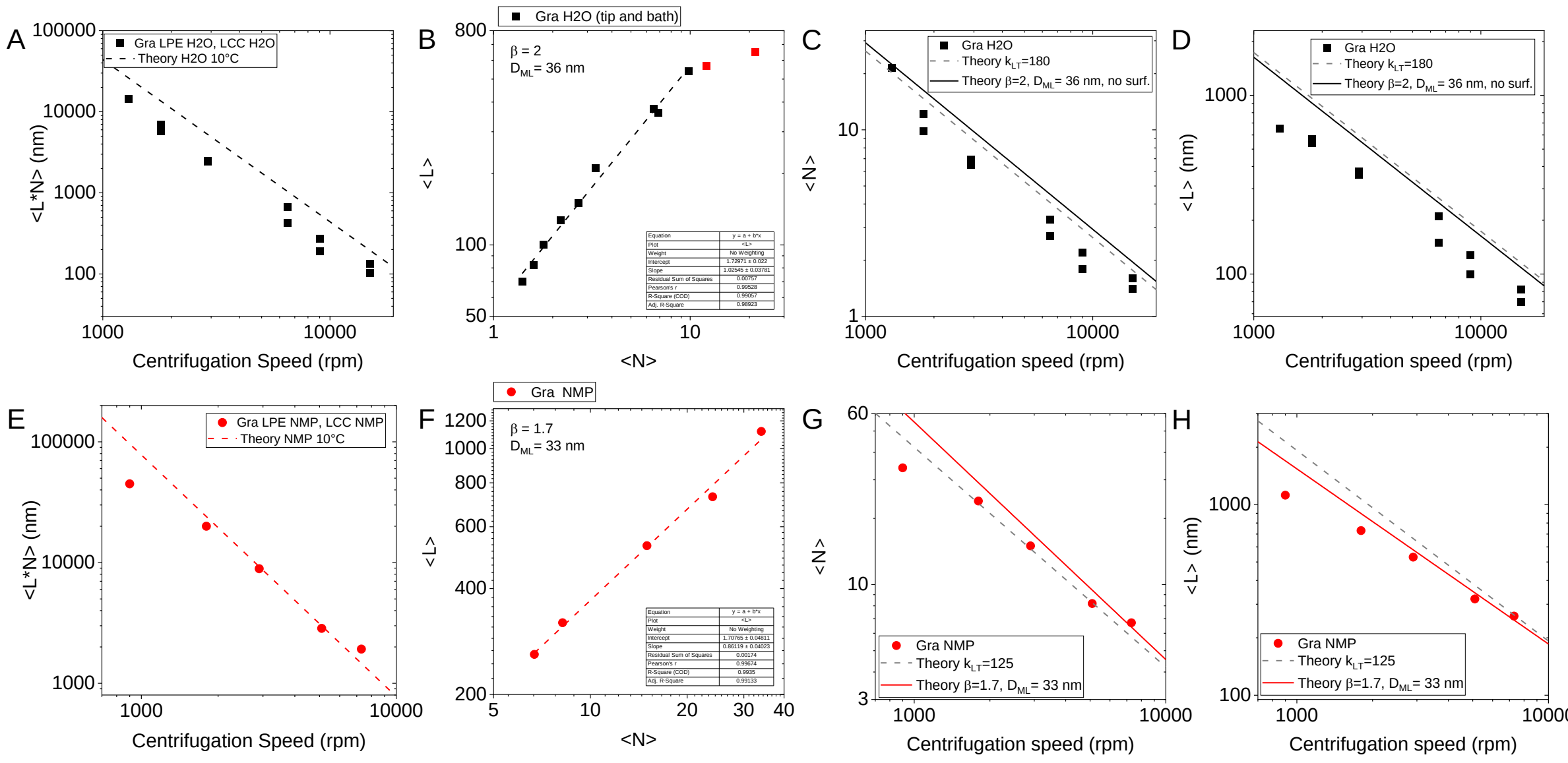

Figure S 29: Comparison experiment and theory in LCC of graphene[6] exfoliated and centrifuged in A-D) aqueous sodium cholate, E-H) NMP. A, E) Plots of experimental ⟨*LN*⟩ versus angular velocity. The dashed line is the most probable nanosheet size from theory according to equation S7. The graphene data in SC is not well described by theory. We attribute this to the neglection of the surfactant coating in the theory which can result in a deviation for low density materials such as graphene (Compare Figure 1G main manuscript). B, F) Plots of ⟨*L*⟩ versus ⟨*N*⟩ from AFM statistics of the two graphene datasets on a double logarithmic scale including fits to a power-law to extract the exponent relating ⟨*L*⟩ and ⟨*N*⟩ ($\beta$) and intercept with 1 layer corresponding to the characteristic monolayer length ($D_{ML}$). These parameters were used as input for the modelling data (solid lines) in C-H). Note that the two datapoints corresponding to the largest nanosheets are excluded from the fit, as they begin to deviate from the power-law scaling as previously observed and attributed to a twinning mechanism in the early stages of exfoliation.[4] C, G) Plots of ⟨*N*⟩ and D,H) ⟨*L*⟩ as function of angular velocity. The dashed lines are the mean aspect ratio approximation according to equation S8,9 and the solid line use the actual relationship of length and thickness as input according to equation S10,11

## 7.4 $MoS_2$ (SC)

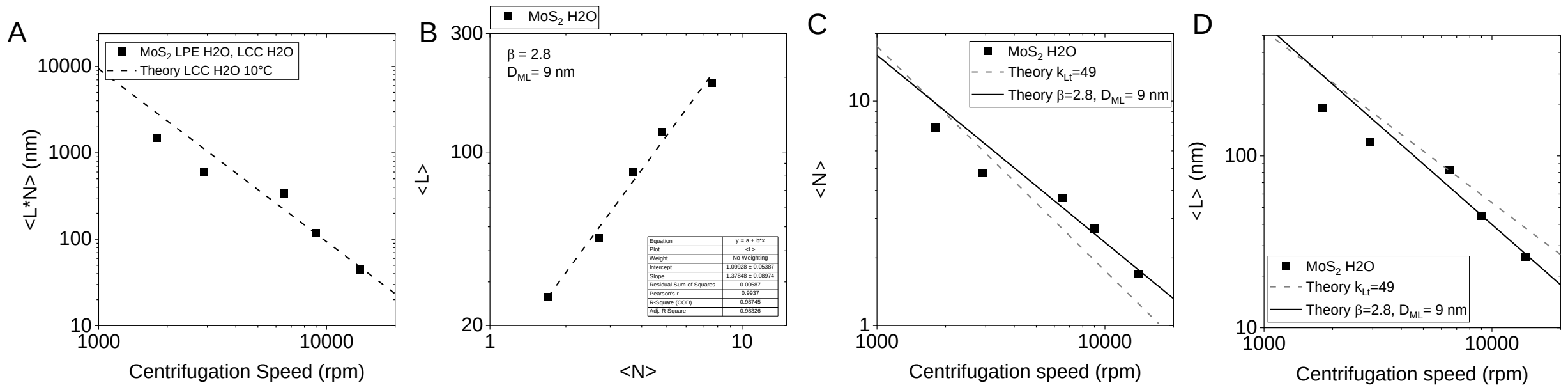

Figure S 30: Comparison experiment and theory in LCC of $MoS_2$[10] exfoliated and centrifuged in aqueous sodium cholate. A) Plot of experimental ⟨*LN*⟩ versus angular velocity. The dashed line is the most probable nanosheet size from theory according to equation S7. B) Plot of ⟨*L*⟩ versus ⟨*N*⟩ from AFM statistics on a double logarithmic scale including fits to a power-law to extract the exponent relating ⟨*L*⟩ and ⟨*N*⟩ ($\beta$) and intercept with 1 layer corresponding to the characteristic monolayer length ($D_{ML}$). These parameters were used as input for the modelling data (solid lines) in C-D. C) Plot of ⟨*N*⟩ and D) ⟨*L*⟩ as function of angular velocity. The dashed lines are the mean aspect ratio approximation according to equation S8,9 and the solid line use the actual relationship of length and thickness as input according to equation S10,11.

## 7.5 h-BN (SC)

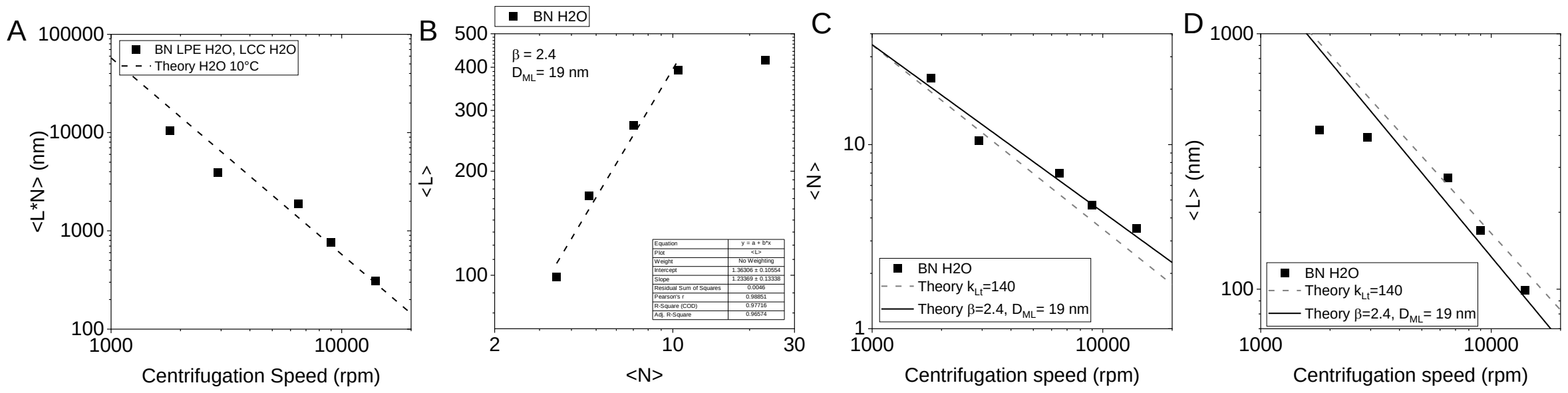

Figure S 31: Comparison experiment and theory in LCC of h-BN[16] exfoliated and centrifuged in aqueous sodium cholate. A) Plot of experimental ⟨*LN*⟩ versus angular velocity. The dashed line is the most probable nanosheet size from theory according to equation S7. B) Plot of ⟨*L*⟩ versus ⟨*N*⟩ from AFM statistics on a double logarithmic scale including fits to a power-law to extract the exponent relating ⟨*L*⟩ and ⟨*N*⟩ ($\beta$) and intercept with 1 layer corresponding to the characteristic monolayer length ($D_{ML}$). These parameters were used as input for the modelling data (solid lines) in C-D). Note that the datapoint corresponding to the largest nanosheets is excluded from the fit, as it begins to deviate from the power-law scaling as previously observed and attributed to a twinning mechanism in the early stages of exfoliation.[4] C) Plot of ⟨*N*⟩ and D) ⟨*L*⟩ as function of angular velocity. The dashed lines are the mean aspect ratio approximation according to equation S8,9 and the solid line use the actual relationship of length and thickness as input according to equation S10,11

## 7.6 GaS (NMP)

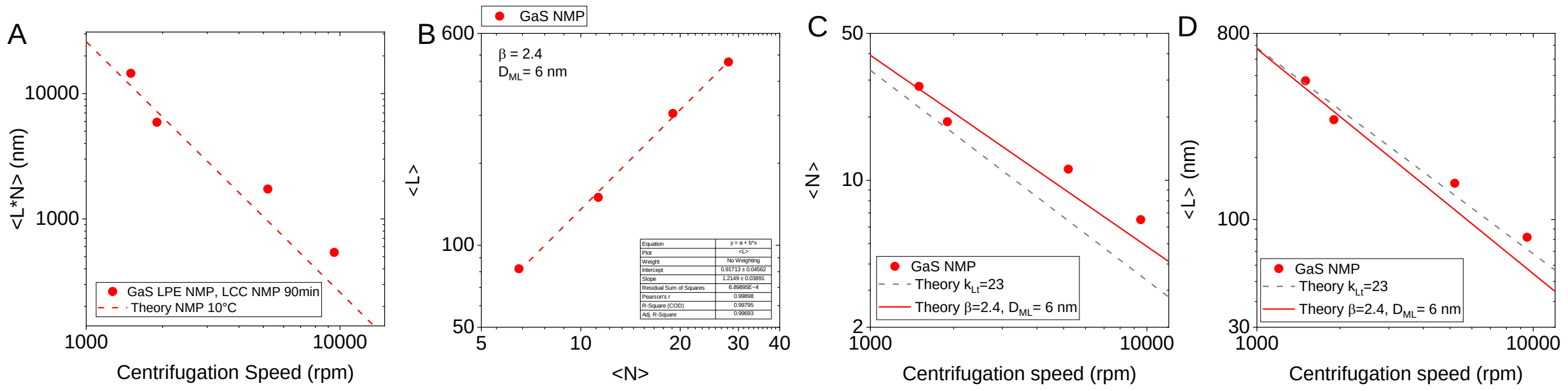

Figure S 32: Comparison experiment and theory in LCC of GaS[6] exfoliated and centrifuged in NMP. A) Plot of experimental ⟨*LN*⟩ versus angular velocity. The dashed line is the most probable nanosheet size from theory according to equation S7. B) Plot of ⟨*L*⟩ versus ⟨*N*⟩ from AFM statistics on a double logarithmic scale including fits to a power-law to extract the exponent relating ⟨*L*⟩ and ⟨*N*⟩ ($\beta$) and intercept with 1 layer corresponding to the characteristic monolayer length ($D_{ML}$). These parameters were used as input for the modelling data (solid lines) in C-D. C) Plot of ⟨*N*⟩ and D) ⟨*L*⟩ as function of angular velocity. The dashed lines are the mean aspect ratio approximation according to equation S8,9 and the solid line use the actual relationship of length and thickness as input according to equation S10,11

## 7.7 $Cu(OH)_2$ (SC)

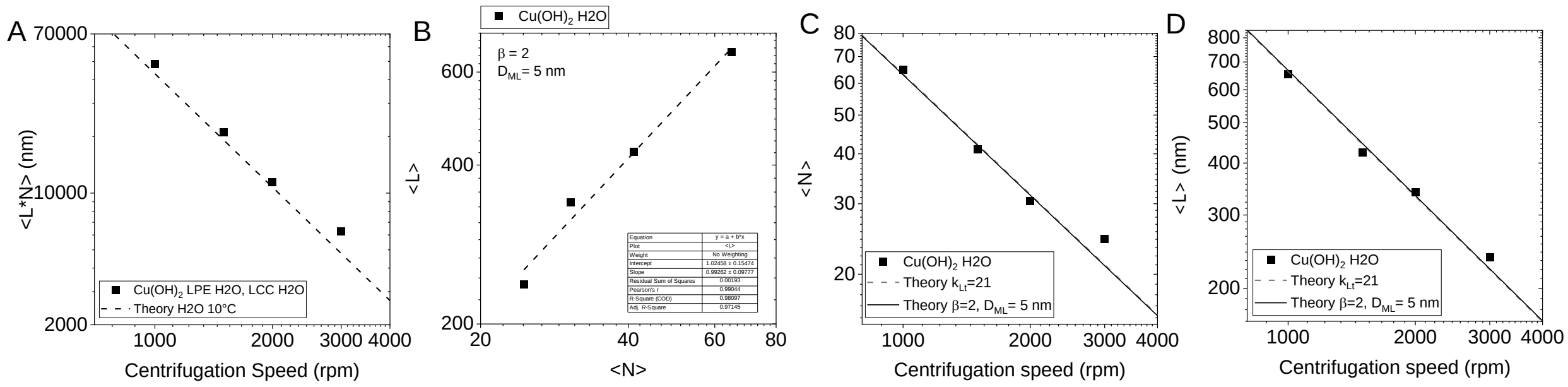

Figure S 33: Comparison experiment and theory in LCC of $Cu(OH)_2$ [12] exfoliated and centrifuged in aqueous sodium cholate. A) Plot of experimental ⟨*LN*⟩ versus angular velocity. The dashed line is the most probable nanosheet size from theory according to equation S7. B) Plot of ⟨*L*⟩ versus ⟨*N*⟩ from AFM statistics on a double logarithmic scale including fits to a power-law to extract the exponent relating ⟨*L*⟩ and ⟨*N*⟩ ($\beta$) and intercept with 1 layer corresponding to the characteristic monolayer length ($D_{ML}$). These parameters were used as input for the modelling data (solid lines) in C-D. C) Plot of ⟨*N*⟩ and D) ⟨*L*⟩ as function of angular velocity. The dashed lines are the mean aspect ratio approximation according to equation S8,9 and the solid line use the actual relationship of length and thickness as input according to equation S10,11

## 7.8 NiPS$_3$ (CHP)

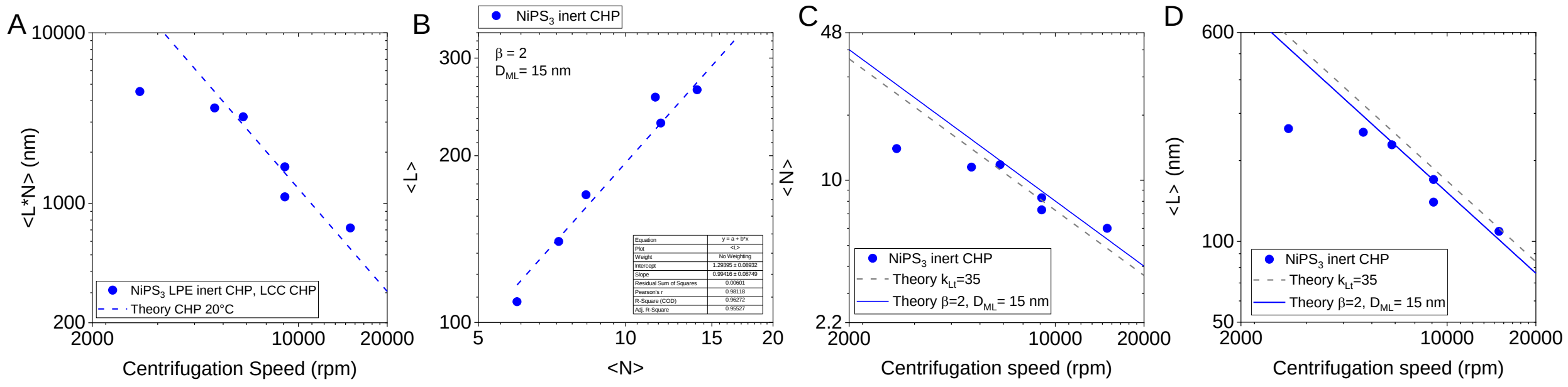

Figure S 34: Comparison experiment and theory in LCC of NiPS$_2$ [18] exfoliated and centrifuged in CHP under inert conditions. A) Plot of experimental ⟨*LN*⟩ versus angular velocity. The dashed line is the most probable nanosheet size from theory according to equation S7. B) Plot of ⟨*L*⟩ versus ⟨*N*⟩ from AFM statistics on a double logarithmic scale including fits to a power-law to extract the exponent relating ⟨*L*⟩ and ⟨*N*⟩ ($\beta$) and intercept with 1 layer corresponding to the characteristic monolayer length ($D_{ML}$). These parameters were used as input for the modelling data (solid lines) in C-D. C) Plot of ⟨*N*⟩ and D) ⟨*L*⟩ as function of angular velocity. The dashed lines are the mean aspect ratio approximation according to equation S8,9 and the solid line use the actual relationship of length and thickness as input according to equation S10,11

## 7.9 CrTe$_3$ (CHP)

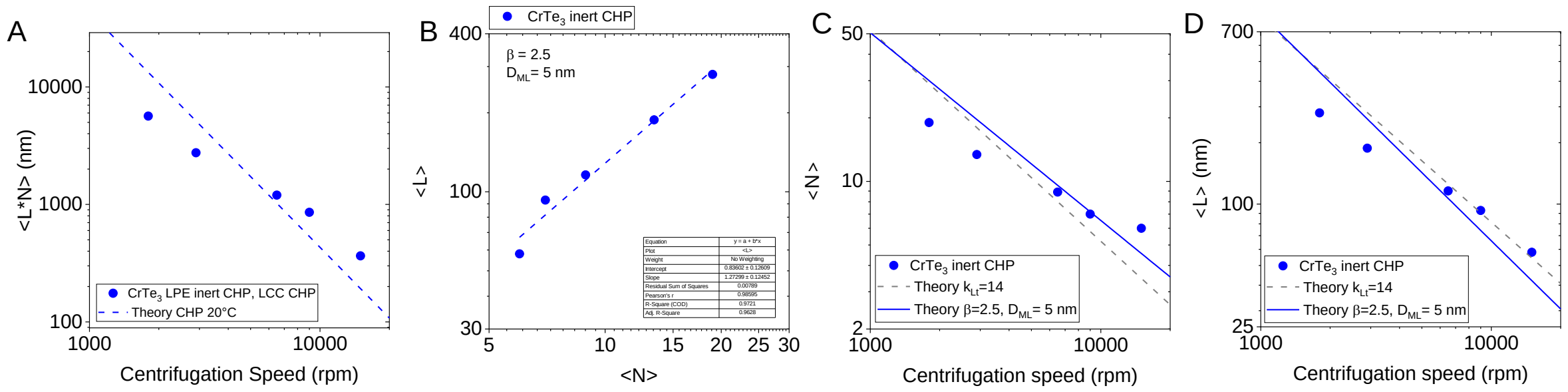

Figure S 35: Comparison experiment and theory in LCC of CrTe$_3$ [17] exfoliated and centrifuged in CHP under inert conditions. A) Plot of experimental ⟨*LN*⟩ versus angular velocity. The dashed line is the most probable nanosheet size from theory according to equation S7. B) Plot of ⟨*L*⟩ versus ⟨*N*⟩ from AFM statistics on a double logarithmic scale including fits to a power-law to extract the exponent relating ⟨*L*⟩ and ⟨*N*⟩ ($\beta$) and intercept with 1 layer corresponding to the characteristic monolayer length ($D_{ML}$). These parameters were used as input for the modelling data (solid lines) in C-D. C) Plot of ⟨*N*⟩ and D) ⟨*L*⟩ as function of angular velocity. The dashed lines are the mean aspect ratio approximation according to equation S8,9 and the solid line use the actual relationship of length and thickness as input according to equation S10,11 .